\documentclass{emulateapj}

\usepackage{amsmath}

\shorttitle{}
\shortauthors{}

\begin{document}

\title{BETHE-Hydro: An Arbitrary Lagrangian-Eulerian
  Multi-dimensional Hydrodynamics Code for Astrophysical Simulations}

\author{Jeremiah W. Murphy\altaffilmark{1,2}}
\author{Adam Burrows\altaffilmark{1}}

\altaffiltext{1}{Steward Observatory, The University of Arizona,
  Tucson, AZ 85721; jmurphy@as.arizona.edu, burrows@as.arizona.edu}
\altaffiltext{2}{JINA Fellow.}

\begin{abstract}
In this paper, we describe a new hydrodynamics code for 1D and 2D
astrophysical simulations, BETHE-hydro, that uses time-dependent,
arbitrary, unstructured grids. The core of the hydrodynamics algorithm
is an arbitrary Lagrangian-Eulerian (ALE) approach, in which the
gradient and divergence operators are made compatible using the
support-operator method.  We present 1D and 2D gravity solvers that
are finite differenced using the support-operator technique, and the
resulting system of linear equations are solved using the tridiagonal
method for 1D simulations and an iterative multigrid-preconditioned
conjugate-gradient method for 2D simulations. Rotational terms are
included for 2D calculations using cylindrical coordinates.  We
document an incompatibility between a subcell pressure algorithm to
suppress hourglass motions and the subcell remapping algorithm and
present a modified subcell pressure scheme that avoids this
problem. Strengths of this code include a straightforward structure,
enabling simple inclusion of additional physics packages, the ability
to use a general equation of state, and most importantly, the ability
to solve self-gravitating hydrodynamic flows on time-dependent,
arbitrary grids.  In what follows, we describe in detail the numerical
techniques employed and, with a large suite of tests, demonstrate that
BETHE-hydro finds accurate solutions with 2$^{nd}$-order convergence.
\end{abstract}

\keywords{hydrodynamics --- instabilities --- methods: numerical ---
  shock waves --- supernovae: general }

\section{Introduction}

The ability to simulate hydrodynamic flow is key to studying most astrophysical objects.
Supernova explosions, gamma-ray bursts, X-ray bursts, classical novae,
the outbursts of luminous blue variables (LBVs), and stellar winds are just a few
phenomena for which understanding and numerical tools evolve in tandem.  This is due
in part to the physical complexity, multi-dimensional character, and instabilities
of such dynamical fluids. Moreover, rotation is frequently a factor in the 
dynamical development of astrophysical transients.  One-dimensional hydrodynamics
is by and large a solved problem, but multi-dimensional hydrodynamics is still a challenge.
In the context of astrophysical theory, this is due primarily to the need to address 
time-dependent gravitational potentials, complicated equations of state (EOSs),
flexible grids, multi-D shock structures, and chaotic and turbulent flows.
As a result, theorists who aim to understand the Universe devote much of their 
time to code development and testing.  

One of the outstanding and complex problems in theoretical astrophysics is the 
mechanism for core-collapse supernovae. For more than two decades, 
the preferred mechanism of explosion has been the delayed neutrino mechanism \citep{bethe85}.
One-dimensional (1D) simulations generally fail to produce explosions
\citep{liebendorfer01a,liebendorfer01b,rampp02,buras03,thompson03,liebendorfer05}.
However, 2-dimensional (2D) simulations, and the accompanying
aspherical instabilities, suggest that the neutrino-mechanism may indeed be viable
\citep{herant94,janka95,burrows95,burrows07a,buras06,kitaura06}, though this
has yet to be proven.  In fact, \citet{burrows06} have recently reported an 
acoustic mechanism, which seems to succeed on long timescales when and if
other mechanisms fail.  These authors identified two primary
reasons why this mechanism might have been missed before:  1) 2D
radiation-hydrodynamic simulations with reasonable approximations are
still expensive to run, and with limited resources simulations are
rarely carried to late enough times; and 2), a noteworthy feature of the code
they used, VULCAN/2D, is its Arbitrary Lagrangian-Eulerian (ALE) structure.
VULCAN/2D incorporates non-standard grids that liberate the inner core from the Courant 
and resolution limitations of standard spherical grids. In this context, 
\citet{burrows06} claim that simulating all degrees of freedom
leads to a new mechanism in which the gravitational energy in aspherical
accretion is converted to explosion energy by first exciting
protoneutron star core g-modes. These modes then radiate acoustic power
and revive the stalled shock into explosion.  Remarkably, the acoustic
mechanism, given enough time, leads to successful
core-collapse supernovae in all progenitors more massive than $\sim$9
M$_{\odot}$ simulated to date \citep{burrows06,burrows07a}.  However, this is a
remarkable claim, and given the implications, must be thoroughly investigated.
For instance, one question to ask is: could the results seen by \citet{burrows06} 
be numerical artifacts of VULCAN?

Therefore, to address this and other issues that surround the acoustic 
supernova mechanism, as well as a host of other outstanding astrophysical puzzles, 
we have designed the new hydrodynamic code,
BETHE-hydro\footnote{Hydrodynamic core of BETHE ({\bf B}asic {\bf E}xplicit/Implicit {\bf
    T}ransport and {\bf H}ydrodynamics {\bf E}xplosion Code)}, from the bottom up.
BETHE-hydro will be coupled with the 2D mixed-frame radiation transport scheme of
\citet{hubeny07} to create the 2D radiation/hydrodynamics code BETHE, and the merger
of these two codes will be the subject of a future paper.  In
this paper, we present and test the hydrodynamic and gravity 
algorithms of BETHE-hydro. Since BETHE-hydro uses arbitrary grids and a general 
gravity solver, we expect it to be a powerful and flexible numerical tool, 
able to configure the grid to suit the computational challenge.

The core of BETHE-hydro is a 1D and 2D ALE
hydrodynamics solver.  First, solutions to the Lagrangian hydrodynamic
equations are obtained on an arbitrary polygonal grid.  Then, to avoid
tangled grids, the hydrodynamic variables can be remapped to another grid.
A unique and powerful feature of ALE schemes is their
flexibility to tailor an arbitrary grid to the computational challenge
and to alter the grid dynamically during runtime.  Hence, purely Lagrangian,
purely Eulerian, or arbitrarily moving grids (chosen to optimize
numerical performance and resolution) are possible.  These grids
can be non-orthogonal, non-spherical, and non-Cartesian.

Some of the earliest 2D hydrodynamic simulations in astrophysics were,
in fact, performed to address the core-collapse problem.  While some
employed standard fixed-grid schemes \citep{smarr81}, other 2D
simulations were calculated with adaptive grids.  Although many
utilized moving grids, for the most part, the differencing
formulations were Eulerian \citep{leblanc70,symbalisty84,miller93}.
On the other hand, \citet{livio80} did employ an ``Euler-Lagrange''
method involving a Lagrangian hydrodynamic solve, followed by a
remapping stage.  Even though these simulations did exploit radially
dynamical grids, they were restricted to be orthogonal and spherical.

Many of the current hydrodynamic algorithms used in
astrophysics are based upon either ZEUS \citep{stone92a} or
higher-order Godunov methods, in particular the Piecewise-Parabolic Method (PPM)
\citep{colella84,woodward84}.  Both approaches have been limited to
orthogonal grids.  One concern with PPM-based codes is that they
solve the hydrodynamic equations in dimensionally-split
fashion.  As a result, there have been concerns that these algorithms
do not adequately resolve flows that are oblique to the grid
orientation.  Rectifying this concern, recent unsplit higher-order
Godunov schemes, or approximations thereof, have been developed
\citep{truelove98,klein99,gardiner05,miniati07}.  Despite this
and the employment of adaptive mesh refinement (AMR), these codes must use
orthogonal grids that are often strictly Cartesian.

In BETHE-hydro, we use 
finite-difference schemes based upon the support-operator method \citep{shashkov95} of
\citet{caramana98c} and \citet{caramana98a}.  Differencing
by this technique enables conservation of energy to roundoff error in the absence of rotation and time-varying
gravitational potentials.  Similarly, momentum is conserved
accurately, but due to the artificial viscosity scheme that we employ,
it is conserved to roundoff error for hydrodynamic simulations
using Cartesian coordinates only.  Moreover, using the support-operator method and borrowing from an
adaptation of the support-operator technique for elliptic equations \citep{morel98},
we have developed a gravity solver for arbitrary grids. Unfortunately, by 
including a general gravity capability, we sacrifice strict energy and momentum conservation.
However, we have performed tests and for most cases the results are
reasonably accurate.  Furthermore, we have included rotational terms and we use a
modified version of the subcell pressure scheme to mitigate hourglass instabilities 
\citep{caramana98a}. 

To resolve shocks, we employ an artificial viscosity method, 
which is designed to maintain grid stability as well \citep{campbell01}, and there are 
no restrictive assumptions made about the equation of state.  
Higher-order Godunov techniques employ Riemann solvers to 
resolve shocks, but frequently need an artificial viscosity scheme 
to eliminate unwanted post-shock ringing.
Moreover, the inner workings of Riemann
solvers often make local approximations that the EOS has a gamma-law
form, which stipulates that as the internal energy goes to zero so
does the pressure. For equations of state appropriate for core-collapse supernovae, this
artificially imposed zero-point energy can pose problems for the
simulations \citep{buras06}.  

Two codes in astrophysics which have already capitalized on the
arbitrary grid formulations of ALE are Djehuty
\citep{bazan03,dearborn05} and VULCAN/2D \citep{livne93,livne04}.  In both
cases, the grids employed were designed to be spherical in
the outer regions, but to transition smoothly to a cylindrical grid near the
center.  These grid geometries reflected the basic structure of stars, while
avoiding the cumbersome singularity of a spherical
grid.  With this philosophy, \citet{dearborn06}, using Djehuty, have
studied the helium core flash phase of
stellar evolution in 3D.  In the core-collapse context,
\citet{burrows06,burrows07c,burrows07a}, using VULCAN/2D, performed 2D
radiation/hydrodynamic and radiation/MHD simulations with rotation.

Several gravity solvers have been employed in
astrophysics.  The most trivial are static or monopole approaches.
For arbitrary potentials, the most extensively used are N-body
schemes, which fit most naturally in Smooth Particle
Hydrodynamics (SPH) codes \citep{monaghan92}.  As a virtue, SPH solves
the equations in a grid-free context, and while SPH
has opened the way to three-dimensional (3D) hydrodynamic simulations
in astrophysics, including core-collapse simulations \citep{fryer02},
the smoothing kernels currently employed pose serious problems for
simulating fundamental hydrodynamic instabilities such as the
Kelvin-Helmholtz instability \citep{agertz07}.  Another approach
to the solution of Poisson's equation for gravity is the use of
multipole expansions of the potential \citep{muller95}.  This
technique achieves its relative speed by calculating simple
integrals on a spherical grid once. Then, the stored integrals are 
used in subsequent timesteps.  In BETHE-hydro, we
construct finite-difference equations for Poisson's equation using the
support-operator method and find solutions to the resulting linear
system of equations for the potentials via an iterative
multigrid-preconditioned conjugate-gradient method \citep{ruge87}.

In \S \ref{section:bethe_hydro}, we give a summary of BETHE-hydro and sketch the flowchart of the
algorithm.  The coordinates and mesh details are discussed in \S \ref{section:coords_mesh}. We then 
describe in \S\ref{section:lag_hydro} the discrete Lagrangian equations, including 
specifics of the 2$^{nd}$-order time-integration and rotational terms.  Section \ref{section:gravity} gives a
complete description of the 1D and 2D gravity solvers.  Hydrodynamic
boundary conditions are discussed in \S \ref{section:hydro_boundary}.
The artificial viscosity algorithm that provides shock resolution and grid stability
is described in \S \ref{section:artificial_viscosity}. In \S \ref{section:hourglass}, we present the
subcell pressure scheme that suppresses hourglass modes. Remapping is described
in depth in \S \ref{section:remap}.  In \S \ref{section:tests}, we
demonstrate the code's strengths and limitations with some test
problems. Finally,  in \S \ref{section:discussion} we summarize 
the central characteristics and advantages of BETHE-hydro.

\section{BETHE-Hydro:  An Arbitrary Lagrangian-Eulerian Code}
\label{section:bethe_hydro}
In ALE algorithms, the equations of hydrodynamics are solved in
Lagrangian form.
Within this framework, equations for the conservation of mass,
momentum, and energy are: 
\begin{equation}
\label{eq:mass_lag}
\frac{d \rho}{d t} = - \rho \vec{\nabla} \cdot \vec{v} \, ,
\end{equation}
\begin{equation}
\label{eq:mom_lag}
\rho \frac{d \vec{v}}{d t} = - \rho \vec{\nabla} \Phi - \vec{\nabla} P
\, ,
\end{equation}
and
\begin{equation}
\label{eq:ene_lag}
\rho \frac{d \varepsilon}{d t} = - P \vec{\nabla} \cdot \vec{v} \, .
\end{equation}
$\rho$ is the mass density (which we refer to simply as
``density'' and is distinct from the energy or
momentum densities), $\vec{v}$ is the fluid velocity, $\Phi$ is the gravitational
potential, $P$ is the isotropic pressure, $\varepsilon$ is the specific
internal energy, and $d/dt = \partial / \partial t + \vec{v} \cdot
\vec{\nabla}$ is the Lagrangian time derivative.  The equation of
state may have the following general form:
\begin{equation}
\label{eq:eos}
P = P(\rho,\varepsilon,X_i) \, ,
\end{equation}
where $X_i$ denotes the mass fraction of species $i$.
Therefore, we also solve the conservation equations:
\begin{equation}
\label{eq:advec}
\frac{d X_i}{dt} = 0 \, .
\end{equation}
Completing the set of equations for self-gravitating astrophysical flows 
is Poisson's equation for gravity:
\begin{equation}
\label{eq:poisson}
\nabla^2 \Phi = 4 \pi G \rho \, ,
\end{equation}
where $G$ is Newton's gravitational constant.

All ALE methods have the potential to solve eqs. (\ref{eq:mass_lag}-\ref{eq:ene_lag})
using arbitrary, unstructured grids.  In this lies the power and functionality of ALE methods.
The solutions involve two steps and are conceptually quite simple: 1) a Lagrangian solver moves the nodes of the mesh in
response to the hydrodynamic forces; and 2) to avoid grid tangling,
the nodes are repositioned, and a remapping algorithm interpolates hydrodynamic quantities from the old grid to this new grid.
Of course, the challenge is to find accurate solutions, while conserving
energy and momentum.  In constructing BETHE-hydro to satisfy these
requirements, we use the ALE hydrodynamic
techniques of \citet{caramana98c}, \citet{caramana98a},
\citet{campbell01}, and \citet{loubere05}.

\begin{figure}
\epsscale{0.8}
\plotone{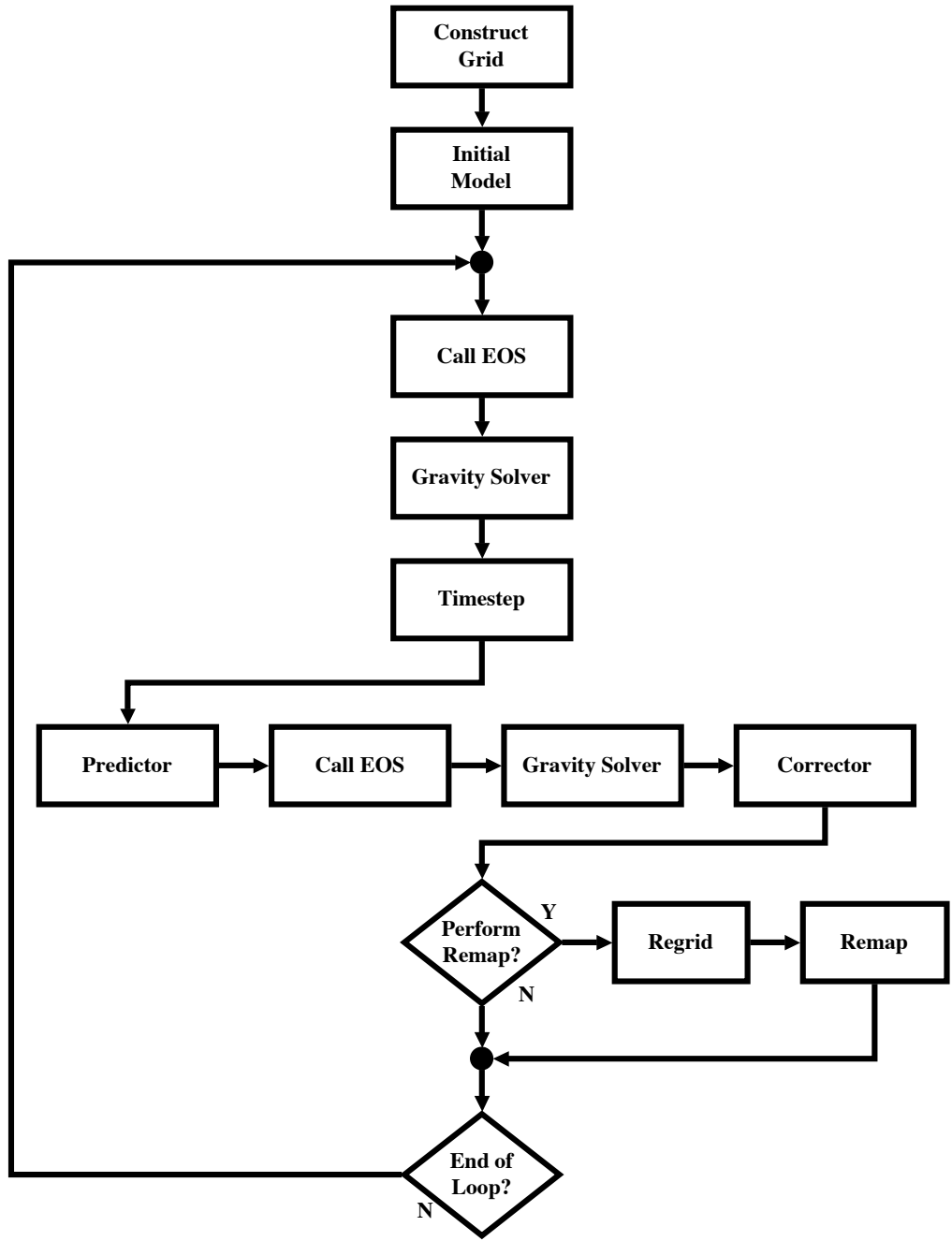}
\caption{The flowchart for BETHE-hydro. See \S \ref{section:bethe_hydro} for a discussion.\label{flowchart}}
\end{figure}

To summarize the overall structure of BETHE-hydro, we present a
schematic flowchart, Fig. \ref{flowchart}, and briefly describe the
key steps.
Establishing the structure for
all subsequent routines, the first step is to construct
the arbitrary, unstructured grid.  This leads to the next step, problem
initialization.  Then, the main loop for timestep integration is
entered.  After a call to the EOS to obtain the pressure, solutions to Poisson's
equation for gravity are calculated using either the 2D or 1D
algorithms in \S \ref{section:gravity}.  Both use the
support-operator method to discretize eq. (\ref{eq:poisson}), and the
resulting system of linear equations is solved by a tridiagonal
solver in 1D and a multigrid pre-conditioned conjugate-gradient
iterative method in 2D (\S \ref{section:gravity}).

After the timestep is calculated (\S
  \ref{section:predictor_corrector}),
the Lagrangian equations of hydrodynamics are solved on
an arbitrary grid using the compatible hydrodynamics algorithm of
\citet{caramana98c} (see \S \ref{section:lag_hydro} for further discussion).  To ensure 2nd-order accuracy in both space and time, we employ a
predictor-corrector iteration (\S \ref{section:predictor_corrector}), in which a second call to the EOS and
the gravity solver are required.
Other than the gravity solver block, the
Lagrangian hydrodynamics solver (\S
\ref{section:lag_hydro}) is represented by the set of steps
beginning with Predictor and ending with Corrector
(Fig. \ref{flowchart}).

After finding Lagrangian hydrodynamic solutions at each timestep, one
could proceed directly to the next timestep, maintaining the
Lagrangian character throughout the simulation.  However, in multiple spatial dimensions,
grid tangling will corrupt flows with significant
vorticity. In these circumstances, we conservatively remap the Lagrangian
solution to a new arbitrary grid that mitigates tangled grids.  The
new grid may be the original grid, in which case we are solving the
hydrodynamic equations in the Eulerian frame, or it may be a new,
arbitrarily-defined, grid.

\section{Coordinate Systems and Mesh}
\label{section:coords_mesh}

In BETHE-hydro, eqs. (\ref{eq:mass_lag}-\ref{eq:poisson}) may be
solved with any of several
coordinate systems and assumed symmetries.  Included are the standard
1D \& 2D Cartesian coordinate systems.  For astrophysical applications, we use 1D
spherical and 2D cylindrical coordinate systems.  Irrespective of the
coordinate system, we indicate a position by $\vec{x}$.  The components
of the Cartesian coordinate system are $x$ and $y$; the spherical
components are $r$, $\theta$, and $\phi$; and the cylindrical
coordinates are $r$, $z$, and $\phi$.

\begin{figure}
\epsscale{1.0}
\plottwo{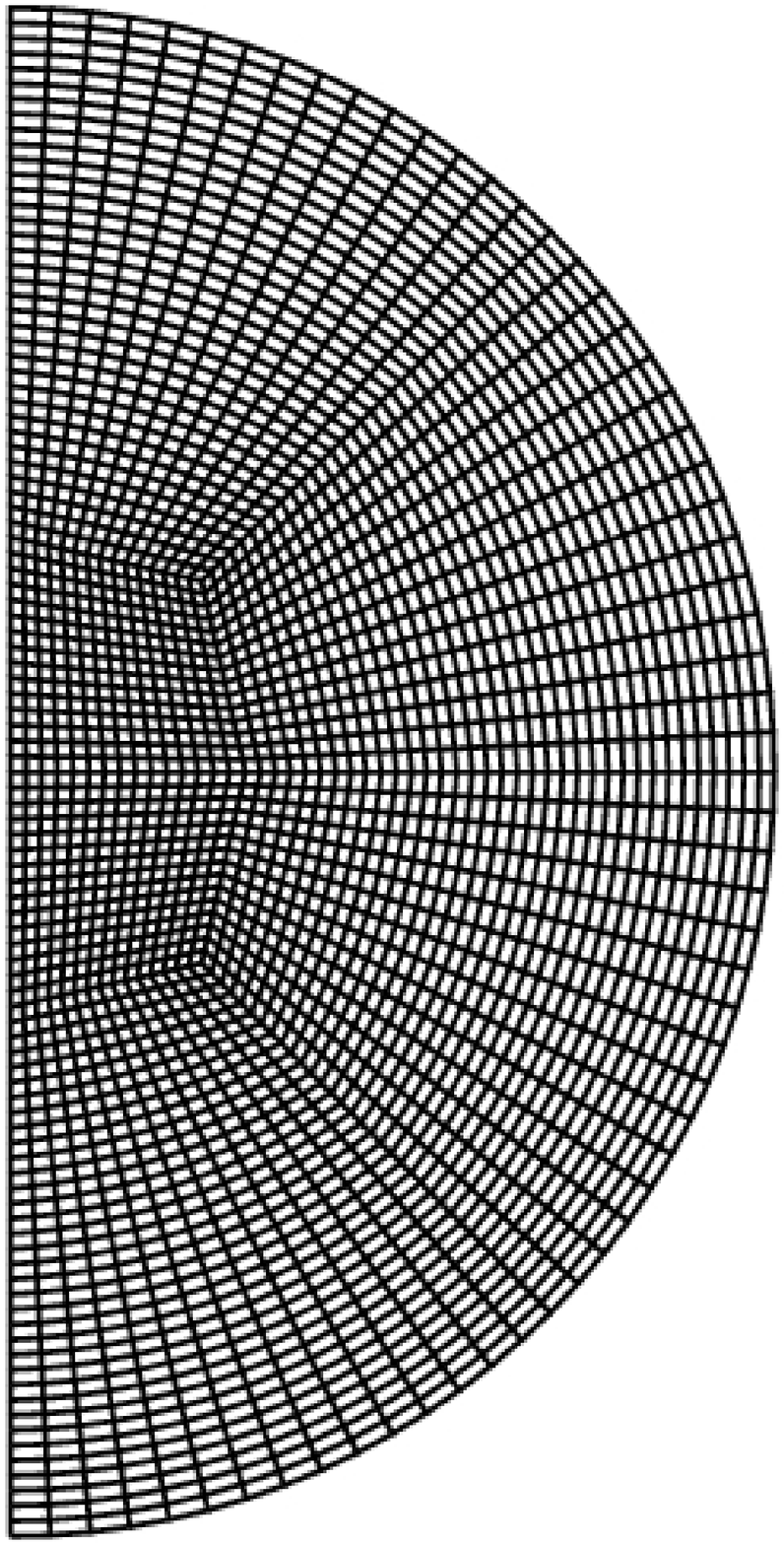}{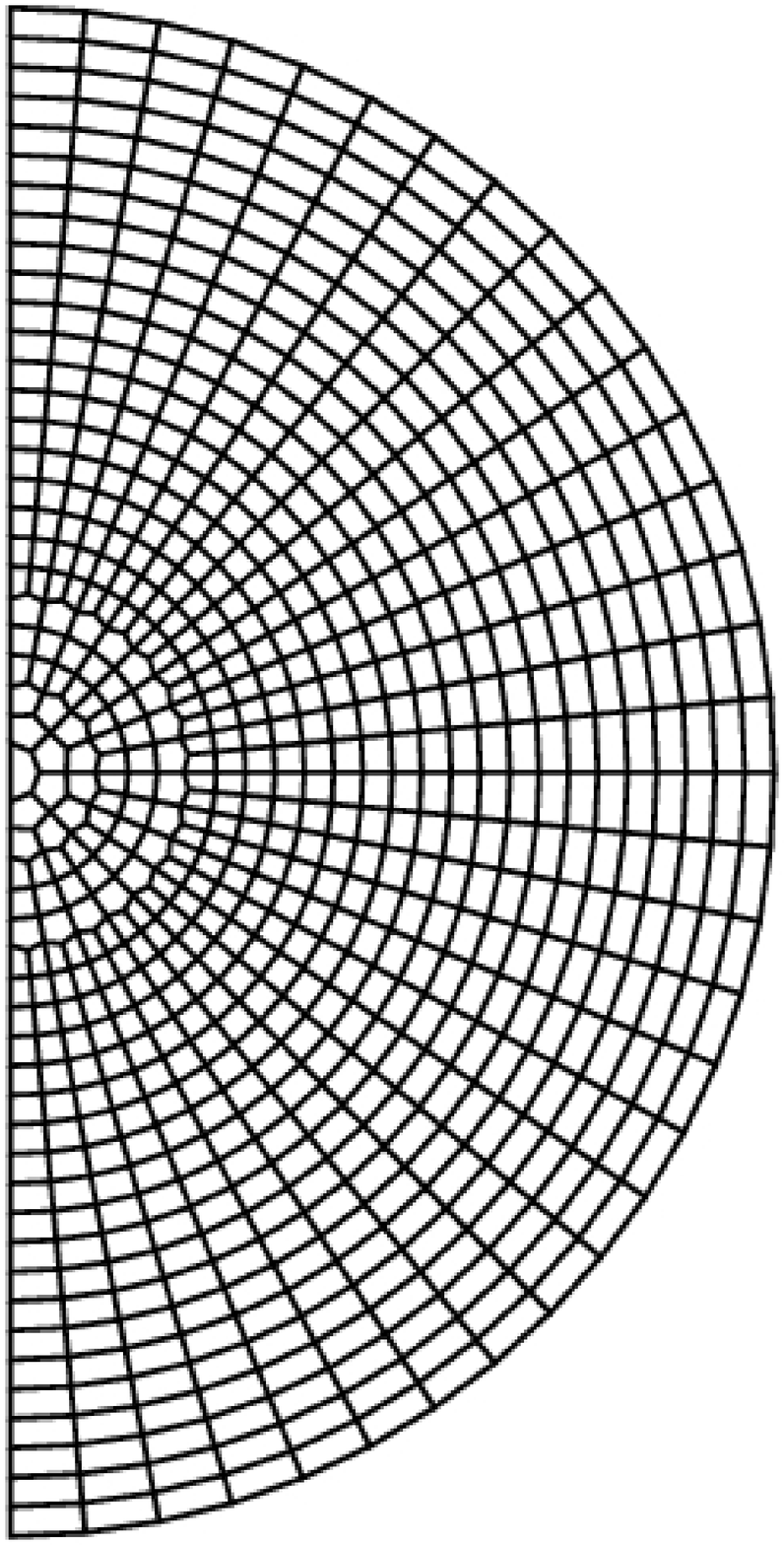}
\caption{On the left, a representative butterfly mesh and on the
  right, a representative spiderweb mesh.  In general, we construct
  meshes with much higher resolution.  However, to easily see the
  features of the mesh we reproduce low resolution versions here.
  The butterfly mesh is a standard grid that captures the benefits of
  a spherical grid near the outer boundary and avoids the singularity
  at the center. 
  The spiderweb mesh captures the benefits of
  a spherical grid throughout, but near the center, the
  angular resolution is modified to avoid extreme Courant-condition limits and the
  singularity.  Construction begins with half of an octagon at the
  center.  Subsequent tiers of nodes are placed $\Delta r$ farther in
  radius from the interior set.  When $r\Delta \theta$ exceeds $\Delta
  r$ the angular size is halved.  For these transition nodes, $\Delta r$
  is alternately multiplied by $1 + \epsilon$ and $1-\epsilon$ to
  exaggerate the concavity of the cells.\label{grid_ex}}
\end{figure}

Distinct from the coordinate system is the grid, or mesh, which
defines an arbitrary discretization of space.  It is this unique
feature and foundation of ALE techniques that provides their flexibility.
Figure \ref{grid_ex} shows two example grids: the butterfly mesh on the
left and a spiderweb mesh on the right.  Either may be used in a 2D
Cartesian or 2D cylindrical simulations, although the placement of the
nodes is neither Cartesian nor cylindrical.  Instead, these meshes have
been designed to simulate spherically or cylindrically convergent
phenomena without the limitation of small zones near the center.

When using cylindrical coordinates in ALE algorithms, one may use control-volume
differencing (CVD) or area-weighted differencing (AWD); we use CVD.
Thorough comparisons of these two differencing schemes may be found in
\citet{caramana98c}.  Here, we give a basic justification for
choosing CVD.  For CVD, the volumes of the subcell, cell, and
node are calculated by straightforward partitioning of these regions by
edges.  Hence, volumes and masses are exact representations of their
respective regions.  While this discretization is natural and easy to
comprehend, it does not preserve strict spherical symmetry when using
a spherical grid and cylindrical coordinates.  On the other hand, AWD is designed
to preserve spherically symmetric flows, but only with the use of
spherical grids that have equal spacing in angle.  Since this
prevents the use of arbitrary grids, and the asymmetries of CVD are
small, we use CVD for discretization.
 
Construction of the grid begins with the arbitrary placement of nodes.  At
these nodes, coordinate positions, velocities, and accelerations are defined.
Simply specifying node locations does not completely define a mesh,
since there is not a unique way to assign cells and masses to these
nodes.  Therefore, the
user must specify the connectivity among nodes, which define the arbitrarily-shaped polygonal cells.  It is within these cells that
cell-centered averages of $\rho$,
$\varepsilon$, $P$, $X_i$, and $\Phi$ are defined.

\begin{figure}
\epsscale{0.9}
\plotone{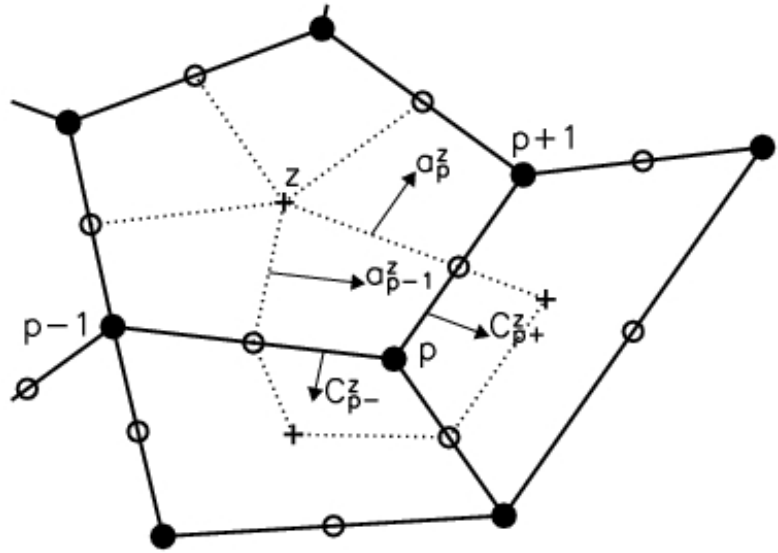}
\caption{A portion of an unstructured, polygonal mesh.  Filled
  circles indicate node positions, $p$.  Crosses mark cell
  centers, $z$.  Open circles show the mid-edge
  locations.  Cells are bounded by the edges (solid lines), and the
  cell is further divided into subcells via dashed
  lines. $\vec{C}^z_{p+}$ and $\vec{C}^z_{p-}$ are the half-edge
area vectors.  $\vec{a}^z_p$ and $\vec{a}^z_{p-1}$ are the area
vectors for the lines connecting the cell centers and mid-edges (see
\S \ref{section:coords_mesh} for a discussion).\label{fig:grid}}
\end{figure}

With the node
positions, their connectivity, and the cells defined, a mesh is completely specified and all other useful
descriptions follow.  Cells are denoted by $z$, and nodes are indexed
by $p$.  The
set of nodes that defines cell $z$ are $p \in S(z)$, where
the nodes are ordered counterclockwise.
Conversely, the set of cells that shares node $p$ is denoted by $z \in S(p)$.
Each cell has $N_p(z)$ nodes that define it, and each node has
$N_z(p)$ cells that share it.  The sample sub-grid depicted in
Fig. \ref{fig:grid} helps to illustrate the nodal and cell structure. The filled circles
indicate node positions, $\vec{x}_p$, and the crosses indicate the
cell-center positions,
$\vec{x}_z = (1/N_p(z))\sum_{p \in S(z)} \vec{x}_p$.  The solid lines
are direction-oriented edges that separate cells from one
another, and the open circles indicate their mid-edge locations.
Partitioning the cell into subcells, the
dashed lines connect the mid-edges with the cell centers.  No matter how many sides a cell
has, with this particular division the subcells are always
quadrilaterals.  Naturally, the cell volumes (see appendix
\ref{section:integrals} for formulae calculating discrete volumes), $V_z$, are related to the
subcell volumes, $V^z_p$, by
\begin{equation}
\label{eq:volume_cell}
V_z = \sum_{p \in S(z)} V^z_p \, .
\end{equation}
Furthermore, each node has a volume, $V_p$, defined by the adjoining subcells
that share the node $p$:
\begin{equation}
\label{eq:volume_node}
V_p = \sum_{z \in S(p)} V^p_z \, .
\end{equation}
For calculating pressure forces and fluxes, 
vectors are assigned to each
half edge on either side of node $p$: $\vec{C}^z_{p+}$ and
$\vec{C}^z_{p-}$, where $+$ indicates the half edge in the
counterclockwise direction around the cell and $-$ indicates its opposite
counterpart.  Their magnitudes are the areas represented by the half edges, and their directions point
outward and normal to the surface of cell $z$.  From these half-edge
area vectors, an area vector, $\vec{S}^z_p$, that is
associated with zone $z$ and node $p$ is then defined:
\begin{equation}
\vec{S}^z_p = \vec{C}^z_{p+} + \vec{C}^z_{p-} \, .
\end{equation}
Similarly, vectors are associated with the lines connecting the
mid-edges and the cell centers: $\vec{a}^z_p$.  Again, their
magnitudes are the corresponding area, but while the vector is
normal to this line, the direction is oriented
counterclockwise around the cell.

\section{Discrete Lagrangian Hydrodynamics}
\label{section:lag_hydro}
The fundamental assumption of Lagrangian algorithms is that the mass,
$m_z$, of a discrete volume $V_z$ is constant
with time.  For staggered-grid methods, in which scalars are defined
as cell-centered averages and vectors are defined at the nodes, it is
necessary to define a
Lagrangian mass, $m_p$, for the nodes as well.  This nodal
mass is associated with the node's volume $V_p$ 
(eq. \ref{eq:volume_node}).  Conservation of mass (eq. \ref{eq:mass_lag}) implies zero mass flux across the boundaries,
$\partial V$, of either the cell volume or nodal volume.  Therefore, the region of overlap for
$V_z$ and $V_p$, which is the subcell volume $V^z_p$, is bounded by
surfaces with zero mass flux.  Consequently, the most elemental
Lagrangian mass is the subcell mass, $m^z_p$, and the cell mass ($m_z$) and node mass ($m_p$) are constructed as appropriate sums
of subcell masses:
\begin{equation}
\label{eq:mass_cell}
m_z = \sum_{p \in S(z)} m^z_p \, ,
\end{equation}
and
\begin{equation}
\label{eq:mass_node}
m_p = \sum_{z \in S(p)} m^p_z \, .
\end{equation}
Hence, we arrive at three discrete forms of mass conservation (eq. \ref{eq:mass_lag}):
\begin{equation}
\label{eq:mass_discrete}
\begin{array}{cccc}
\rho^z_p = \frac{m^z_p} {V^z_p} \, , &
\rho_p = \frac{m_p} {V_p} \, ,&
\mbox{and} &
\rho_z = \frac{m_z} {V_z} \, .
\end{array}
\end{equation}

In defining the discrete momentum and energy equations, we use the
compatible hydrodynamics algorithms developed by
\citet{caramana98c}. Specifically, the discrete divergence and
gradient operators are {\it compatible} in that
they faithfully represent their analog in continuous space
{\it and} their definitions are expressly related to one another using
the hydrodynamic expressions for conservation of momentum and
energy.  As a result, this approach leads to discretizations that satisfy momentum and energy conservation to machine
accuracy.  This is accomplished with the support-operator method
\citep{shashkov95}.  Given
the integral identity,
\begin{equation}
\label{eq:intidentity}
\int_V \Phi (\vec{\nabla} \cdot \vec{H}) dV 
+ \int_V \vec{H} \cdot \vec{\nabla} \Phi dV 
= \oint_{\partial V} \Phi \vec{H} \cdot d\vec{S} \, ,
\end{equation}
where $\Phi$ is any scalar and $\vec{H}$ is some vector,
there is an incontrovertible connection between the divergence and
gradient operators.  For many choices of discretization, the discrete
counterparts of these operators could violate this integral identity.
Simply put, the goal of the support-operator method is to define the discrete operators so
that they satisfy eq. (\ref{eq:intidentity}).  The first step is to
define one of the discrete operators.  It is often, but
not necessary, that the discrete
divergence operator is defined via Gauss's Law:
\begin{equation}
\label{eq:gausslaw}
\int_V \vec{\nabla} \cdot \vec{v} dV = \oint_{\partial V} \vec{v}
\cdot d\vec{S} \, ,
\end{equation}
and then eq. (\ref{eq:intidentity}) is used to compatibly define the other
discrete operator.  Discretizing the hydrodynamic
equations, 
\citet{caramana98c} begin by defining
$(\vec{\nabla}P)_p$ at the nodes and use the integral form of energy
conservation and an equation equivalent to eq. (\ref{eq:intidentity})
to define for each cell the discrete divergence of the velocity, $(\vec{\nabla} \cdot \vec{v})_z$.

To begin, \citet{caramana98c} integrate eq. (\ref{eq:mom_lag}) (excluding gravity and rotational
terms) over the volume of node $p$, producing the discrete form for
$(\vec{\nabla P})_p$ and the momentum equation:
\begin{eqnarray}
\label{eq:mom_discrete1}
m_p \frac{\Delta \vec{v}_p}{\Delta t} & = & - \int_{V_p}\vec{\nabla} P \,
dV \nonumber \\
& = & - \oint_{\partial V_p} P \, d\vec{S} \nonumber \\ 
& = & - \sum_{z \in S(p)} P_z \vec{S^p_z} \, ,
\end{eqnarray}
where $\Delta \vec{v}_p = \vec{v}_p^{n+1} - \vec{v}^n_p$ is the change
in velocity from timestep $n$ to the next timestep $n+1$, the
timestep is $\Delta t$, and $P_z$ is the pressure in cell $z$.  In other words, the net force
on node $p$ is a sum of the pressure times the
directed zone areas that share node $p$.  Hence, the subcell force
exerted by zone $z$ on point $p$ is $\vec{f}^p_z = P_z \vec{S}^p_z$.  A more complete description of the
subcell force, however, must account for artificial
viscosity (\S \ref{section:artificial_viscosity}):
\begin{equation}
\vec{f}^p_z = P_z \vec{S}^p_z + \vec{f}^p_{z,\rm{visc}} \, ,
\end{equation}
where $\vec{f}^p_{z,\rm{visc}},$ is the subcell force due to
artificial viscosity.  Furthermore, in this work, we include gravity
and rotation for 2D axisymmetric simulations, and the full discrete
momentum equation becomes
\begin{equation}
\label{eq:mom_discrete2}
m_p \frac{\Delta \vec{v}_p}{\Delta t} =
\sum_{z \in S(p)} \vec{f}^p_z + m_p \vec{g}_p + m_p\vec{A}_p \, ,
\end{equation}
where $\vec{g}_p$ and $\vec{A}_p$ are the gravitational and rotational
accelerations, respectively.  For simplicity, and to parallel the discussion in
\citet{caramana98c}, we ignore these terms in the momentum equation
and proceed with the compatible construction of the energy equation
(see \S \ref{section:consenerot} and \S \ref{section:consenegravity}
for discussions of total energy conservation including rotation and
gravity, respectively).

To construct the discrete energy equation, \citet{caramana98c} integrate
eq. (\ref{eq:ene_lag}) over the discrete volume of cell $z$:
\begin{equation}
m_z \frac{\Delta \varepsilon_z}{\Delta t} = - \int_{V_z} P \vec{\nabla} \cdot
\vec{v} \, dV = - P_z ( \vec{\nabla} \cdot
\vec{v} )_z V_z \, ,
\end{equation}
where $\Delta \varepsilon_z = \varepsilon_z^{n+1} - \varepsilon_z^n$.
Then, the objective is to determine a discrete form for the right hand
side of this equation that conserves energy and makes the discrete
gradient and divergence operators compatible.
\citet{caramana98c} accomplish this with the integral for conservation
of energy (neglecting gravity):
\begin{eqnarray}
\int \left ( \rho \frac{d \varepsilon}{dt} + \frac{1}{2} \rho \frac{d
    v^2}{dt}\right ) dV 
& = & -\int_V (P \vec{\nabla} \cdot \vec{v} + \vec{v} \cdot \vec{\nabla} P
) dV \nonumber \\
& = & -\oint_{\partial V} P\vec{v} \cdot d\vec{S}
\, ,
\end{eqnarray}
where the second expression is the integral identity,
eq. (\ref{eq:intidentity}), that defines the physical relationship
between the gradient and divergence operators.
Neglecting boundary terms, the discrete form of this integral is
\begin{equation}
\label{eq:enecons_discrete}
\sum_z \left ( m_z \frac{\Delta \varepsilon_z}{\Delta t} + \sum_{p \in S(z)}
\vec{v}^{n+1/2}_p \cdot \vec{f}^p_z \right ) =
0 \, ,
\end{equation}
where we used $\Delta \vec{v}^2_p = (\vec{v}^{n+1}_p)^2 -
(\vec{v}^n_p)^2 = 2 \vec{v}^{n+1/2}_p \cdot \Delta \vec{v}_p$, $\vec{v}^{n+1/2}_p = 1/2 ( \vec{v}^{n+1}_p + \vec{v}^n_p)$,
 and
substituted in eq. (\ref{eq:mom_discrete1}).  If we set the expression
for each zone in eq. (\ref{eq:enecons_discrete}) to zero, we
arrive at the compatible energy equation:
\begin{equation}
\label{eq:ene_discrete}
m_z \frac{\Delta \varepsilon_z}{\Delta t} = - \sum_{p \in S(z)}
\vec{v}^{n+1/2}_p \cdot \vec{f}^z_p \, .
\end{equation}
The RHS of
eq. (\ref{eq:ene_discrete}) is merely the $PdV$ work term of the first
law of thermodynamics.  Its unconventional form is a consequence of
the support-operator method.

Thus, we derive two significant results.  First, 
inspection of eqs. (\ref{eq:ene_discrete}) and
(\ref{eq:mom_discrete1}) leads to compatible definitions for the
discrete gradient and divergence operators. Specifically, the discrete
analogs of the gradient and divergence operators are
\begin{equation}
(\vec{\nabla} P)_p = \frac{1}{V_p} \sum_{z \in S(p)} P_z \vec{S}^p_z
\end{equation}
and 
\begin{equation}
(\vec{\nabla} \cdot \vec{v})_z = \frac{1}{V_z} \sum_{p \in S(z)}
\vec{v}_p \cdot \vec{S}^p_z \, .
\end{equation}
Second, eqs. (\ref{eq:mass_discrete}), (\ref{eq:mom_discrete1}), and
(\ref{eq:ene_discrete}) form the compatible discrete equations of
Lagrangian hydrodynamics.

\subsection{Momentum Conservation}
\label{section:momcons}

In its current form, BETHE-hydro strictly conserves momentum for
simulations that use Cartesian coordinates and not cylindrical
coordinates.  \citet{caramana98c} show that the requirement for strict
conservation is the use of control-volume differencing, or that exact
representations of cell surfaces are employed.  As a simple example,
consider momentum conservation with pressure forces only:
\begin{equation}
\sum_p \vec{F}_p = \sum_p \sum_{z \in S(p)} \vec{f}^p_z = \sum_z
P_z \sum_{p \in S(z)} \vec{S}^z_p \, .
\end{equation}
For Cartesian coordinates, the sum $\sum_{p \in S(z)} \vec{S}^z_p$ is exactly
zero for each cell as long as the surface-area vectors are exact
representations of the cell's surface.  This is not the case for
area-weighted differencing, and hence momentum is not conserved in
that case.  For cylindrical coordinates, the sum of surface-area
vectors gives zero only for the $z$-component.  However, the assumption
of symmetry about the cylindrical axis ensures momentum conservation
for the $r$-component.  Therefore, control-volume differencing ensures
momentum conservation, even with the use of cylindrical coordinates.

Based upon similar arguments, the subcell-pressure forces that
eliminate hourglass motions (\S \ref{section:hourglass}) are formulated to conserve momentum for
both Cartesian and cylindrical coordinates.  On the other hand, because
we multiply the Cartesian artificial viscosity force by $2 \pi r_p$ to
obtain the force appropriate for cylindrical coordinates, the
artificial viscosity scheme (\S \ref{section:artificial_viscosity})
that we employ is conservative for Cartesian coordinates only.

We test momentum conservation with the Sedov blast wave problem (\S
\ref{section:sedov}) using Cartesian and cylindrical coordinates.
Symmetry dictates conservation in the $r$-component of the momentum.
Consequently, we compare the $z$-component of the total momentum at
$t=0$ s with subsequent times.  Since the initial momentum of this
test is zero, we obtain a relative error in momentum by $P_z(t)/P_{z,z >
0}(t)$, where $P_z(t)$ is the $z$-component of the total momentum at
time $t$ and $P_{z,z > 0}(t)$ is a similar quantity for $z > 0$.  As
expected, the relative error using Cartesian coordinates reflects
roundoff error.  For cylindrical coordinates and 35,000 cells, the
relative error in momentum is $\sim 3 \times 10^{-7}$ after 0.8 s and
71,631 timesteps.  Although momentum is strictly conserved for
simulations using Cartesian coordinates only, the Sedov blast wave
problem demonstrates that momentum is conserved very accurately, even
when cylindrical coordinates are used.

\subsection{Second-Order Time Integration: Predictor-Corrector}
\label{section:predictor_corrector}

Our method is explicit in time.  Therefore, for numerical convergence, we limit the timestep, $\Delta t$, to be smaller than
three timescales.  By the Courant-Friedrichs-Levy (CFL) condition we
limit $\Delta t$ to be smaller than the shortest sound-crossing time
and the time it takes flow to traverse a cell.  The latter condition
is useful to avoid tangled meshes and is necessary when remapping is
used.  In addition, it is
limited to be smaller than the fractional change in the volume
represented by $(\vec{\nabla} \cdot \vec{v})_z$.  In calculating
these timescales a characteristic length for each cell is needed.
Given arbitrary polygonal grids, we simply use the shortest edge for
each cell.  In practice, the timestep is the shortest of these
timescales multiplied by a scaling factor, CFL.  In our
calculations, we set ${\rm CFL}= 0.25$.

To ensure second-order accuracy and consistency with the time
levels specified in eqs. (\ref{eq:mom_discrete2}) and
(\ref{eq:ene_discrete}), we employ predictor-corrector
time integration using the finite difference stencil from \citet{caramana98c}.
For each timestep, the
current time level of each quantity is identified with $n$ in
superscript.  The time-centered quantities are indicated with a
$n+1/2$ superscript, while the predicted values are labeled
$n+1,pr$, and the fully updated values for the next timestep are
labeled with a $n+1$ superscript.

Table \ref{table:predictor-corrector} lists the steps for the
predictor-corrector integration. 
The predictor step begins by defining the forces on nodes, $\vec{f}^n$,
using current pressures $P^n$, areas $\vec{S}^n$, and artificial
viscosity forces $\vec{f}^n_{\rm{visc}}$.
Eq. (\ref{eq:mom_discrete2}) is then used to calculate the predicted
velocity, $\vec{v}^{n+1,pr}$.  The predicted node coordinate is then
calculated by $\vec{x}^{n+1,pr} = \vec{x}^n + \Delta t
\vec{v}^{n+1/2}$, where $\vec{v}^{n+1/2}$ is the time-centered
velocity computed as $\vec{v}^{n+1/2} = \frac{1}{2}(\vec{v}^{n+1,pr} +
\vec{v}^n)$. Then, using $\vec{f}^n$ and $\vec{v}^{n+1/2}$ in
eq. (\ref{eq:ene_discrete}), the predicted specific internal energy,
$\varepsilon^{n+1,pr}$, is obtained.  Using the predicted node positions,
the predicted rotational acceleration ($\vec{A}^{n+1,pr}_p$), volume ($V^{n+1,pr}$), and density ($\rho^{n+1,pr}$), are
computed, which in turn are used to calculate the predicted gravitational
acceleration, $\vec{g}^{n+1,pr}$.  Calling the EOS with $\varepsilon^{n+1,pr}$ and
$\rho^{n+1,pr}$ gives the predicted pressure $P^{n+1,pr}$.  We finish
the predictor step and prepare for the corrector step by calculating the time-centered positions
$\vec{x}^{n+1/2}$, pressures $P^{n+1/2}$, rotational accelerations $\vec{A}^{n+1/2}_p$, and gravitational
accelerations $\vec{g}^{n+1/2}$ as simple
averages of their values at the $(n+1,pr)$ and $n$ time levels.  

The corrector step then proceeds similarly to the predictor step.  $P^{n+1/2}$ and
$\vec{S}^{n+1/2}$ are used to compute $\vec{f}^{n+1/2}$, which in
turn is used to compute the updated velocity, $\vec{v}^{n+1}$.  After
computing the new time-centered velocity, the node coordinates are
finally updated using the time-centered velocity.  Then, the specific internal energy, $\varepsilon^{n+1/2}$, is updated using
$\vec{f}^{n+1/2}$ and $\vec{v}^{n+1/2}$.  Finally, the volume,
$V^{n+1}$, and
density $\rho^{n+1}$ are computed for the next timestep.

\subsection{Rotation in 2D cylindrical coordinates}
\label{section:rotation}

In 2D simulations using cylindrical coordinates, we have extended
BETHE-hydro to include rotation about the axis of symmetry.
For azimuthally symmetric configurations, all partial derivatives
involving $\phi$ are zero.  Consequently, the conservation of mass and
energy equations are
\begin{equation}
\label{eq:mass_lag_rot}
\frac{D \rho}{Dt} = - \rho \vec{\nabla}_{rz} \cdot \vec{v}
\end{equation}
and
\begin{equation}
\label{eq:ene_lag_rot}
\rho \frac{D \varepsilon}{Dt} = -P \vec{\nabla}_{rz} \cdot \vec{v} \, ,
\end{equation}
where the $rz$ subscript on the del operator reminds us that the
derivatives with respect to $\phi$ vanish.  For
concise notation, we define a pseudo Lagrangian derivative:
\begin{equation}
\frac{D}{Dt} = \frac{\partial}{\partial t} + v_z
\frac{\partial}{\partial z} + v_r\frac{\partial}{\partial r} \, .
\end{equation}
While eqs. (\ref{eq:mass_lag_rot}) and (\ref{eq:ene_lag_rot}) are
simple extensions of the true mass (eq. \ref{eq:mass_lag}) and
energy (eq. \ref{eq:ene_lag}) equations, the presence of ``Christoffel''
terms in the momentum equation give it extra terms:
\begin{equation}
\label{eq:mom_lag_rot}
\rho \frac{D \vec{v}}{Dt} = - \vec{\nabla}_{rz} P + \rho \vec{A} \, .
\end{equation}
where the extra rotational terms are represented by $\vec{A}$:
\begin{equation}
\vec{A} = 
\left\{
\begin{array}{l}
0 \, \hat{e}_z\\
\frac{v^2_{\phi}}{r} \, \hat{e}_r\\
- \frac{v_{\phi}v_r}{r} \, \hat{e}_{\phi}\\
\end{array}
\right.
\end{equation}

To see the origin of these extra terms, one must begin with the
momentum equation in cylindrical coordinates and Eulerian form:
\begin{equation}
\label{eq:mom_cyl}
\begin{array}{rcl}
\rho \left (
\frac{\partial v_z}{\partial t} + v_z \frac{\partial v_z}{\partial z}
+  v_r \frac{\partial v_z}{\partial r} + v_{\phi}\frac{1}{r}
\frac{\partial v_z}{\partial \phi}
\right) &
= &
- \frac{\partial P}{\partial z} \\

\rho \left (
\frac{\partial v_r}{\partial t} + v_z \frac{\partial v_r}{\partial z}
+  v_r \frac{\partial v_r}{\partial r} + v_{\phi}\frac{1}{r}
\frac{\partial v_r}{\partial \phi} - \frac{v^2_{\phi}}{r}
\right) &
= &
- \frac{\partial P}{\partial r} \\

\rho \left (
\frac{\partial v_{\phi}}{\partial t} + v_z \frac{\partial v_{\phi}}{\partial z}
+  v_r \frac{\partial v_{\phi}}{\partial r} + v_{\phi}\frac{1}{r}
\frac{\partial v_{\phi}}{\partial \phi} + \frac{v_{\phi}v_r}{r}
\right) &
= &
- \frac{1}{r}\frac{\partial P}{\partial \phi} \, . \\
\end{array}
\end{equation}
Again, dropping terms involving partial derivatives of $\phi$, and
moving all terms not involving partial derivatives to the right-hand
side (RHS), eq. (\ref{eq:mom_cyl}) reduces to 
\begin{equation}
\label{eq:mom_azsymmetric}
\begin{array}{rcl}
\rho \left (
\frac{\partial v_z}{\partial t} + v_z \frac{\partial v_z}{\partial z}
+  v_r \frac{\partial v_z}{\partial r}
\right) &
= &
- \frac{\partial P}{\partial z} \\

\rho \left (
\frac{\partial v_r}{\partial t} + v_z \frac{\partial v_r}{\partial z}
+  v_r \frac{\partial v_r}{\partial r}
\right) &
= &
- \frac{\partial P}{\partial r}  + \rho \frac{v^2_{\phi}}{r}\\

\rho \left (
\frac{\partial v_{\phi}}{\partial t} + v_z \frac{\partial v_{\phi}}{\partial z}
+  v_r \frac{\partial v_{\phi}}{\partial r}
\right) &
= &
- \rho \frac{v_{\phi}v_r}{r} \, .\\
\end{array}
\end{equation}
Hence, using our definition for the pseudo Lagrangian derivative,
$D/Dt$, and $\vec{\nabla}_{rz}$ in eq. (\ref{eq:mom_azsymmetric}) we
obtain eq. (\ref{eq:mom_lag_rot}).  It is now apparent why $D/Dt$ is a
pseudo Lagrangian derivative.  Strictly, the ``extra'' rotational
terms are a result of ``Christoffel'' terms in the true Lagrangian
derivative.  Therefore, in defining $D/Dt$ without them, we've defined
a pseudo Lagrangian derivative.

Conveniently, the $\phi$ component of eq. (\ref{eq:mom_lag_rot}),
\begin{equation}
\rho \frac{D v_{\phi}}{D t}= 
- \rho \frac{v_{\phi}v_r}{r} \, ,
\end{equation}
is simply a statement of the conservation of angular momentum about the
axis of symmetry:
\begin{equation}
\label{eq:cons_ang}
\frac{D J}{D t} = 0 \, ,
\end{equation}
where $J$ is the angular momentum about the $z$-axis of a Lagrangian
mass.

Other than the appearance of an extra term on the RHS of the $r$-component
and of an equation for the $\phi$ component in
eq. (\ref{eq:mom_lag_rot}), the 2D hydrodynamic equations including
rotation about the $z$-axis are similar to the equations without
rotation.  Therefore, the algorithms for dynamics in the 2D plane
remain the same, except for the inclusion of the conservation of angular
momentum equation and the fictitious force term.

It seems natural to define $J = m_p v_{\phi,p} r_p$ and $v_{\phi,p} =
\omega_p r_p$, where $\omega_p$ is the angular velocity of node $p$.
In fact, our first implementation of rotation did just this.  However,
results indicated that this definition presents problems near the axis of rotation.  Whether
$v_{\phi,p}$ or $\omega_p$ is defined, the angular momentum at node
$p$ is associated with the region of subcells that comprise the node's
volume.
For the nodes on the axis, this definition states that $J = 0$
since $r_p =
0$, which implies that the angular momentum for the region of subcells
near the axis is zero.  This awkward feature
causes no problems
for Lagrangian calculations with no remapping.  However, during the subcell remapping step, the
subcells near the axis contain no angular momentum, yet physically
there should be some angular momentum to be remapped. 

Therefore, we have devised an alternative method for including rotational terms in 2D
simulations.
Just as for mass, the angular momentum of a subcell, $J^z_p$, is a primary
indivisible unit, which by eq. (\ref{eq:cons_ang}) is constant during
a Lagrangian hydrodynamic timestep.  The angular momentum for a cell is
\begin{equation}
J_z = \sum_{p \in S(z)} J^z_p \, ,
\end{equation}
and for each node is
\begin{equation}
\label{eq:define_jp}
J_p = \sum_{z \in S(p)} J^z_p \, .
\end{equation}
To relate angular velocity and $v_{\phi}$ to the angular momentum, we
treat each subcell as a unit with constant angular velocity.
For such regions of space the angular momentum is related to the
angular velocity by
\begin{equation}
J^z_p = \omega^z_p \int_{V^z_p} \rho^z_p r^2 dV = \omega^z_p I^z_p \, ,
\end{equation}
where $I^z_p = \int_{V^z_p} \rho^z_p r^2 dV$ is the moment of inertia
for subcell $\{z,p\}$.  
Since we are interested in accelerations and velocities at the nodes,
we choose that $\omega^z_p = \omega_p$, which naturally leads to
\begin{equation}
\label{eq:define_jzp}
J^z_p = \omega_p I^z_p \, .
\end{equation}
Substituting eq. (\ref{eq:define_jzp}) into eq. (\ref{eq:define_jp}),
we define a relationship between the angular momentum at the nodes and
the angular velocity at the nodes:
\begin{equation}
J_p = \omega_p \mathcal{I}_p \, ,
\end{equation}
where
\begin{equation}
\mathcal{I}_p = \sum_{z \in S(p)} I^z_p \, .
\end{equation}
By conservation of angular momentum, $J^{n+1}_p = J^n_p = J_p =
\omega^{n+1}\mathcal{I}^{n+1}_p$, the angular velocity at the $n+1$
timestep , $\omega^{n+1}$, is
\begin{equation}
\omega^{n+1} = \frac{J_p}{\mathcal{I}^{n+1}_p} \, .
\end{equation}

Finally, a new form for the $r$-component of $\vec{A}_p$ is
constructed.  The previously defined form is $v_{\phi}^2/r$, and
having $r$ in the denominator poses numerical problems near and at
$r=0$.  Since $v_{\phi}^2/r = \omega^2r$, the new half timestep
acceleration is
\begin{equation}
\label{eq:arot}
\vec{A}_{r,p} = \left ( \omega^2_p r_p \right )^{n+1/2} \, .
\end{equation}

\subsubsection{Conservation of Energy with Rotation}
\label{section:consenerot}

Without rotation and time-varying gravity, the equations for 2D
simulations using cylindrical coordinates conserve energy to machine
precision.  However, with rotation, strict energy
conservation is lost.  Using the equations for energy and momentum
including rotation, eqs. (\ref{eq:ene_lag_rot}) and
(\ref{eq:mom_lag_rot}) the energy integral becomes
\begin{multline}
\int_v \left ( \rho \frac{D \varepsilon}{Dt} + \frac{1}{2} \rho
  \frac{D v^2}{Dt}\right ) dV
= \int_v \left ( - P \vec{\nabla}_{rz} \cdot \vec{v} \right . \\
\left . - \vec{v} \cdot
  \vec{\nabla}_{rz} P  
+ \rho (v_{r}A_r + v_{\phi}A_{\phi}) \right ) dV \, .
\end{multline}
Since the terms $v_rA_r$ and $v_{\phi}A_{\phi}$ are equal in magnitude
and opposite in sign, the rate of change for the total energy reduces
to the usual surface integral:
\begin{equation}
\frac{\partial E}{\partial t}= \oint P\vec{v} \cdot d \vec{S} \, .
\end{equation}
If the form of the energy equation is to remain the same, discrete
versions of $v_rA_r$ and $v_{\phi}A_{\phi}$ must be derived that
cancel one another.

In the previous section, we made
arguments not to use $J = m_p v_{\phi} r$.  However, it
is simpler and more intuitive to illustrate why the 2D equations, including rotation, can
not satisfy strict energy conservation, while satisfying angular
momentum conservation.  With conservation of angular
momentum we have $v^{n+1}_{\phi}r^{n+1} = v^n_{\phi}r^n$ or
\begin{equation}
v^{n+1}_{\phi} = v^n_{\phi} \left ( \frac{r^{n}}{r^{n+1}} \right ) \, .
\end{equation}
The discrete time derivative of $v_{\phi}$ is then
\begin{equation}
\label{eq:aphi_timelevels}
\frac{\Delta v_{\phi}}{\Delta t} = - \frac{v^n_{\phi}
  v^{n+1/2}_r}{r^{n+1}} \, ,
\end{equation}
where we have used $\Delta r = v^{n+1/2}_r \Delta t$.  This specifics
the required time levels for the components of $A_{\phi}$ to conserve
angular momentum.  Considering time integration, the discrete
analog of $v_rA_r + v_{\phi}A_{\phi}$ is $v^{n+1/2}_rA_r +
v^{n+1/2}_{\phi}A_{\phi}$.  To ensure that this sums to zero, the discrete analog of $A_r$ is
\begin{equation}
A_{r} = \frac{v^{n+1/2}_{\phi} v^n_{\phi}}{r^{n+1}} \, .
\end{equation}
With these time levels, the 2D equations, which include rotation,
would satisfy energy and angular momentum conservation to round-off error.  However,
$r^{n+1}$ and $v^{n+1/2}_{\phi}$ are not known at the beginning of the
timestep.  It is only
after the predictor step of the predictor-corrector method that
$v^{n+1/2}_{\phi}$ and $r^{n+1,pr}$ are obtained. As a result, $A_r$,
including the correct time levels and predictor values, is
\begin{equation}
A_{r} = \frac{v^{n+1/2}_{\phi} v^n_{\phi}}{r^{n+1,pr}} \, ,
\end{equation}
but to conserve angular momentum and energy, $r^{n+1}$ appears in the denominator.
Therefore, the terms $v^{n+1/2}_{\phi}A_{\phi}$ and $v^{n+1/2}_rA_{r}$
will not be exactly opposite and equal in magnitude, and angular
momentum and energy conservation can not be enforced to machine
precision at the same time.  

In the previous section, we motivated a different discretization of
angular momentum, angular velocity, and $v_{\phi}$.  However, the
basic results of this section do not change.  There is still the fundamental
problem that to strictly conserve angular momentum, positions at the
$n+1$ time level are required, while one must use the
predicted, $n+1,pr$, values in updating the radial velocity.
Sacrificing strict energy conservation, we have chosen to strictly
conserve angular momentum for simulations including rotation.

\subsubsection{Simple Rotational Test}
\label{section:rot_test}

We have designed a test that isolates and tests rotation.
Using cylindrical coordinates, the calculational domain lies in the
$r-z$ plane and extends
from the symmetry axis, $r = 0$, to $r = 1.0$ cm and from $z=-0.1$ cm to
$z=0.1$ cm .  The initial density is homogeneous with $\rho = 1.0$ g cm$^{-3}$.
Isolating the rotational forces, we eliminate pressure gradients by setting the
internal energy everywhere to zero.  Consequently,
the trajectory of a Lagrangian mass is a straight line in 3D and has a
very simple analytic evolution for $r(t)$, determined only by its initial
position and velocity.  We set $v_z = 0$ and give $v_{\phi}$ and $v_r$
homologous profiles: $v_r = -3 r / r_0$ cm s$^{-1}$ and $v_{\phi} = r
/r_0$ cm s$^{-1}$.  
This produces a homologous self-similar solution for $v_r$ and
$v_{\phi}$ that our algorithm for rotation should reproduce.

\begin{figure}
\epsscale{0.6}
\plotone{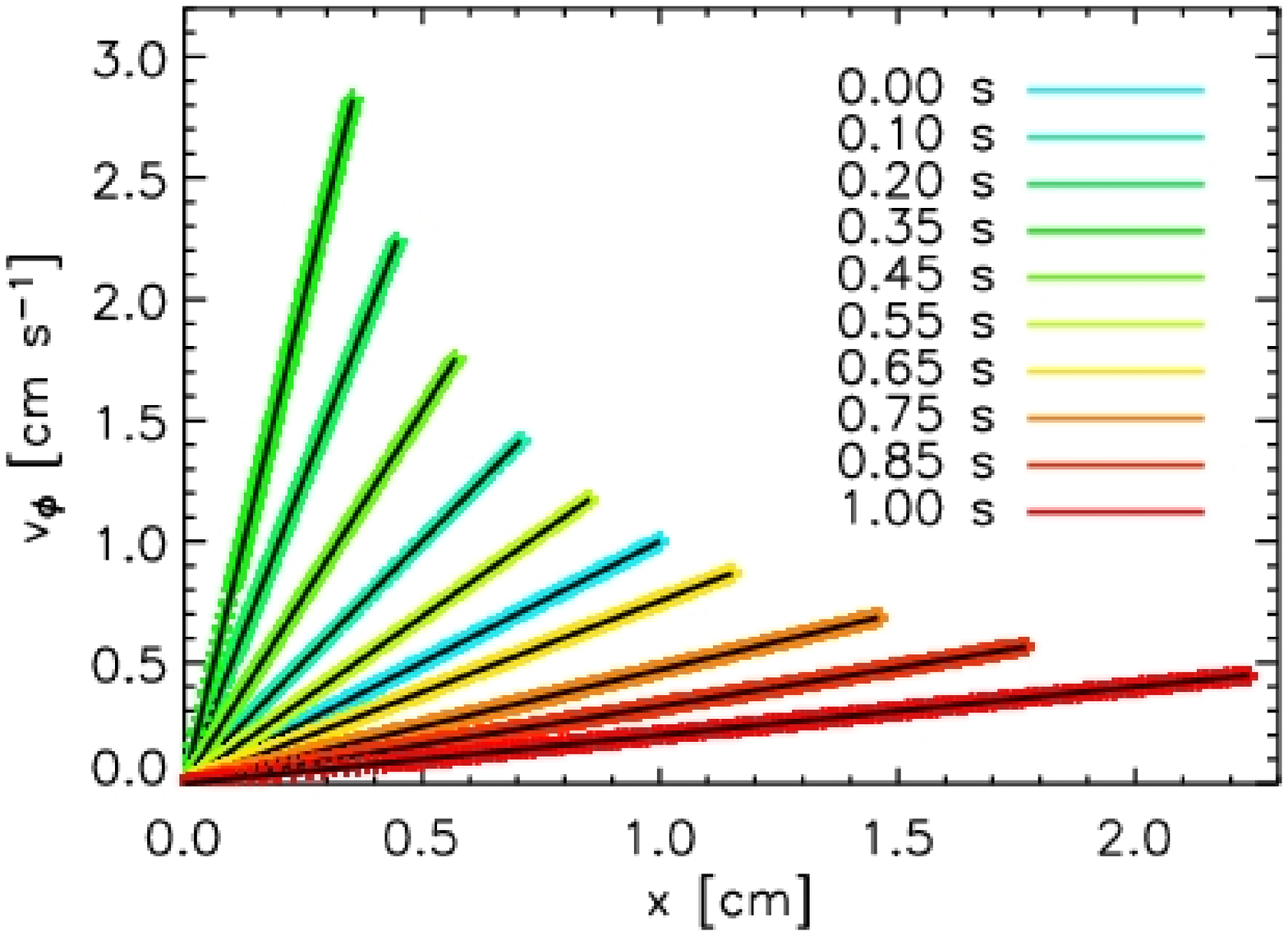} \\
\plotone{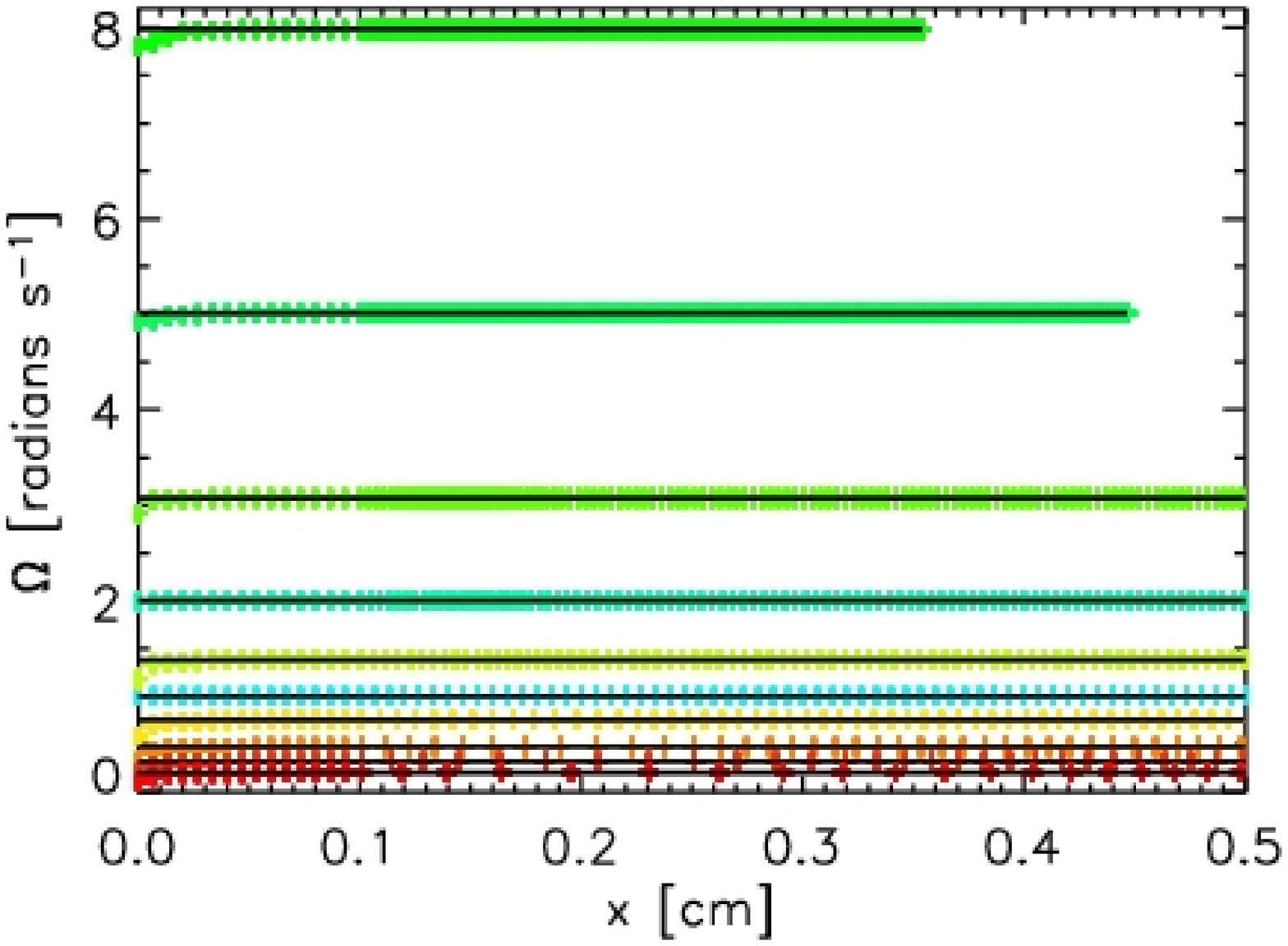}
\caption{Plots of $v_{\phi}$ vs. $r$ (top
panel) and $\Omega$ vs. $r$ (bottom panel) for the simple rotational
test discussed in \S \ref{section:rot_test}.  We employ three remapping regions.  The inner 0.1 cm is
Eulerian, zones exterior to 0.2 cm follow Lagrangian dynamics, and the
region in between provides a smooth transition between the two domains.
$v_{\phi}$ and  $\Omega$ profiles are presented at $t =
  0.0, 0.1, 0.2, 0.35, 0.45, 0.55, 0.65, 0.75, 0.85$, and $1.0$ s.
In the top panel, the deviation of the simulation (crosses) from the analytic solution (solid lines) is not
discernible.  The maximum error as measured by
$(v_{\phi,{\rm ana}} - v_{\phi})/ {\rm max}(v_{\phi,{\rm ana}})$ is
$\sim 4 \times 10^{-5}$ near the beginning of the simulation and
$\sim 8 \times 10^{-4}$ at the end.  The $\Omega$ vs. $r$ panel (bottom) shows similar accuracy, except at the center, where
$(\Omega_{\rm ana} -\Omega)$ reaches a maximum value of 0.2 rads
s$^{-1}$ at $t = 0.55$ s.
\label{rottestplot}}
\end{figure}

In Figure \ref{rottestplot}, plots of $v_{\phi}$ vs. $r$ (top
panel) and $\Omega$ vs. $r$ panel (bottom) show that BETHE-hydro
performs well with this test.  Simultaneously testing the rotational remapping (\S
\ref{section:ang_vel_remap}) and hydrodynamic algorithms, we employ three remapping regions.  The inner 0.1 cm is
Eulerian, zones exterior to 0.2 cm follow Lagrangian dynamics, and the
region in between provides a smooth transition between the two domains.
Both $v_{\phi}$ and  $\Omega$ profiles are presented at $t =
  0.0, 0.1, 0.2, 0.35, 0.45, 0.55, 0.65, 0.75, 0.85$, and $1.0$ s.
As designed the angular momentum is conserved to machine accuracy.
In the top panel ($v_{\phi}$ vs. $r$), it is hard to discern any
deviation of the simulation (crosses) from the analytic
solution (solid line).  In fact, the maximum error as measured by
$(v_{\phi,{\rm ana}} - v_{\phi})/ {\rm max}(v_{\phi,{\rm ana}})$
ranges from
$\sim 4 \times 10^{-5}$ near the beginning of the simulation to
$\sim 8 \times 10^{-4}$ at the end.  $\Omega$ vs. $r$ (bottom panel) shows similar
accuracy, except for the region near the axis.  At the center $(\Omega_{\rm ana} -\Omega)$
reaches a maximum value of 0.2 rads s$^{-1}$ at $t = 0.55$ s.  On the
whole, Fig. \ref{rottestplot} demonstrates that this code reproduces the analytic result with and without remapping.

\section{Gravity}
\label{section:gravity}

There are two general strategies for solving Poisson's equation for
gravity on
arbitrary grids.  The first is to define another regular grid,
interpolate the density from the hydro grid to this new gravity grid, and
use many of the standard Poisson solvers for regular grids.  However,
in our experience this approach can lead to numerical errors, which
may lead to unsatisfactory results.  The second
approach is to solve eq. (\ref{eq:poisson}) explicitly on the
arbitrary hydro grid, eliminating an interpolation step.  We prefer the
latter approach, giving potentials and accelerations that are more
consistent with the hydrodynamics on an unstructured mesh. 

In determining the discrete analogs of eq. (\ref{eq:poisson}), we use the support-operator method.  After a bit of algebra these discrete equations
may be expressed as a set of linear equations whose unknowns are the
discretized gravitational potentials.  Solving for these potentials is
akin to solving
a matrix equation of the form, ${\bf A}\vec{x} = \vec{b}$, and since the solutions change little from
timestep to timestep, we use an iterative matrix inversion approach
in which the initial guess is the solution from the previous timestep. 

In solving Poisson's equation for gravity, we seek an algorithm
satisfying the following conditions:  1) the potential should be defined at zone
centers and the gravitational acceleration should be defined at the
nodes;  2) we should conserve momentum and energy as accurately
as possible;  and 3) the solver must be fast.  To this end, we employ
the method described in \citet{shashkov95},
\citet{shashkov96}, and \citet{morel98} for elliptic solvers.
Specifically, we follow the particular implementation of
\citet{morel98}.

Applying the support-operator method, we first define the divergence operator and, in
particular, define $(\vec{\nabla} \cdot \vec{g})_z$.
Poisson's equation for gravity may
be written as an equation for the divergence of the acceleration,
\begin{equation}
\label{eq:poisson_div}
- \vec{\nabla} \cdot \vec{g} = 4 \pi G \rho \, ,
\end{equation}
 where $\vec{g}$, the gravitational acceleration, is the negative of the
 gradient of the potential.
\begin{equation}
\label{eq:gacc}
\vec{g} = - \vec{\nabla} \Phi \, .
\end{equation}
Integrating eq. (\ref{eq:poisson_div}) over a finite volume gives
\begin{equation}
\label{eq:divtheorem}
\int_V \vec{\nabla} \cdot \vec{g} dV = \oint_{\partial V} \vec{g}
\cdot d\vec{S} = - \int_V 4 \pi G \rho dV \, ,
\end{equation}
and if the volume of integration, $V$, is chosen to be that of a cell, the RHS
of eq. (\ref{eq:divtheorem}) becomes $- 4 \pi G m_z$.
In discretizing the middle term in eq. (\ref{eq:divtheorem}),
\begin{equation}
\label{eq:discretegauss}
\oint_{\partial V_z} \vec{g} \cdot d\vec{S} \longrightarrow
\sum_e g_e A_e \, ,
\end{equation}
an expression for the divergence of $\vec{g}$ follows naturally:
\begin{equation}
\label{eq:discretediv}
( \vec{\nabla} \cdot \vec{g})_z =
\frac{1}{V_z} \sum_{e \in S(z)} g_e A_e \, ,
\end{equation}
where 
$A_e$ is the area of edge $e$, and $g_e$ is the component of the
acceleration parallel to the direction of the edge's area vector.
Therefore, substituting eq. (\ref{eq:discretediv}) into
eq. (\ref{eq:divtheorem}) suggests a discrete expression for eq. (\ref{eq:poisson_div}):
\begin{equation}
\label{eq:discretepoisson_div}
- \sum_e g_e A_e = 4 \pi G m_z \, .
\end{equation}

Having defined
the discrete divergence of a vector, we then use the support-operator method to write the
discrete equivalent of the gradient.  Once again, the integral identity relating
the divergence and gradient operators is given by eq. (\ref{eq:intidentity}).
Applying the definition for
the discrete divergence, the first term on the LHS of
eq. (\ref{eq:intidentity}) is approximated by
\begin{equation}
\label{eq:intidentity_lhs1}
\int_V \Phi (\vec{\nabla} \cdot \vec{H}) dV
\sim \Phi_z \sum_e h_e A_e \, ,
\end{equation}
where $h_e$ is the component of $\vec{H}$ perpendicular to edge $e$,
and the RHS of eq. (\ref{eq:intidentity}) is approximated by
\begin{equation}
\label{eq:intidentity_rhs}
\oint_{\partial V} \Phi \vec{H} \cdot d\vec{S}
\sim \sum_e \Phi_e h_e A_e \, .
\end{equation}
It is already apparent that the discrete potential is
defined for both cell centers ($\Phi_z$) and edges ($\Phi_e$).  This
fact seems cumbersome.  However, we will demonstrate below that in the
final equations the number of unknowns may be reduced by eliminating
the cell-centered potentials, $\Phi_z$, in favor of the edge-centered
potentials, $\Phi_e$.
Completing the discretization of eq. (\ref{eq:intidentity}), the second
term on the LHS is
\begin{equation}
\label{eq:intidentity_lhs2}
\int_V \vec{H} \cdot \vec{\nabla} \Phi dV
\sim - \sum_p \vec{H}_p \cdot \vec{g}_p W^z_p V_z \, ,
\end{equation}
where $W^z_p$ is a volumetric weight and $\vec{H}_p$ and $\vec{g}_p$
are defined at the nodes.  The volumetric weight associated with each
corner is defined as one quarter of the area of the parallelogram
created by the sides that define the corner.  Additionally, the
weights are normalized so that $\sum_{p \in N(z)} W^z_p = 1$.  The
remaining task is to define the dot product $\vec{H}_p \cdot
\vec{g}_p$ at each corner when the vector is expressed in components
of a nonorthogonal basis set:
\begin{equation}
\label{eq:h_basis}
\vec{H}_p = h_e \hat{e}_e+ h_{e-1} \hat{e}_{e-1} \, ,
\end{equation}
where $\hat{e}_{e-1}$ and $\hat{e}_e$ are basis vectors located at the
center of the edges on either side of the corner associated with node
$p$ with orientations perpendicular to the edges, and $h_{e-1}$ and
$h_e$ are the corresponding magnitudes in this basis set.  With this
basis set, the inner product is
\begin{equation}
\vec{H}_p \cdot \vec{g}_p = \langle S^p_{e,e-1}\vec{H}_p,\vec{g}_p \rangle \, ,
\end{equation}
where
\begin{equation}
\label{eq:smatrix}
S^p_{e,e-1} = 
\frac{1}{\sin^2 \theta^p_{e,e-1}}
\left [
\begin{array}{cc}
1 & \cos \theta^p_{e,e-1} \\
\cos \theta^p_{e,e-1} & 1 \\
\end{array}
\right ] \, ,
\end{equation}
and $\theta^p_{e,e-1}$ is the angle formed by edges $e$ and $e-1$.
Using eqs. (\ref{eq:h_basis})-(\ref{eq:smatrix}) in
eq. (\ref{eq:intidentity_lhs2}), the second term on the LHS
of eq. (\ref{eq:intidentity}) is
\begin{multline}
- \sum_p \vec{H}_p \cdot \vec{g}_p W^z_p V_z 
= - \sum_e h_e \left ( (\alpha_p +\alpha_{p+1})g_e \right . \\
\left . + \beta_p g_{e-1} + \beta_p g_{e+1} \right )
\end{multline}
where $\alpha_p = W^z_p V_z / \sin^2 \theta_p$ and $\beta_p = W^z_p
V_z \cos \theta_p / \sin^2 \theta_p$.
Combining eqs. (\ref{eq:intidentity_lhs1}), (\ref{eq:intidentity_rhs}), and (\ref{eq:intidentity_lhs2}), we have the discrete form of the
identity eq. (\ref{eq:intidentity}):
\begin{multline}
\label{eq:intidentity_discrete}
\sum_e \left [
(\Phi_e - \Phi_z)h_eA_e - h_e \left ( (\alpha_p +\alpha_{p+1})g_e \right . \right . \\
\left . \left . + \beta_p g_{e-1} + \beta_p g_{e+1} \right )
\right] = 0 \, .
\end{multline}

To find the equation for $g_e$ on each edge we set the
corresponding $h_e = 1$ and set all others to zero.  This gives an
expression for $g_e$ in terms of $\Phi_e$, $\Phi_z$, and the other
edge-centered accelerations of cell $z$: 
\begin{multline}
\label{eq:ge_equation}
(\Phi_e - \Phi_z)A_e - \left [ (\alpha_p +\alpha_{p+1})g_e \right . \\
\left . + \beta_p g_{e-1} + \beta_p g_{e+1} \right ] = 0
\end{multline}
or
\begin{multline}
\left [ (\alpha_p +\alpha_{p+1})g_e 
+ \beta_p g_{e-1} + \beta_p g_{e+1} \right ] \\
= (\Phi_e - \Phi_z)A_e \, .
\end{multline}
Together, the set of equations for the edges of cell $z$ forms a
matrix equation of the form\footnote{
On a practical note, all terms involving $g_e$ on the $z$-axis are
zero because the areas on the axis of symmetry are zero.
Therefore, for the zones along the axis, one less equation and variable appear in the set of equations
for $g_e$.}
\begin{equation}
{\bf Q} \vec{g} = \vec{b} \, ,
\end{equation}
where $\vec{b}$ is a vector based upon the edge values, $b_e =
A_e(\Phi_z - \Phi_e)$.  Inverting this equation gives an expression
for the edge accelerations as a function of the cell- and
edge-centered potentials:
\begin{equation}
\label{eq:gacc_phiz_phie}
g_e = \sum_{e'} a_{ee'} A_{e'}(\Phi_z - \Phi_{e'}) \, .
\end{equation}
After finding the edge acceleration for each zone,
these values are inserted into eq. (\ref{eq:discretepoisson_div}).
The stencil of potentials for this equation for each cell includes the
cell center and all edges.  While eq. (\ref{eq:gacc_phiz_phie}) gives
the edge accelerations associated with cell $z$, there is no
guarantee that the equivalent equation for a neighboring cell will
give the same acceleration for the same edge.
To ensure continuity, we must include an equation which equates
an edge's acceleration vectors from the neighboring cells:
\begin{equation}
\label{eq:gacc_continuity}
-A_e g_{e,L} - A_e g_{e,R} = 0 \, ,
\end{equation}
where $g_{e,L}$ and $g_{e,R}$ are the edge accelerations as determined
by eq. (\ref{eq:gacc_phiz_phie}) for the left ($L$) and right ($R$)
cells.  While we could have chosen a simpler expression for
eq. (\ref{eq:gacc_continuity}), the choice of sign and coefficients
ensures that the eventual system of linear equations involves a
symmetric positive-definite matrix.  This enables the use of fast and
standard iterative matrix inversion methods such as the
conjugate-gradient method. The stencil for this equation involves the
edge in question, the 
cell centers on either side, and the rest of the edges associated with
these cells.  At first glance, it seems that we need to invert the
system of equations that includes the cell-centered equations,
eq. (\ref{eq:gacc_phiz_phie}), and the edge-centered equations, eq. (\ref{eq:gacc_continuity}).  Instead, we use
eq. (\ref{eq:discretepoisson_div}) to express the cell-centered
potentials in terms
of edge-centered potentials and the RHS of the equation:
\begin{equation}
\label{eq:phizeq}
\Phi_z = \frac{1}{\zeta_z}
\left (
\sum_{e'} A_{e'} \Phi_{e'} \psi_{e'} - 4 \pi G m_z
\right ) \, ,
\end{equation}
where $\psi_e = \sum_{e'} A_e'a_{e'e}$, and $\zeta_z = \sum_e
\sum_{e'} A_e a_{ee'}A_{e'} = \sum_{e'}A_{e'}\psi_{e'}.$  Substituting
eq. (\ref{eq:phizeq}) into the expression for the edge acceleration,
eq. (\ref{eq:gacc_phiz_phie}),
\begin{equation}
\label{eq:gacc_phie}
g_k = - \frac{\eta_k}{\zeta_z} 4 \pi G m_z + \sum_e \Phi_e A_e 
\left (
\frac{\eta_k}{\zeta_z} \psi_e - a_{ke}
\right ) \, ,
\end{equation}
where $\eta_k = \sum_e a_{ke} A_e$.  Finally, substituting
eq. (\ref{eq:gacc_phie}) into eq. (\ref{eq:gacc_continuity}) gives the matrix equation for gravity with
only edge-centered potential unknowns:
\begin{multline}
A_*\sum_{e^R} \Phi_e A_e(a_{*e}^R - \frac{\eta^R_* \psi^R_e}{\zeta_R})
+ A_*\sum_{e^L} \Phi_e A_e(a_{*e}^L - \frac{\eta^L_*
  \psi^L_e}{\zeta_L}) \\
= -4 \pi G A_* (\frac{\eta^R_*}{\zeta_R} m_R + \frac{\eta^L_*}{\zeta_L} m_L) \, .
\end{multline}
The stencil for this equation involves the edge and all other edges
associated with the zones on either side of
the edge in question.  Taken together, the equations result in a set of linear equations for the edge-centered
potentials, $\Phi_e$.  

Since this class of matrix equations is ubiquitous,
there exists many accurate and fast linear system
solvers we can exploit.  To do so, we recast our discrete
form of Poisson's equation as
\begin{equation}
\label{eq:linear}
\mathbf{A} \vec{\Phi} = \vec{s} \, ,
\end{equation}
where $\vec{\Phi}$ is a vector of the edge-centered potentials,
 and $\vec{s}$
is the corresponding vector of source terms.
We solve the matrix equation using the conjugate-gradient method with
a multigrid preconditioner.  In particular, we use the algebraic
multigrid package, AMG1R6\footnote{See \S \ref{section:corecollapse}
  for details on performance.} \citep{ruge87}.

\subsection{Gravitational Acceleration}
\label{section:gravity_acceleration}

Unfortunately, this discretization scheme does not adequately define
the accelerations at the nodes where they are needed in eq. (\ref{eq:mom_discrete2}).  Therefore, we use the least-squares
minimization method to determine the gradient on the unstructured
mesh.
Assuming a linear
function for the potential in the neighborhood of node $i$, $\hat{\Phi}_i(x,y)$, we seek to minimize the difference between
$\Phi_k$ and $\hat{\Phi}_i(x_k,y_k)$, where $\Phi_k$ is the neighbor's
value and $\hat{\Phi}_i(x_k,y_k)$ is the evaluation of
$\hat{\Phi}_i(x,y)$ at
the position of the neighbor, $(x_k,y_k)$.
More explicitly, we minimize the following equation:
\begin{equation}
\label{eq:least_gacc_eq1}
\sum_{k \in \mathcal{N}(i)} \omega^2_{ki} E^2_{ik} \, ,
\end{equation}
where the ``neighbors'' are the nearest cell-centers and
edges to node $i$ and are denoted by $k \in \mathcal{N}(i)$,
\begin{equation}
E^2_{ik} = \left (
\Phi_i - \Phi_k + \Phi_{x,i} \Delta x_{ik} + \Phi_{y,i} \Delta y_{ik}
\right )^2 \, ,
\end{equation}
$\Delta x_{ik} = x_k - x_i$, $\Delta y_{ik} = y_k - y_i$, and $\omega^2_{ik} = 1/(\Delta x_{ik}^2 + \Delta y_{ik}^2)$.

Usually, minimization of
eq. (\ref{eq:least_gacc_eq1}) leads to a matrix equation for two
unknowns: $\Phi_{x,i}$ and $\Phi_{y,i}$, the gradients for $\Phi$ in
the $x$ and $y$ directions, respectively (\S \ref{section:swept}).  However, $\Phi_i$, the potential at the node
is not defined and is a third unknown with which
eq. (\ref{eq:least_gacc_eq1}) must be minimized.
Performing the least-squares minimization process with respect to the
three unknowns, $\Phi_i$, $\Phi_{x,i}$, and $\Phi_{y,i}$, leads to the
following set of linear equations for each node:
\begin{equation}
\begin{array}{rcl}
a \Phi_i + b \Phi_{x,i} + c \Phi_{y,i} & = & k\\
d \Phi_i + e \Phi_{x,i} + g \Phi_{y,i} & = & l\\
h \Phi_i + i \Phi_{x,i} + j \Phi_{y,i} & = & m\\
\end{array}
 \, ,
\end{equation}
where 
\begin{equation}
\begin{array}{rcl}
a & = & \sum_{k} \omega^2_{ik}\\
b & = & \sum_{k} \omega^2_{ik} \Delta x_{ik}\\
c & = & \sum_{k} \omega^2_{ik} \Delta y_{ik}\\
d & = & \sum_{k} \omega^2_{ik} \Delta x_{ik}\\
e & = & \sum_{k} \omega^2_{ik} \Delta x^2_{ik}\\
g & = & \sum_{k} \omega^2_{ik} \Delta x_{ik} \Delta y_{ik}\\
h & = & \sum_{k} \omega^2_{ik} \Delta y_{ik}\\
i & = & \sum_{k} \omega^2_{ik} \Delta x_{ik} \Delta y_{ik}\\
j & = & \sum_{k} \omega^2_{ik} \Delta y^2_{ik}\\
k & = & \sum_{k} \omega^2_{ik} \Phi_k\\
l & = & \sum_{k} \omega^2_{ik} \Phi_k \Delta x_{ik}\\
m & = & \sum_{k} \omega^2_{ik} \Phi_k \Delta y_{ik}\\
\end{array}
\end{equation}
Inversion of this linear system gives the potential and the gradient
at the node.  Specifically, we use the adjoint matrix inversion method
to find the inverse matrix and the three unknowns, including $\Phi_{x,i}$ and $\Phi_{y,i}$.

\subsection{Gravitational Boundary Conditions}
\label{section:gravity_boundary}

The outer boundary condition we employ for the Poisson solver is a
 multipole expansion for the gravitational potential at a spherical
 outer boundary.  Assuming no material exists outside the calculational
 domain and that the potential asymptotes to zero, the potential at the
 outer boundary, ${\bf r}$, is given by
\begin{equation}
\label{eq:boundary_eq1}
\phi({\bf r}) = - G \sum_{n=0}^{\infty} \frac{1}{r^{n+1}} \int
(r')^nP_n(\cos \theta') \rho({\bf r}')dV' \, ,
\end{equation}
where $P_n(\cos \theta')$ is the usual Legendre polynomial.  In
discretized form, eq. (\ref{eq:boundary_eq1}) becomes
\begin{equation}
\phi({\bf r}) = - G \sum_{n=0}^{N_n} \frac{1}{r^{n+1}} \sum_z
(r_z)^nP_n(\cos \theta_z) m_z \, ,
\end{equation}
where $N_n$ is the order at which the multipole expansion is truncated.

\subsection{Gravity: 1D Spherical Symmetry}
\label{section:gravity_1d}

For 1D spherically symmetric simulations, calculation of gravity can
be straightforward using
\begin{equation}
\label{eq:gacc_enclosed}
g_p = - \frac{G m_{p,{\rm enclosed}}}{r_p^2} \, ,
\end{equation}
where $m_{p,{\rm enclosed}}$ is all of the mass enclosed by point
$p$.  Instead, we solve Poisson's equation for
gravity using similar methods to those employed for the 2D gravity.  By doing
this, we remain consistent when comparing our 1D and 2D results.
Furthermore, we show that deriving a Poisson solver based upon
the support-operator method is equivalent to the more traditional form, eq. (\ref{eq:gacc_enclosed}).

The inherent symmetries of 1D spherical simulations reduce
eq. (\ref{eq:ge_equation}) to the following form:
\begin{equation}
(\Phi_e - \Phi_z)A_e - V^z_p g_{e,z} = 0 \, .
\end{equation}
Unlike the full 2D version, one may eliminate the edge potentials in favor
for the cell-centered potentials and the accelerations at the edges.
Since each edge $e$ will have an equation for its value associated with
cells $z$ and $z+1$, this leads to
\begin{equation}
\Phi_z A_e + V^z_p g_{e,z} = \Phi_{z+1} A_e + V^{z+1}_p g_{e,z+1} \, .
\end{equation}
Using the continuity of accelerations at the edges,
\begin{equation}
g_{e,z} = - g_{e,z+1} \, ,
\end{equation}
the expression for gravitational acceleration at edge $e$ in terms of
the cell-centered potentials is
\begin{equation}
\label{eq:gacc_1dsph}
g_{e,z} = \frac{1}{V_p}(\Phi_{z+1} - \Phi_z) A_e \, .
\end{equation}
Substituting this expression into
\begin{equation}
\label{eq:1dgaussgravity}
\sum_e A_e g_{e,z} = - 4 \pi G m_z \, ,
\end{equation}
leads to a tridiagonal matrix equation for the cell-centered
potential, which can be inverted in O(N) operations to give the cell-centered potentials.  Substituting these cell-centered potentials back into
eq. (\ref{eq:gacc_1dsph}) gives the gravitational acceleration needed
for the momentum equation (eqs. \ref{eq:mom_lag} and \ref{eq:mom_discrete2}).

It should be noted that eq. (\ref{eq:1dgaussgravity}) can be rewritten
as a recursion relation for the gravitational acceleration, $g_p$, at node
$p$:
\begin{equation}
\label{eq:recursive}
A_p g_p = A_{p-1} g_{p-1} - 4 \pi G m_{z,{\rm interior}} \, ,
\end{equation}
where $m_{z,{\rm interior}}$ is the mass of the cell
interior to point $p$.  With the boundary condition $g_1 = 0$ and
then recursively solving eq. (\ref{eq:recursive}) for $g_p$ from the
center outward, the expression for the
gravitational acceleration simplifies to eq. (\ref{eq:gacc_enclosed}).
This is the traditional form by which the gravitational acceleration
is calculated for 1D spherically symmetric simulations and for which
the potential is usually not referenced.  The beauty of our approach
is that 1D and 2D simulations are consistent, and that it
self-consistently gives the gravitational potential.  

\subsection{Conservation of Energy with Gravity}
\label{section:consenegravity}

For self-gravitating hydrodynamic systems, the total energy is
\begin{equation}
\label{eq:totalenergy}
E = \mathcal{E} + K + U \, ,
\end{equation}
where $\mathcal{E}$ is the total internal energy, $K$ is the total
kinetic energy, and $U$ is the total potential energy:
\begin{equation}
\label{eq:totpot1}
U = \frac{1}{2}\int_V \rho \Phi dV \, ,
\end{equation}
An alternative form is possible with the substitution of
eq. (\ref{eq:poisson_div}) into eq. (\ref{eq:totpot1}) and integration
by parts:
\begin{equation}
\label{eq:totpot2}
U = -\frac{1}{8 \pi G} \left [ \oint_S \vec{g}\Phi \cdot d\vec{S}
 + \int_V |\vec{g}|^2 dV
\right ] \, .
\end{equation}
The integral form of eq. (\ref{eq:totalenergy}) is
\begin{equation}
E = \int_V (\rho e + \frac{1}{2}\rho v^2 + \frac{1}{2}\rho \Phi) dV \, ,
\end{equation}
and total energy conservation is given by
\begin{equation}
\label{eq:dedt}
\frac{\partial E}{\partial t} = \frac{\partial}{\partial t}\int_V
(\rho e + \frac{1}{2}\rho v^2 + \frac{1}{2}\rho \Phi) dV = 0\, .
\end{equation}
Since the solution to Poisson's equation for
gravity, using Green's function, is
\begin{equation}
\Phi(\vec{x}) = -G \int{\frac{\rho(\vec{x}') dV'}{|\vec{x} - \vec{x}'|}} \, ,
\end{equation}
the time derivative of the total potential energy may take three forms:
\begin{equation}
\label{eq:dudt}
\int_V \rho \frac{\partial \Phi}{\partial t}dV = \int_V \Phi
\frac{\partial \rho}{\partial t}dV = \frac{1}{2} \int_V \frac{\partial (\rho
  \Phi)}{\partial t}dV \, .
\end{equation}
Note that this is true only for the total potential energy and does
not imply $\rho \frac{\partial \Phi}{\partial t} = \Phi \frac{\partial \rho}{\partial t} = \frac{1}{2} \frac{\partial (\rho
  \Phi)}{\partial t}$.
Using eq. (\ref{eq:dudt}) and the Eulerian form of the hydrodynamics equations  (see appendix \ref{section:eulerian_hydro}),
eq. (\ref{eq:dedt}) becomes a surface integral:
\begin{equation}
\frac{\partial E}{\partial t} = 
- \oint_S (\rho e + P + \frac{1}{2}\rho v^2
  + \rho \Phi) \vec{v}
\cdot d \vec{S} \, ,
\end{equation}
which simply states that energy is conserved in the absence of a flux
of energy through the bounding surface.

The discrete analogs of $\mathcal{E}$ and
$K$ are trivially obtained using $\mathcal{E} = \sum_z m_z \varepsilon_z$
and $K = \frac{1}{2}\sum_p m_p v^2_p$, respectively.
The discrete total potential energy may take two forms, the analogs of
eqs. (\ref{eq:totpot1}) and (\ref{eq:totpot2}):
\begin{equation}
U = 
\left \{
\begin{array}{l}
\frac{1}{2}\sum_z m_z \Phi_z \\
-\frac{1}{8 \pi G} \left [
\sum_{pb} \Phi_{pb} \vec{g}_{pb} \cdot \vec{S}_{pb}
+ \sum_p \vec{g}_p \cdot \vec{g}_p V_p
\right ] \, ,\\
\end{array}
\right . 
\end{equation}
where $pb$ indicates outer-boundary values.
Upon trying both forms, we get similar results.  Therefore, we use the
simpler form involving the cell-center potential: $U = \frac{1}{2}\sum_z m_z \Phi_z$.

Because our discrete hydrodynamics equations including gravity do not give strict
energy conservation, we use the
core-collapse simulation (\S \ref{section:corecollapse}) to gauge how
well energy is conserved.  
Rather than measuring the relative error in total energy by $\Delta E /
E_{\rm ref}$, where $\Delta E = E_n - E_{\rm ref}$, $E_{\rm ref}$ is
the initial total energy, and $E_n$ is the
total energy at timestep $n$, we use $\Delta E / |U_n|$.  For stellar
profiles, the kinetic energy is small and the internal and gravitational
energies are nearly equal in magnitude, but opposite in sign.
Since the total energy is roughly zero
in comparison to the primary constituents, $\mathcal{E}$ and $U$, of the total energy, we use the
gravitational potential energy as a reference.  For example, the internal
and gravitational energies start at $\sim 4 \times
10^{51}$ erg and reach $\sim 1 \times 10^{53}$ ergs at the end of the
run.  However, the total energy is a small fraction of these energies
initially, $\sim 5\times 10^{50}$,
and after core bounce, $\sim 1.5 \times 10^{51}$.  Hence, we measure
the relative error in total energy as $\Delta E / |U_n|$.

The total energy for the 1D and 2D core-collapse simulations evolve similarly and is
conserved quite well.  For all times except a few milliseconds around
bounce, total energy is conserved better than $\sim$1 $\times
10^{-3}$.  During collapse, from 0 to 147 ms, the total energy deviates by only $\sim$3
$\times 10^{-4}$.  Over a span of 5 ms around bounce, $t = 148$ ms, the total energy changes by
$\sim$2.2 $\times 10^{-2}$.  For 100s of milliseconds afterward, the
relative error in the total energy reduces to $\sim$7 $\times 10^{-4}$.
Hence, for all but about 5 ms of a simulation that lasts many hundreds
of milliseconds the relative error in total energy is better than $\sim$1 $\times 10^{-3}$.

\subsection{Tests of Gravity}
\label{section:gravity_tests}

We include here several assessments of the 1D and 2D Poisson solvers.  First, we calculate, using the
butterfly mesh, the potential
for a homogeneous sphere and compare to the
analytic solution.  This tests overall accuracy and
the ability of our solver to give spherically symmetric potentials
when a non-spherical mesh is employed.  Similarly, we substantiate the
algorithm's ability to produce aspherical potentials of homogeneous oblate spheroids.  Then, we verify that the
hydrodynamics and gravity solvers give accurate results for stars in
hydrostatic equilibrium, and for a dynamical problem, the Goldreich-Weber
self-similar collapse \citep{goldreich80}.

\begin{figure}
\epsscale{0.6}
\plotone{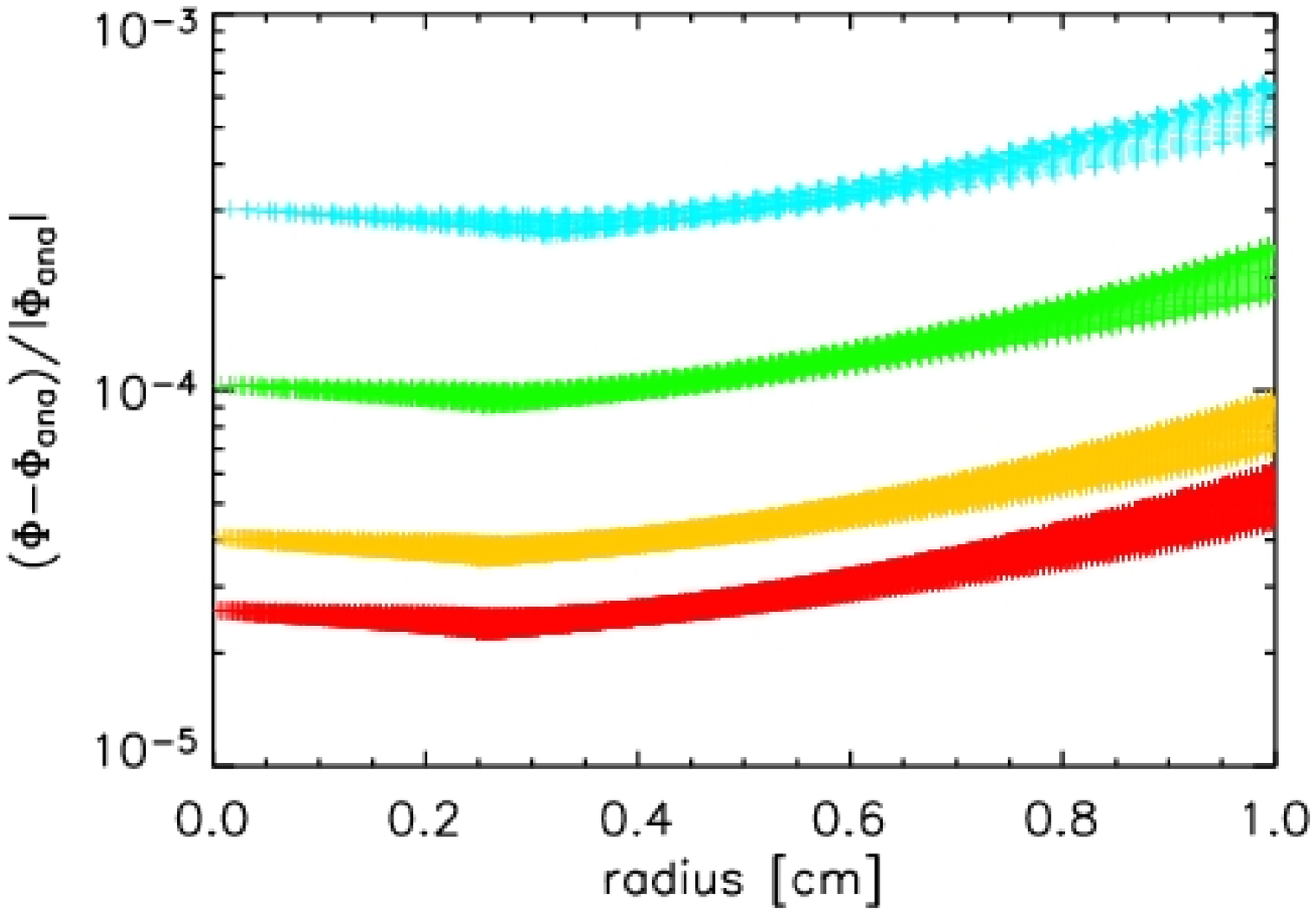}\\
\plotone{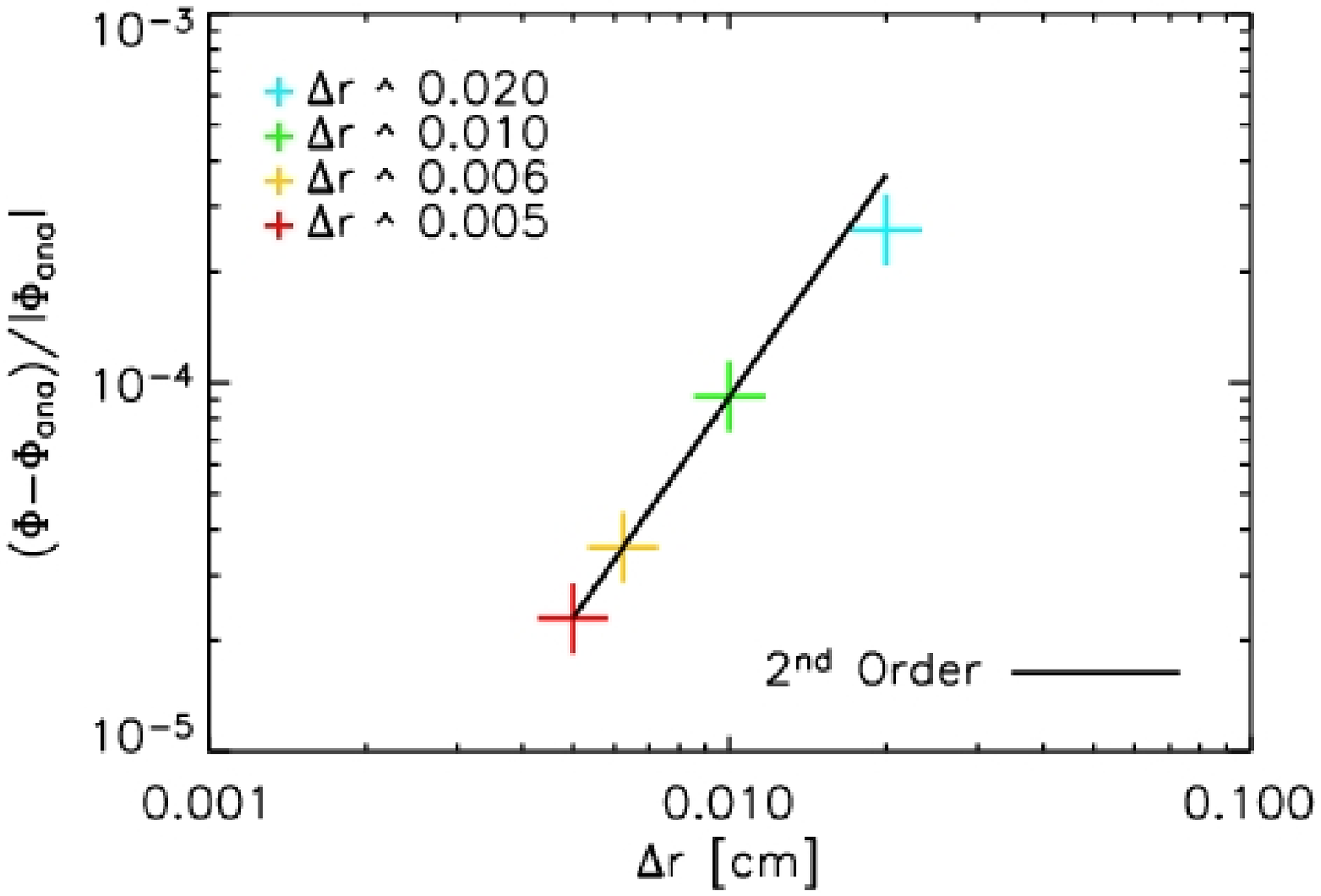}
\caption{Potential of a uniform density sphere.  The density of the
  sphere is 1 g cm$^{-3}$ and the radius is 1 cm.  Presented are
  results for four
  resolutions of the butterfly mesh (see left panel of Fig. \ref{grid_ex}):
  2550 cells with an effective radial resolution $\Delta r \sim$ 0.02
  cm (blue); 8750 cells with $\Delta r \sim$ 0.01 cm (green); 15,200 cells with
  $\Delta r \sim$ 0.006 cm (yellow); and 35,000 cells with $\Delta r
  \sim$ 0.005 cm (red).  
  The top panel
  shows $(\Phi - \Phi_{\rm{ana}})/|\Phi_{\rm{ana}}|$, where $\Phi$ are
  the cell-center potentials as determined by the Poisson solver, and
  $\Phi_{\rm{ana}}$ is the analytic potential.  The bottom panel shows the
  minimum error for each resolution as a function of the effective
  radial resolution.  The solid line illustrates the fact that the 2D Poisson
  solver converges with 2$^{\rm{nd}}$-order accuracy.
\label{maclaurin_sphere}}
\end{figure}

Figure \ref{maclaurin_sphere} compares the simulated and analytic
solutions for a homogeneous
sphere, having density $\rho_0 =
1$ g cm$^{-3}$, and maximum radius $r_a = 1$ cm.  The analytic
potential inside the sphere is $\Phi_{\rm{ana}} = G \rho_0 \frac{2}{3}
\pi (r^2 + z^2 - 3 r_a^2)$, and the top panel of Fig. \ref{maclaurin_sphere}
  shows the relative difference of the analytic potential and the
  numerical solution, $\Phi_z$, as a function of radius.  Results for four
  resolutions of the butterfly mesh are shown:
  2550 cells with an effective radial resolution $\Delta r \sim$ 0.02
  cm (blue); 8750 cells with $\Delta r \sim$ 0.01 cm (green); 15,200 cells with
  $\Delta r \sim$ 0.006 cm (yellow); and 35,000 cells with $\Delta r
  \sim$ 0.005 cm (red).  
  Two facts are obvious:  1) the solutions are accurate at a level of
  $\sim 3 \times 10^{-5}$ for $\Delta r/r_a \sim 0.005$ and $\sim 3
  \times 10^{-4}$ for $\Delta r/r_a \sim 0.02$; and 2) the degree to
  which the solution is spherically symmetric is similarly a few times
  $10^{-5}$ for $\Delta r/r_a \sim 0.005$ and a few times $10^{-4}$
  for $\Delta r/r_a \sim 0.02$.  Plotting the minimum error as a function of the effective
  radial resolution, the bottom panel of
  Fig. \ref{maclaurin_sphere}  verifies that the 2D Poisson
  solver convergences with 2$^{\rm{nd}}$-order accuracy (solid line).

Next, we calculate the aspherical potential for an oblate spheroid .
The homogeneous oblate spheroid has an elliptic meridional cross section, and the minor-
and major-axes of the ellipse are $r_b$ and $r_a$,
where $r_a$ is the equatorial radius.  Thus, the
eccentricity of the spheroid is $e = \sqrt{1 -
  (r_b/r_a)^2}$.  Given a uniform density, $\rho_0$, the potential for
a spheroid is
\begin{equation}
\Phi_{\rm{ana}}(r,z) = - \pi G \rho_0 \left ( 
I_{BT} r_a^2 - [ a_1 r^2 + a_3 z^2]
\right ) \, ,
\end{equation}
where, for oblate spheroids,
\begin{equation}
a_1 = \left [
\frac{\sin^{-1} e}{e} - (1 - e^2)^{1/2}
\right ] \frac{(1 - e^2)^{1/2}}{e^2} \, ,
\end{equation}
\begin{equation}
a_3 = 2 \left [
(1 - e^2)^{-1/2} -  \frac{\sin^{-1} e}{e}
\right ] \frac{(1 - e^2)^{1/2}}{e^2} \, ,
\end{equation}
and $I_{BT} = 2a_1 + a_3 (1 - e^2)$ \citep{chandrasekhar69}.

\begin{figure}
\epsscale{0.5}
\plotone{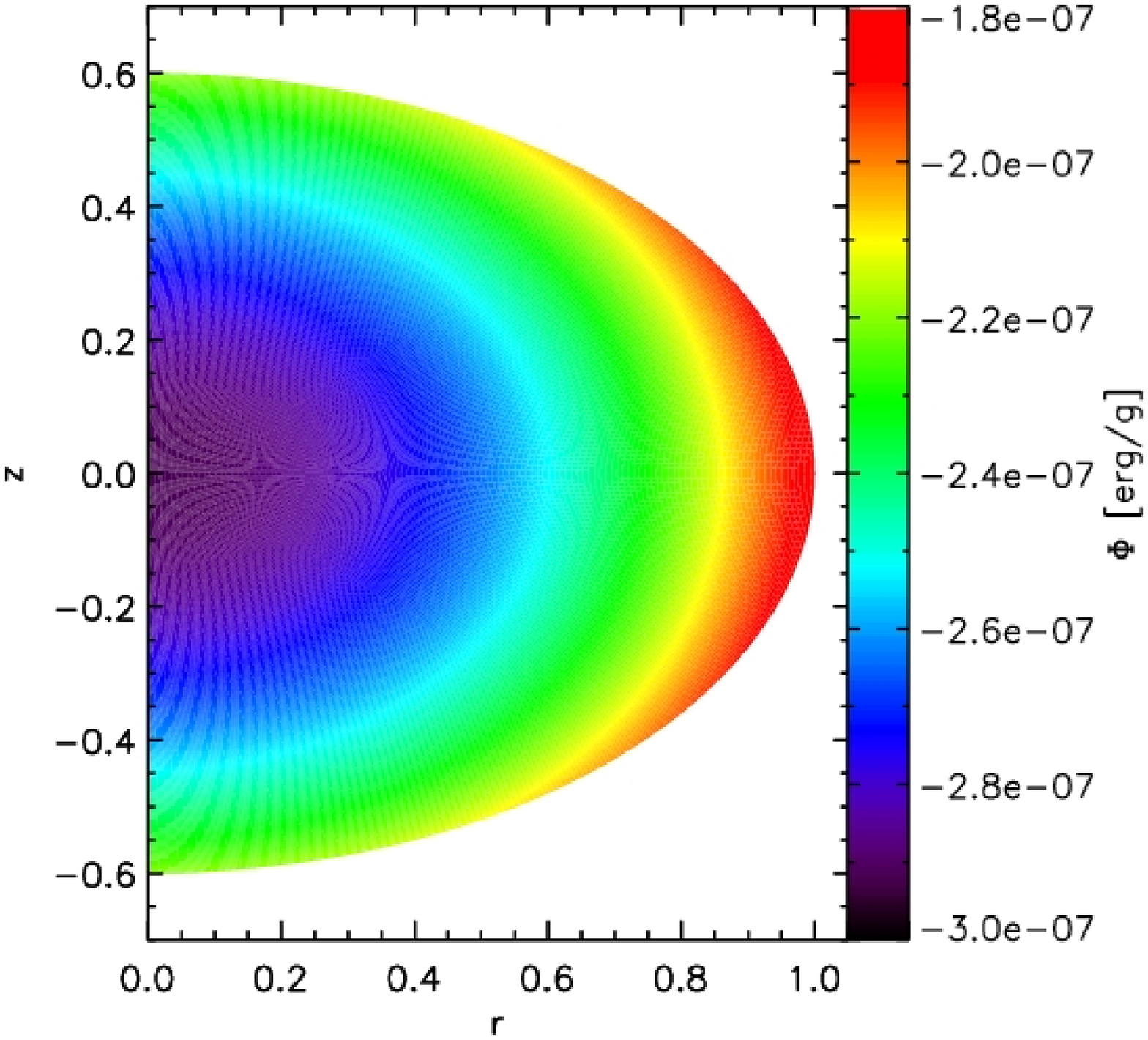}\\
\plotone{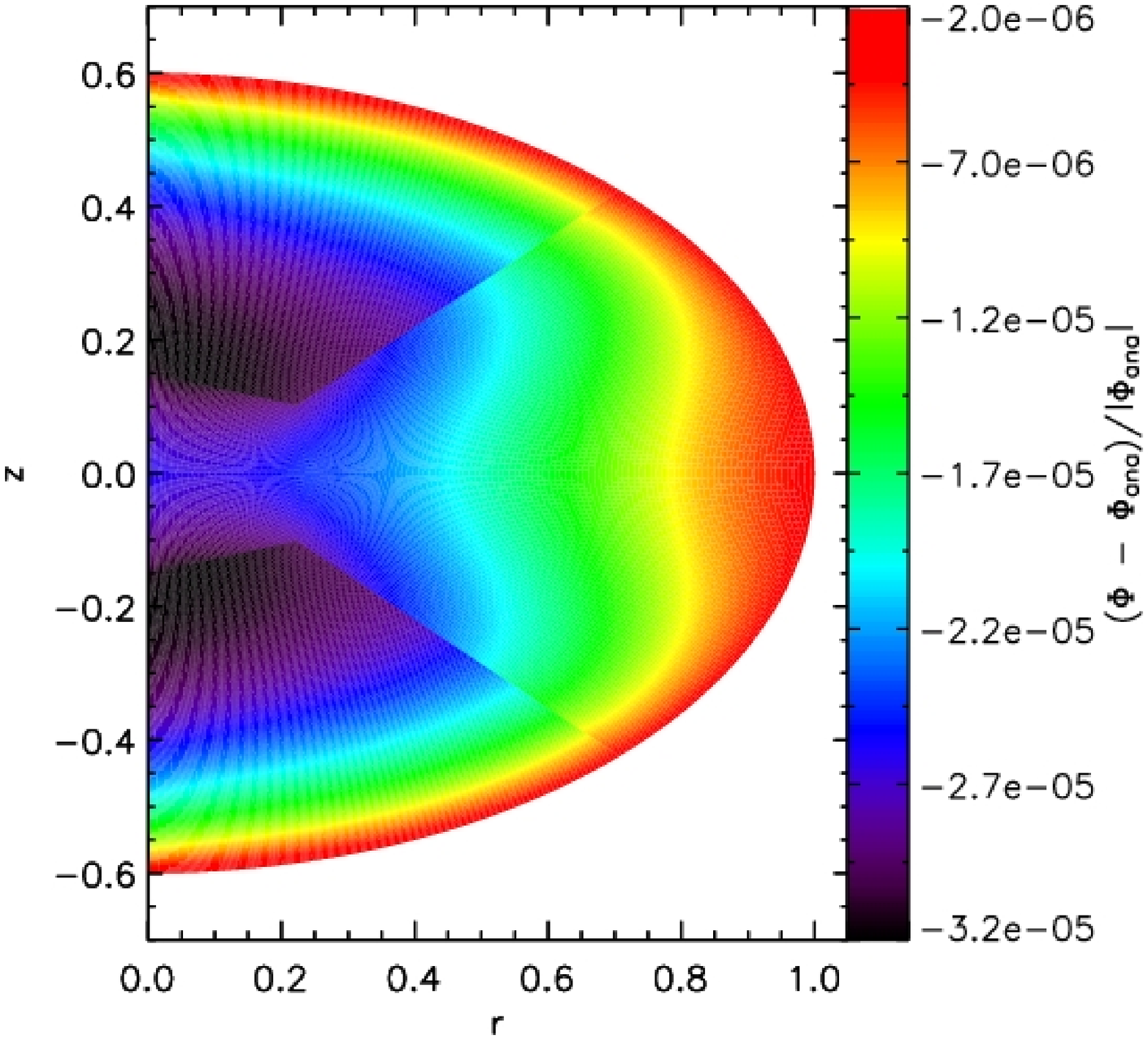}
\caption{The potential of a homogeneous
spheroid with $e = 0.8$ (top panel) and the relative accuracy of the
potential, $(\Phi - \Phi_{\rm ana})/|\Phi_{\rm ana}|$ (bottom panel).
$\rho_0 = 1$ g cm$^{-3}$ and the equatorial radius is $r_a = 1$ cm.  For the given eccentricity, the polar-axis
radius, $r_b$, is 0.6 cm.  The grid is a butterfly mesh, but
the outer boundary follows the ellipse defining the
surface of the spheroid.  With $N_{\rm{cell}} =
35,000$ and $\Delta r/r_a \sim 0.005$, the relative error in the
potential ranges from $\sim$2 $\times 10^{-6}$ near the outer boundary
to $\sim$3 $\times 10^{-5}$ in interior regions.  
Features in the relative error that track abrupt grid orientation changes in
the mesh are apparent, but the magnitude of these features does not
dominate the errors.
\label{maclaurin_spheroid}}
\end{figure}

Numerical results with $e = 0.8$ are shown in Fig. \ref{maclaurin_spheroid}.
As with the sphere, $\rho_0 = 1$ g cm$^{-3}$
and $r_a = 1$ cm.  However, for the given eccentricity, the polar-axis
radius, $r_b$, is 0.6 cm.  Once again, the grid is a butterfly mesh, but
this time the outer boundary follows the ellipse defining the
surface of the spheroid.  The top panel of
Fig. \ref{maclaurin_spheroid} presents the spheroid's potential, and the degree of
accuracy is presented in the bottom plot.  With $N_{\rm{cell}} =
35,000$ and $\Delta r/r_a \sim 0.005$, the relative error in the
potential ranges from $\sim 2 \times 10^{-6}$ near the outer boundary
to $\sim 3 \times 10^{-5}$ in interior regions.
Conspicuous are
features in the relative error that track abrupt grid orientation changes in
the mesh.  Fortunately, these features have
magnitudes smaller than or similar to the relative error in the local
region.
The relative error in the gravitational acceleration magnitude, $(|\vec{g}_p| -
|\vec{g}_{\rm ana}| / {\rm max}(|\vec{g}_{\rm ana}|)$ ranges from
$\sim$$-10^{-4}$ to $\sim$$10^{-4}$, where $\vec{g}_{\rm ana}$
is the analytic acceleration and ${\rm max}(|\vec{g}_{\rm ana}|)$ is
the maximum magnitude on the grid.  Typical errors in the acceleration
direction range from $\sim$$-10^{-4}$ to
$\sim$$10^{-4}$ radians with rare deviations as large as
$\sim$$10^{-3}$ radians near the axis and abrupt grid orientation changes.

\begin{figure}
\epsscale{.80}
\plotone{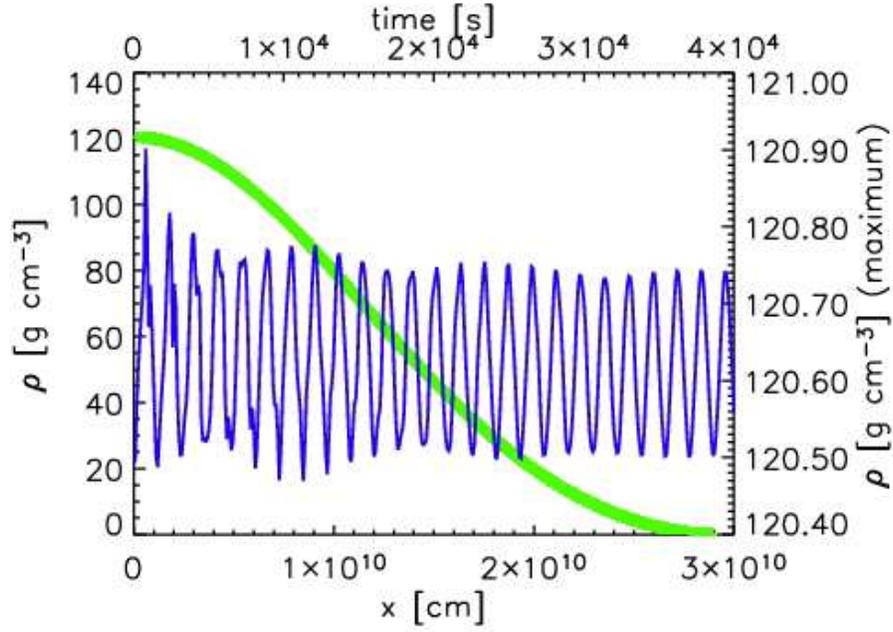}
\caption{Hydrostatic Equilibrium.  The grid is the butterfly mesh
  with 8750 zones.  The initial model is a Lane-Emden polytrope with
  $\gamma = 5/3$, $M = 1$ M$_{\sun}$, and $R = 2.9 \times 10^{10}$
  cm.  The green crosses show the density profile at $t = 1 \times 10^{4}$
  s.  The solid line shows the maximum density as a function of time.
  The oscillations are due to the slight difference between an analytic
  hydrostatic equilibrium structure and a discretized hydrostatic
  equilibrium structure.  This figure shows the code's ability to follow
  oscillations for many periods with very little or no attenuation. 
\label{laneemden_plot}}
\end{figure}

In Fig. \ref{laneemden_plot}, we demonstrate that the ALE
algorithm in combination with our
gravity solver produces reasonably accurate hydrostatic equilibria.
The grid is the butterfly mesh with 8750 zones, and the initial model is a Lane-Emden polytrope with
  $\gamma = 5/3$, $M = 1$ M$_{\sun}$, and $R = 2.9 \times 10^{10}$
  cm.  Crosses show the density profile at $t = 1 \times 10^{4}$
  s, while the solid line shows the maximum density
  as a function of time and that the star pulsates.  These oscillations result from the slight difference between an analytic
  hydrostatic equilibrium structure and a discretized hydrostatic
  equilibrium structure, and with increasing resolution, they decrease in
  magnitude.  Interestingly, the oscillations continue for many cycles with very little attenuation.

Simulating the
Goldreich-Weber self-similar collapse \citep{goldreich80} in 1D and
2D is a good test of dynamic simulations including gravity. The analytic profile is similar to a Lane-Emden polytrope with $\gamma =
4/3$, and in fact, we use the gamma-law EOS with $\gamma = 4/3$.
While the Lane-Emden
polytropes are assumed to be in hydrostatic equilibrium, the
Goldreich-Weber self-similar collapse has a homologous velocity
profile and a self-similar density profile.
The physical dimensions have been scaled so that $M = 1.3$ M$_{\sun}$,
the initial central density is $10^{10}$ g cm$^{-3}$, and the maximum
radius of the profile is $1.66 \times 10^8$ cm.  For the 1D
simulation, we initiate the grid with 200 evenly-spaced zones, and for
2D, a butterfly mesh with
35,000 zones (effectively with 200 radial zones) is used to initiate the
grid.  Subsequent evolution for both simulations uses the Lagrangian
configuration.

\begin{figure}
\epsscale{.60}
\plotone{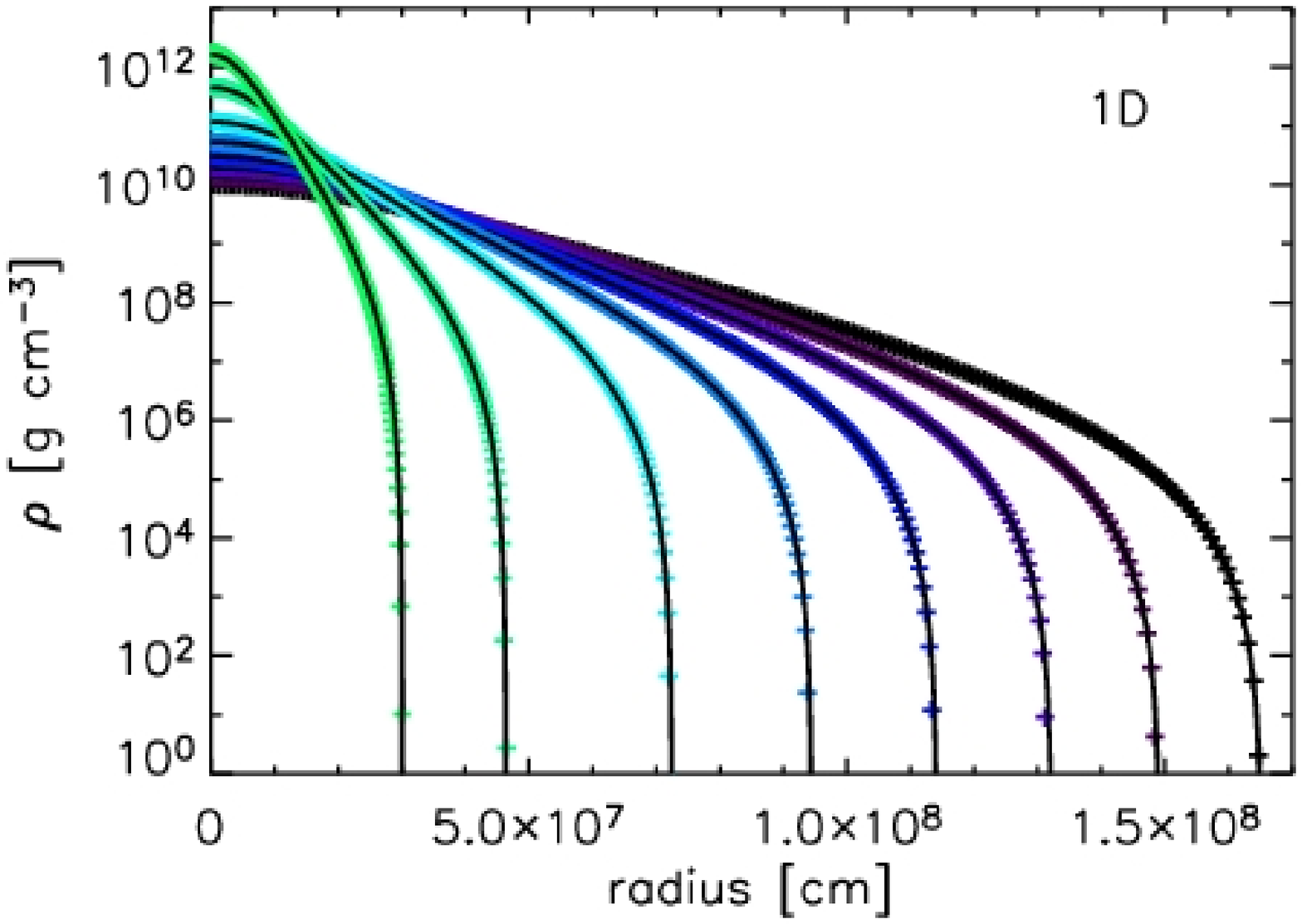} \\
\plotone{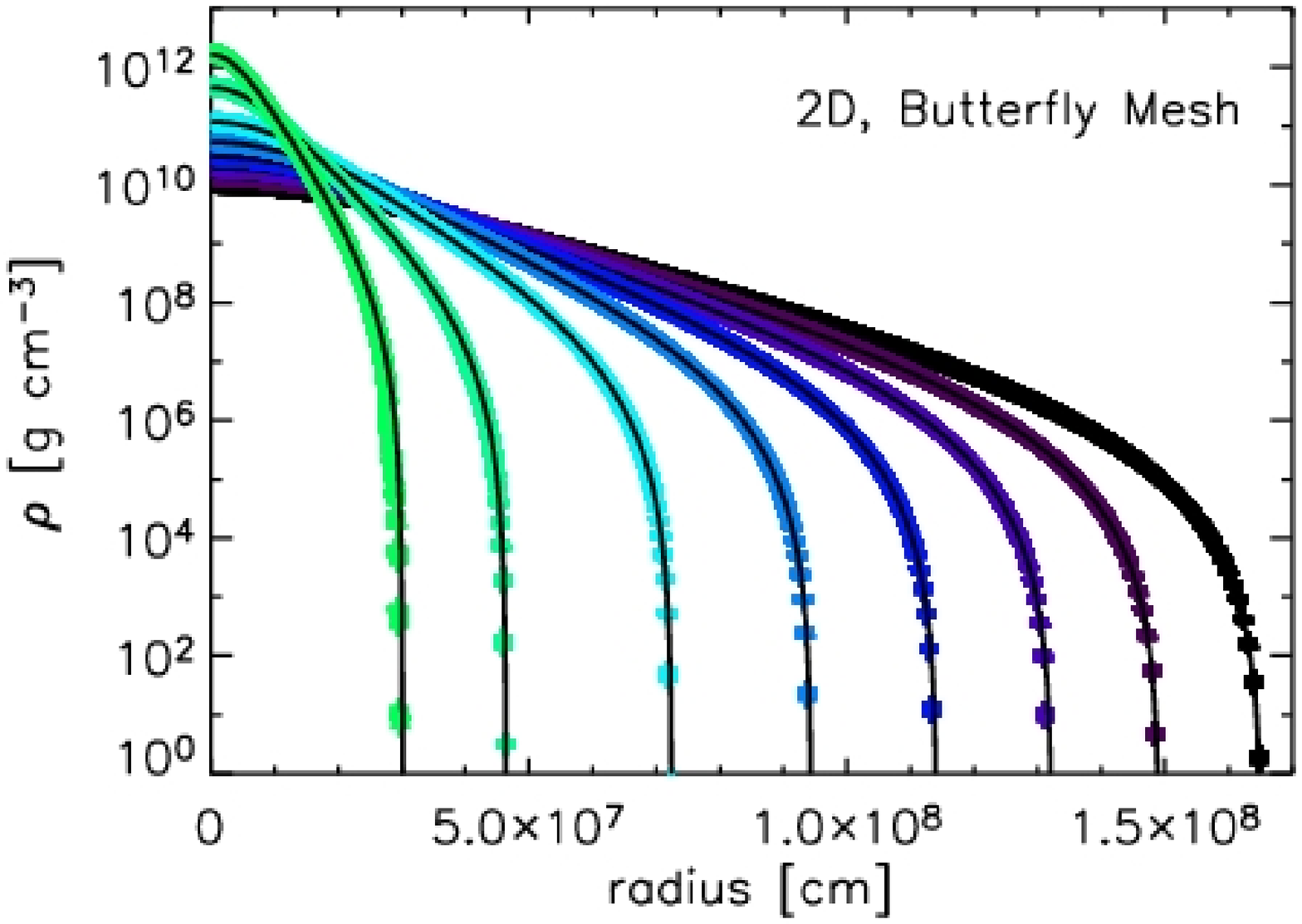}
\caption{Density profiles
for Goldreich-Weber self-similar collapse in 1D (top
panel) and 2D (bottom panel) and at $t = $0, 20, 40, 60, 80,
100, 120, and 130 ms.  Simulation results (crosses) are compared with
analytic solutions (solid lines).  
A gamma-law EOS is used with $\gamma = 4/3$.
The physical
  dimensions have been scaled so that $M = 1.3$ M$_{\sun}$, the
  initial central density is $10^{10}$ g cm$^{-3}$, and the maximum
  radius of the profile is $1.66 \times 10^8$ cm.  The initial grid
  for the 1D simulation
  uses 200 evenly-spaced zones. For the 2D simulation, a butterfly mesh with 35,000 zones with
  effectively 200 radial zones is used.  These tests are calculated
  using the Lagrangian configuration.
See \S \ref{section:gravity_tests}
and Table \ref{table:gw} for quantitative discussion of the accuracy.
\label{gwplot}}
\end{figure}

Figure \ref{gwplot} shows snapshots of density vs. radius
for 1D (top
panel) and 2D (bottom panel) simulations at $t = $0, 20, 40, 60, 80,
100, 120, and 130 ms.  Both plots indicate that the simulations
(crosses) track the analytic solution (solid lines).  Quantitatively,
we measure a relative difference, $(\rho_z - \rho_{\rm ana})/(\max
(\rho_{\rm ana} ) )$, for the reported times, where $\rho_z$ are the simulated
cell-center densities and $\rho_{\rm ana}$ are the analytic values.
Consistently, the largest deviations are at the center and Table
\ref{table:gw} gives these values for the 1D simulation (2$^{nd}$
column) and the 2D simulation (3$^{rd}$ column).  The relative
differences range from $-9.2 \times 10^{-8}$ at 0 ms to $-3.6 \times
10^{-2}$ at 130 ms for the 1D simulation and $-1.2 \times 10^{-7}$ at
0 ms to $-6.3 \times 10^{-2}$ at 130 ms for the 2D simulation.  At all times, the departure from spherical
symmetry is no more than $\sim 1 \times 10^{-4}$.

\section{Hydrodynamic Boundary Conditions}
\label{section:hydro_boundary}

Boundary conditions are implemented in one of two ways.  Either an
external pressure is specified or the velocities at the nodes are
fixed.  For external pressures, ghost cells are
defined that have no true volume or mass associated with them.
Their only function is to apply an external force to the boundary
cells equal to an external pressure times the boundary surface area.
Specifying nodal velocities on the boundary
accomplishes the same task.  In this case, there is an implied
external force and pressure.  In practice, generic dynamical
boundaries use the external pressure boundary condition.  On the other
hand, pistons, reflecting walls, and the azimuthal axis have
the velocity perpendicular to the boundary specified, while the
parallel component executes unhindered hydrodynamic motions.

\section{Artificial Viscosity}
\label{section:artificial_viscosity}

To resolve shocks over just a few zones, we include artificial
viscosity in the equations of hydrodynamics.  For 1D simulations, we
add a viscous-like term to the pressure \citep{vonneumann50}.  We denote this viscous
pressure by $q$.  The original realization of $q$ employed one term proportional to
$(\Delta v)^2$, where $\Delta v$ is the difference in velocity from one
zone to the next.  While this form adequately
resolved shocks, unphysical oscillations were observed in the
post-shock flow.  A second term, linear in $\Delta v$, was then added that
effectively damped these oscillations \citep{landshoff55}.  Another form which we employ
for our 1D simulations is
\begin{multline}
\label{eq:viscosity}
q = \rho \left (
c_2 \frac{\gamma + 1}{4} |\Delta v| \right. \\
\left. +
\sqrt{c^2_2 \left ( \frac{\gamma + 1}{4}\right )^2 (\Delta v)^2 +
  c^2_1 c^2_s
}
\right ) |\Delta v| \, ,
\end{multline}
where $c_s$ is the sound speed, $c_1$ is the parameter associated with the
linear term, and $c_2$ is the parameter associated with the quadratic term.
This form has the appealing attribute that it is motivated by the
expression for the shock-jump condition for pressure in an ideal gas \citep{wilkins80}.

For 2D simulations, the artificial viscosity scheme we have settled upon
is the tensor artificial viscosity algorithm of \citet{campbell01}.
Here, we do not re-derive the artificial viscosity scheme, but highlight
some of its salient features and practical implementations.  The useful feature of the tensor algorithm is its ability to calculate artificial viscosity on an
arbitrary grid, while suppressing artificial grid buckling when the
flow is not aligned with the grid.  In the strictest sense, the tensor
artificial viscosity does not employ the simple viscous pressure
described above.  Instead, \citet{campbell01} assume that the
artificial viscosity tensor is a combination of a scalar coefficient,
\begin{equation}
\label{eq:mu_av}
\mu \propto \rho \left (
c_2 \frac{\gamma + 1}{4} |\Delta v| +
\sqrt{c^2_2 \left ( \frac{\gamma + 1}{4}\right )^2 (\Delta v)^2 +
  c^2_1 c^2_s
}
\right ) \, ,
\end{equation}
and the gradient of the velocity tensor, $\vec{\nabla} \vec{v}$.
Therefore, references to the parameters $c_1$ and $c_2$ in 1D and 2D
refer to the same linear or quadratic dependence on $|\Delta \vec{v}|$.  With this assumption, the
momentum equation, eq. (\ref{eq:mom_lag}), becomes
\begin{equation}
\label{eq:mom_lag_av}
\rho \frac{d \vec{v}}{d t} = - \rho \vec{\nabla} \Phi - \vec{\nabla} P
+ \vec{\nabla} \cdot ( \mu \vec{\nabla} \vec{v} ) \, ,
\end{equation}
and the energy equation, eq. (\ref{eq:ene_lag}), is
\begin{equation}
\label{eq:ene_lag_av}
\rho \frac{d \varepsilon}{d t} = - P \vec{\nabla} \cdot \vec{v} 
+ \mu (\vec{\nabla} \vec{v}) : (\vec{\nabla} \vec{v})\, ,
\end{equation}
where $(\vec{\nabla} \vec{v}) : (\vec{\nabla} \vec{v}) = (\vec{\nabla}
\vec{v})_{ij} (\vec{\nabla} \vec{v})_{ij}$, using the normal Einstein
summation convention.  With the method of support operators,
\citet{campbell01} derive discrete forms of these equations.
Analogous to the gradient of the pressure term, the discrete artificial
viscosity term in the momentum equation becomes a summation of corner
forces, 
\begin{equation}
\vec{\nabla} \cdot ( \mu \vec{\nabla} \vec{v} ) \longrightarrow \sum_{z \in S(p)}\vec{f}^p_{z,\rm{visc}} \, ,
\end{equation}
and the
corresponding term in the energy equation is the usual
force-dot-velocity summation: 
\begin{equation}
\mu (\vec{\nabla} \vec{v}) : (\vec{\nabla} \vec{v}) \longrightarrow
-\sum_{p \in S(z)} \vec{f}^z_{p,\rm{visc}} \cdot \vec{v}_p.
\end{equation}
Therefore, implementation of the artificial viscosity scheme is
straightforward and similar to that for the pressure forces.

In practice, this artificial viscosity scheme is
formulated for Cartesian coordinates, and to obtain the equivalent
force for cylindrical coordinates, we multiply the Cartesian subcell
force by $2 \pi r_p$.  In general, this works quite well,
but sacrifices strict momentum conservation (See \S
\ref{section:momcons}).

\section{Subcell Pressures: Eliminating Hourglass Grid
  Distortion}
\label{section:hourglass}
A problem that can plague Lagrangian codes using cells with 4 or more sides is the
unphysical hourglass mode \citep{caramana98a}.  To suppress this problem, we employ a modified version of the
subcell pressure algorithm
of \citet{caramana98a}.  For all
physical modes (translation, shear, and extension/contraction) the
divergences of the velocity on the subcell and cell levels are equal,
\begin{equation}
(\nabla \cdot {\bf v})^z_p = (\nabla \cdot {\bf v})_z \, .
\end{equation}
Hence, as long as the calculation is initiated such that the subcell
densities of a cell are equal to the cell-averaged density, then they
should remain equal at all subsequent times.  Any deviation of the
subcell density from the cell density, $\delta \rho^z_p = \rho^z_p -
\rho_z$, is a direct result of the hourglass mode.  The scheme of
\citet{caramana98a} uses this deviation in subcell density to define a
subcell pressure that is related to the deviation in subcell and cell
density, $\delta \rho^z_p$, by
\begin{equation}
P^z_p = P_z + c^2_s \delta \rho^z_p
\end{equation}
with the corresponding deviation in pressure: $\delta P^z_p = c^2_s
\delta \rho^z_p$.

Application of these pressures as subcell forces counteracts the hourglass distortion.
However, the pressure throughout the cell is no longer uniform.  As
a result, the subcell pressures exert forces on the cell centers
and mid-edge points.  These locations in the grid are not subject to physical
forces as are the nodes.  Instead, their movement is tied to the
motions of the nodes.  To conserve momentum, the forces
acting on these enslaved points must be redistributed to the dynamical
nodes.  Though the choice of redistribution is not unique, we follow
the  procedure established by \citet{caramana98a} for the form of the
subcell force (for the definitions of $\vec{S}^z_p$ and
  $\vec{a}^z_p$, see \S \ref{section:coords_mesh} and Fig. \ref{fig:grid}.):
\begin{equation}
\vec{f}^z_p = \delta P^z_p \vec{S}^z_p + \frac{1}{2}[
(\delta P^z_p-\delta P^z_{p+1}) \vec{a}^z_p
+(\delta P^z_{p-1}-\delta P^z_p) \vec{a}^z_{p-1}] \, .
\end{equation}

This particular scheme works quite well for Lagrangian
calculations.  However, it can be incompatible with
subcell remapping algorithms.  The hourglass suppression scheme
described above assumes that only hourglass motions result in nonzero
values for $\delta \rho^z_p$.  Even in calculations that are
completely free of hourglass motions, the subcell remapping scheme can
and will produce subcell densities within a cell that are different
from the cell density.  Since this difference of subcell and cell densities
has nothing to do with hourglass motions, any subcell forces that
arise introduce spurious motions that don't correct the
hourglass motions.

Investigating several alternative schemes, we settled on a
modification of the \citet{caramana98a} approach.  After each remap, 
we define a tracer density whose only purpose is to
track the relative changes in the subcell and cell volumes during the
subsequent hydrodynamic solve.  For
convenience, and to keep the magnitude of the subcell pressures about
right, we set the subcell tracer density equal to the cell density.  After
replacing $\delta \rho^z_p$ with the difference between the subcell tracer
density and the cell density, $\delta \rho^z_{p,tr}$, we proceed with
the scheme prescribed by \citet{caramana98a}.  We have found that
the subcell forces of \citet{caramana98a} significantly
resist hourglass motions only after $\delta \rho^z_p$ achieves
significant magnitude.  For our new hourglass scheme, if a remap is implemented
after each hydrodynamic solve, $\delta \rho^z_{p,tr}$ does not
have a chance to achieve large values, and hence, the
subsequent subcell forces are not very resistant to hourglass
distortions.  However, for many
simulations, we have found that it is not necessary to remap after
every Lagrangian hydrodynamic solve.  Rather, the remap may be performed after $N$
timesteps.  In these circumstances, the subcell tracer density is
allowed to evolve continuously as determined by the hydrodynamic
equations.  With large $N$, $\delta \rho^z_{p,tr}$ does develop
significant amplitude and the subcell forces become effective in
suppressing the hourglass modes.  In fact, in the limit that $N \gg 1$,
this scheme becomes the hourglass suppression scheme of \citet{caramana98a}.

In practice, we multiply this subcell force by a scaling factor.  To
determine the appropriate magnitude of this scaling factor, we
executed many test problems and found that problems that
involved the perturbation of hydrostatic equilibrium provide a good
test of the
robustness of the hourglass fix.  During the testing protocol, we ran a simulation for at least 200,000 timesteps, ensuring that
the hourglass fix remains robust for long calculations.  For
Lagrangian calculations, a scaling factor $\sim 1$ produced
reasonable results.  For runs in which remapping
occurred after $N$ timesteps, we found that larger values of the
scaling factor were required.  However, for very large values of $N$, such as 64 or
greater, such large values compromised
small structures in the flow.  Hence, we settled on scaling factors
near 1 for large $N$.  For small $N$, we found that the difference in
tracer density and the cell-centered density had little time to build
to significant amplitudes, so larger values of the scaling factor are
required, but anything above 4 produced noticeable problems in
flows with many timesteps.  In summary, the scaling factor should be
adjusted to be $\sim 1$ for large $N$ and $\sim 4$ for small $N$.

\subsection{Tests of the Hourglass Elimination Algorithm}

\begin{figure}
\epsscale{.90}
\plotone{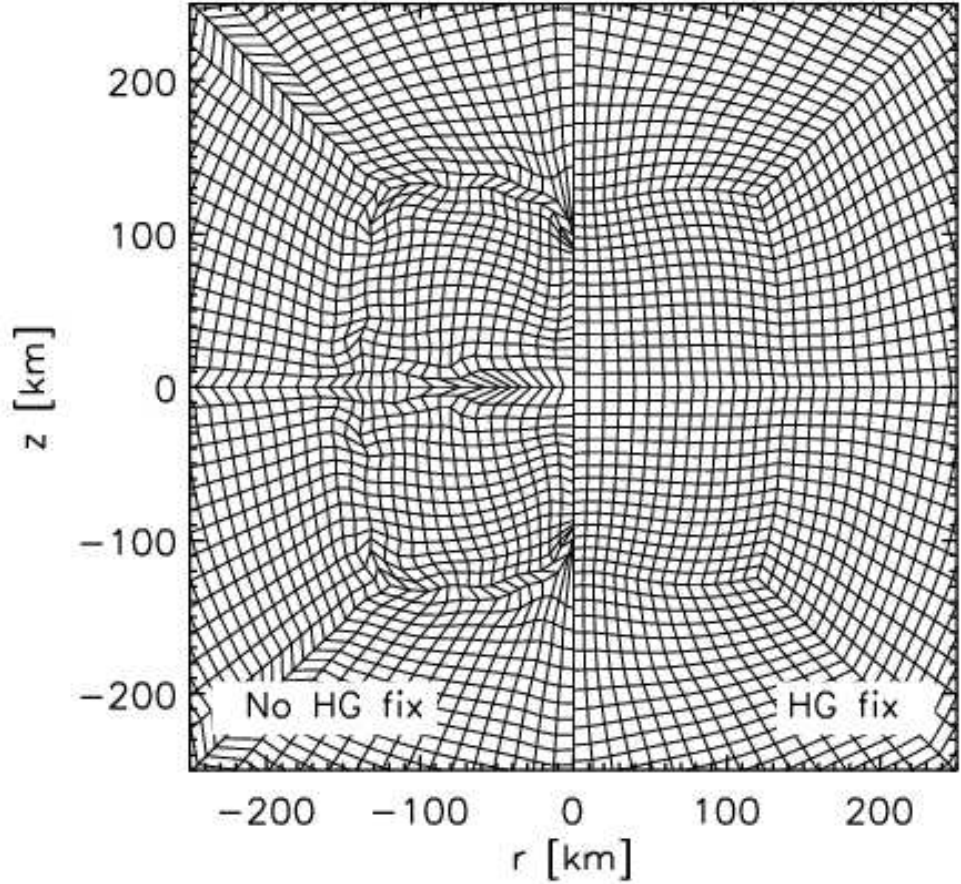}
\caption{Comparison when hourglass suppression is turned off (left)
  and on (right) for the Goldreich-Weber self-similar collapse.  The
  simulations in this figure were run in the Lagrangian configuration
  with the butterfly mesh.  Displayed in the
  region ``$r > 0$'' are results at $t = 0.118$ s when the hourglass suppression scheme of
  \S \ref{section:hourglass} is used and the scale factor has been
  set to 2.0.  The results shown in the region ``$r < 0$'' represent what
  happens when the hourglass suppression scheme is turned off. \label{gw_fmerit_comp}}
\end{figure}

To demonstrate the need for an hourglass suppression scheme, and that
our algorithm to address the hourglass distortions works, we show in
Fig. \ref{gw_fmerit_comp} results for the Goldreich-Weber self-similar
collapse in 2D (\S \ref{section:gravity_tests}), with and without the hourglass
suppression (\S \ref{section:hourglass}).  These results
represent a Lagrangian simulation of the self-similar collapse at $t =
118$ ms.  On the left (``$r < 0$'') the grid clearly shows problematic
grid buckling when the hourglass suppression is turned off.  The grid
on the right (``$r > 0$'') shows the elimination of grid buckling with
the use of the hourglass suppression scheme and a scaling factor of 2.0. 

\begin{figure}
\epsscale{.80}
\plotone{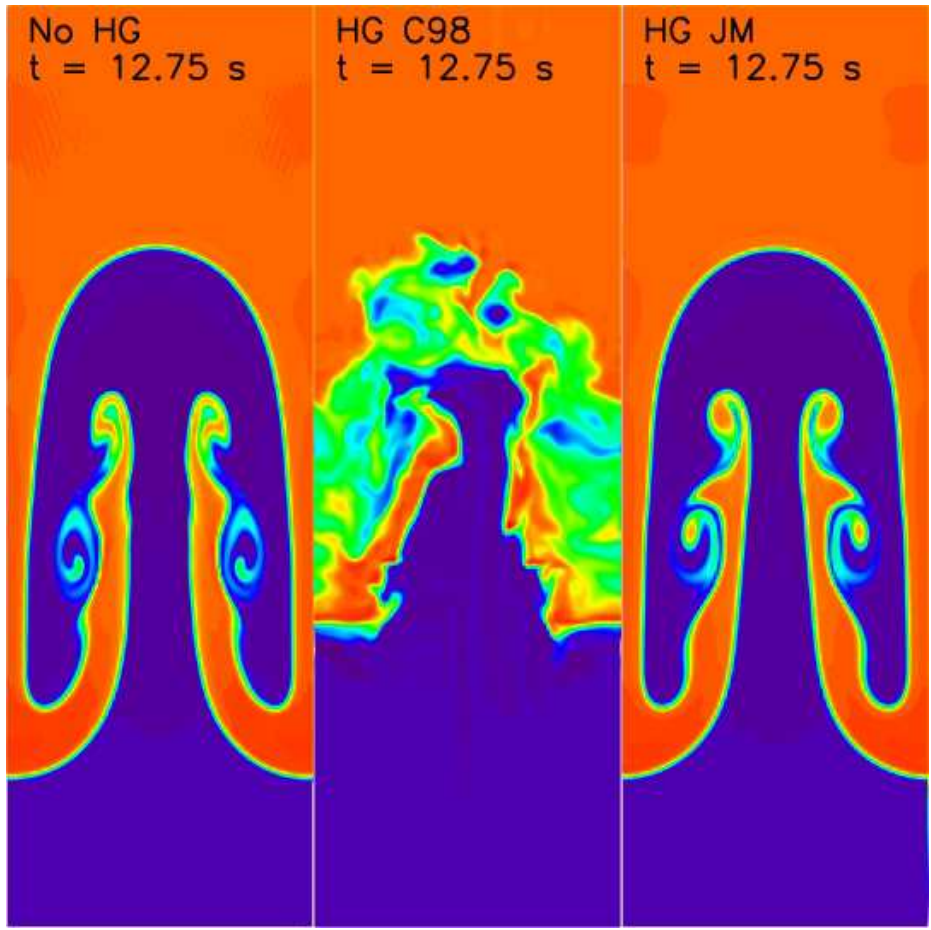}
\caption{Single-mode Rayleigh-Taylor instability: investigating
  hourglass suppression schemes (see \S \ref{section:rt} for a
  description of the setup).  We present the
  results of the single-mode Rayleigh-Taylor instability with no
  hourglass fix (left panel), the hourglass subcell pressure scheme of
  \citet{caramana98a} (center panel), and the modified subpressure
  scheme described in this paper (right panel).  While the results of
  the left panel are acceptable, there are hints of hourglass patterns
  in the contours.  The central panel is a consequence of
  the incompatible nature of the subcell remapping scheme and the
  subcell pressure scheme of \citet{caramana98a}, and our modified
  subpressure scheme (right panel) suppresses the hourglass
  distortions, while preserving the expected flow.
\label{rt_stills_fmerit}}
\end{figure}

Figure \ref{rt_stills_fmerit} illustrates the problem of using the
hourglass suppression scheme of \citet{caramana98a} in combination
with the subcell remapping algorithm (\S \ref{section:remap}).  All
three panels show the results of the single-mode Rayleigh-Taylor
instability at $t = 12.75$ s.  For the left
panel, no hourglass suppression is employed.  This panel
represents our control.  While the hourglass instability does not cause serious
problems for the calculation, one can see evidence of slight hourglass
patterns at the scale of the grid resolution.  In particular, two
blobs near the top and on the edges form distinct patterns.  The
central panel demonstrates the problem of using the subcell remapping scheme and the
  subcell pressure method of \citet{caramana98a}.  On the other hand,
  the right panel demonstrates that the modified subpressure scheme we have developed
  suppresses the hourglass distortions, while preserving the expected flow.

\section{Remapping}
\label{section:remap}
Fluid flows with large vorticity quickly tangle a Lagrangian
mesh, presenting severe problems for Lagrangian
hydrodynamic codes.  To avoid entangled grids, it is common to remap the state
variables after each Lagrangian hydrodynamic advance to another less
tangled mesh.  Even flows that exhibit very
little vorticity, but extreme compression or expansion along a
particular direction, leading to very skewed cells, can
limit the accuracy of Lagrangian calculations.  In this case as well, one can
employ a remapping scheme which is designed to minimize the
calculational error.

To remap, a new grid must be established.
One choice is to remap to the original
grid, thereby effectively solving the hydrodynamics equations on an
Eulerian mesh.  Another is to use a reference Jacobian matrix
rezone strategy, which establishes a new mesh close to the
Lagrangian mesh, while reducing numerical error by reducing unnecessarily
skewed zones.  The arbitrary nature of the
remapping algorithm even allows the freedom of choosing one
rezoning scheme in one sector of the grid and another
rezoning scheme in other sectors.  For example, one could remap in an
Eulerian fashion in one sector, let the grid move in a
Lagrangian manner in another, and in the intervening region smoothly
match these two regions.

Generally, there are two options for remapping
schemes.  One can remap from one arbitrary grid to another
completely unrelated arbitrary grid.  For these unrelated grids, one needs to
determine the overlap of the zones of the first grid with the zones
of the second.  In general, this can be a very cumbersome and
expensive process.  As a result most ALE codes, remap to an arbitrary grid
that is not too different from the first.  Specifically, the usual
stipulation employed is that the face of each cell does not traverse
more than one cell during a timestep, and that the connectivity among nodes,
faces, edges, and cells remains the same from the first grid to the
remapped grid.  The regions swept by the faces then contain the mass,
momentum, or energy which is added to the cell on one side of the face
and subtracted from the cell on the other side.

It is in this context that we use the swept-region remapping
algorithms of \citet{loubere05}, \citet{margolin03,margolin04}, \citet{kucharik03}, and \citet{loubere06}.  This
remapping scheme may be described by four
stages:
\begin{enumerate}
\item The first is the gathering stage.  This is a
  stage in which subcell quantities of the mass, momenta, and
  energies are defined in preparation for the bulk of the remapping
  process.
\item The second is to remap the subcell quantities from the
  Lagrangian grid to the new rezoned grid using the swept-region
  approximation.  In doing so, the remapping algorithm remains 2nd-order
  accurate, but avoids time intensive routines to calculate the overlap
  regions of the old and new grids.
\item The third is to repair the subcell densities.  Because exact
  spatial integration is avoided with the swept-region
  approximation, local bounds of subcell densities may be violated.
  Therefore, a
  repair algorithm which redistributes mass, momentum, and energies to
  preserve the local bounds is implemented.
\item Finally, there is the
  scattering stage, in which the primary variables of the hydrodynamics
  algorithm are recovered for the new rezoned grid.
\end{enumerate}

\subsection{Gathering Stage}
\label{section:gather}

In order to remap the primary quantities using the subcell remapping
algorithm, mass, momenta, internal energy, and kinetic energy must be
defined at the subcell level.  In the Lagrangian stage, the subcell mass, $m^z_p$, is already
defined.  The subcell density which is a consequence of
the change in the subcell volume ($V^z_p$) during the Lagrangian hydro step is
\begin{equation}
\rho^z_p = \frac{m^z_p}{V^z_p} \, .
\end{equation}
Unfortunately, there is no equivalent quantity for the subcell
internal energy.  Instead, we define the internal energy density for
each zone, $e_z = \varepsilon_z \rho_z$, fit a linear function for the
internal energy density, $\hat{e}_z = e_z + (\vec{\nabla}e)_z \cdot
(\vec{x} - \vec{x}_c)$, and integrate this function over the volume of
each subcell to determine the total subcell internal energy and,
consequently, the average subcell internal energy density:
\begin{equation}
\label{eq:gather_energy}
\begin{array}{ccc}
E^z_p = \int_{v^z_p} \hat{e} dV  &
\mbox{and}&
 e^z_p = \frac{E^z_p}{V^z_p} \, .
\end{array}
\end{equation}
Performing the integration of eq. (\ref{eq:gather_energy}) involves a
volume integral and volume integrals weighted by $x$ and $y$ in
Cartesian coordinates or $r$ and $z$ in cylindrical coordinates (see appendix
\ref{section:integrals} for formulae calculating these discrete integrals).
Note, that, by construction, these newly defined internal energies
satisfy conservation of internal energy for each zone, $m_z
\varepsilon_z = \sum_{p \in S(z)} E^z_p$.

Similarly, there are no readily defined subcell-centered averages for the
momenta, $\vec{\mu}^z_p$, or velocities, $\vec{u}^z_p$.  However, they
should be related to one another by
\begin{equation}
\vec{\mu}^z_p = m^z_p \vec{u}^z_p \, .
\end{equation}
If the gathering stage is to preserve momentum conservation, then the
following equality must hold:
\begin{equation}
\label{eq:momconssubcell}
\sum_{p \in S(z)} m^z_p \vec{u}^z_p = \sum_{p \in S(z)} m^z_p
\vec{v}_p \, .
\end{equation}
By inspection, it might seem natural to set $\vec{u}^z_p =
\vec{v}_p$.  However, we follow the more accurate suggestion made by
\citet{loubere05} to define the subcell-averaged velocity as follows:
\begin{equation}
\label{eq:subcellvel}
\vec{u}^p_z = \frac{ 
\vec{u}_z + \vec{v}_p + \vec{v}_{p+1/2} + \vec{v}_{p-1/2}}{4} \, ,
\end{equation}
where $\vec{u}_z$ is obtained by averaging over all nodal velocities associated
with zone $z$, and $\vec{v}_{p+1/2}$ and $\vec{v}_{p-1/2}$ are edge-averaged velocities given by $\vec{v}_{p+1/2} = 1/2 \left (
  \vec{v}_p + \vec{v}_{p+1} \right )$ and $\vec{v}_{p-1/2} = 1/2 \left
  ( \vec{v}_p + \vec{v}_{p-1} \right )$.
Substituting eq. (\ref{eq:subcellvel}) into
eq. (\ref{eq:momconssubcell}) and rearranging terms, we get an
expression for the subcell velocity that depends on known subcell masses
and nodal velocities:
\begin{multline}
\label{eq:vsubcellexplicit1}
\vec{u}^p_z = \frac{1}{4} \left (
2\vec{v}_p + \frac{\vec{v}_{p+1}}{2} + \frac{\vec{v}_{p-1}}{2}
\right ) \\
+ \sum_{p' \in S(z)} \frac{m^z_{p'}}{8 m_z} (4\vec{v}_{p'} -
\vec{v}_{p'+1} - \vec{v}_{p'-1}) \, .
\end{multline}
We now rewrite
eq. (\ref{eq:vsubcellexplicit1}) in a form that obviously makes it
easy to write the
gathering operation in matrix form:
\begin{multline}
\label{eq:vsubcellexplicit}
\vec{u}^p_z = \frac{1}{4} \left (
2\vec{v}_p + \frac{\vec{v}_{p+1}}{2} + \frac{\vec{v}_{p-1}}{2}
\right ) \\
+ \sum_{p' \in S(z)} \vec{v}_{p'} \left (
\frac{-m^z_{p'-1} + 4 m^z_{p'} - m^z_{p'+1}}{8 m_z}
\right ) \, .
\end{multline}
If ${\bf U}_z$ is a vector of one component of all
velocities associated with cell $z$, and ${\bf U}^{SC}_z$ is the
equivalent for subcell velocities, then
\begin{equation}
{\bf U}^{SC}_z = {\bf M}_z {\bf U}_z \, ,
\end{equation}
where the matrix, ${\bf M}_z$, for each zone is given by the
coefficients in eq. (\ref{eq:vsubcellexplicit}).

To conserve total energy in the remap stage, we define the subcell
kinetic energy and follow the same gather procedure
as for the velocity by simply replacing
velocity with the specific kinetic energy:
\begin{equation}
k_p = \frac{|\vec{v}|^2}{2} \, .
\end{equation}
We
ensure conservation of total kinetic energy for a cell:
\begin{equation}
\sum_{p \in S(z)} m^z_p k^z_p = \sum_{p \in S(z)} m^z_p k_p = K_z \, ,
\end{equation}
where $k^z_p$ is the subcell-averaged specific kinetic energy.
Then, the transformation from nodal specific
kinetic energies to subcell-averaged specific kinetic energies is
given by
\begin{equation}
{\bf k}^{SC}_z = {\bf M}_z {\bf k}_z \, ,
\end{equation}
where ${\bf k}^{SC}_z$ is a vector of the subcell-averaged specific
kinetic energies, ${\bf k}_z$ is a vector of the nodal specific
kinetic energies, and ${\bf M}_z$ is the same transformation matrix
used for the velocity transformations.  Having found the specific
kinetic energies, the calculation of the kinetic
energy for each subcell is then straightforward: $K^z_p =
m^z_p k^z_p$.

Upon completion of the gathering stage, each relevant quantity, $m^z_p$, $\vec{\mu}^z_p$, $E^z_p$, and $K^z_p$, is
expressed as a fundamentally conserved quantity for each subcell.  From these conserved
quantities and the subcell volumes, we have the corresponding
densities, which, as is explained in \S \ref{section:swept}, are
important components of the remapping process.

\subsection{Swept-edge Remap}
\label{section:swept}

Having gathered all relevant subcell quantities, we begin
the bulk of the remapping process.  Since all variables are now expressed in terms of
a conserved quantity (subcell mass, energy, and momentum) and
a density (mass density, energy density, and momentum density), for clarity of exposition, we focus on representative variables, subcell mass and density, to explain
the generic remapping algorithm.  When there are differences in the algorithm for the other
variables, we note them.

We have instituted the remapping algorithm of \citet{kucharik03}.  To avoid the
expensive process of finding the overlap of the Lagrangian grid with
the rezoned grid, this algorithm modifies the mass of each subcell
with a swept-edge approximation for the rezoning process.  As
long as the connectivity and neighborhood for each cell remain the
same throughout the calculation, time spent in connectivity overhead
is greatly reduced.  The objective of the remapping algorithm is then to
find the amount of mass, $\delta m_e$, that each edge effectively sweeps up due to the rezoning
process and to add or subtract this change
in mass to the old subcell mass to obtain the new subcell mass, $\tilde{m}^z_p$:
\begin{equation}
\label{eq:swept_eq1}
\tilde{m}^z_p = m^z_p + \sum_{e \in S(z)} \delta m_e \, .
\end{equation}
Of course, the change in mass is obtained by integration of a density
function over the volume of the swept region:
\begin{equation}
\label{eq:swept_eq2}
\delta m_e = \int_{\delta e} \hat{\rho} dV \, .
\end{equation}
The density function, $\hat{\rho}(x,y)$, may take on any functional
form.  To maintain second-order accuracy of the algorithm we use a linear density function;
\begin{equation}
\hat{\rho} = \rho^p_z + (\vec{\nabla}\rho)^p_z \cdot (\vec{x} - \vec{x}^p_z) \, ,
\end{equation}
where $\vec{x}^p_z$ is the average of the nodal positions defining the subcell.

In determining the gradient, $(\vec{\nabla}\rho)^p_z$, we employ one of
two standard methods for calculating the gradient using the densities
from subcell $\{z,p\}$ and its neighbors.
The first method uses Green's theorem to rewrite a bounded volume integral as a
boundary integral around the same volume.  We begin
with a definition of the average gradient of $\rho$ for a given
region.  For each component, the
average gradient is
\begin{equation}
\begin{array}{ccccc}
\left < \frac{\partial \rho}{\partial x} \right > &
= &
\frac{1}{V}\int_{V} \rho_x dV &
= &
\frac{1}{V}\oint_{\partial V} \rho dy\\
\left < \frac{\partial \rho}{\partial y} \right >&
= &
\frac{1}{V}\int_{V} \rho_y dV &
= &
-\frac{1}{V}\oint_{\partial V} \rho dx \, , \\
\end{array}
\end{equation}
where $\rho_x$ and $\rho_y$ are the gradients in the x- and y-directions, respectively.
The region over which the integrals are evaluated is defined by
segments connecting the centers of the neighboring subcells.  Hence, 
the discrete form of the average gradient is
\begin{equation}
\begin{array}{rcl}
\left < \frac{\partial \rho}{\partial x} \right >^p_z&
 = &
\frac{1}{V}\sum_e \frac{\rho_1 + \rho_2}{2} \Delta y_{12}\\
\left < \frac{\partial \rho}{\partial y} \right >^p_z&
 = &
-\frac{1}{V}\sum_e \frac{\rho_1 + \rho_2}{2} \Delta x_{12} \, .\\
\end{array}
\end{equation}
While this method works reasonably well, an unfortunate consequence
of the integration and division by volume is that the value of the gradient is
directly influenced by the shape and size of the cell.

An alternative method for determining gradients, which is not as easily influenced by the shape
of the mesh, is a least-squares procedure.  For simplicity of notation,
we describe this method in the context of cells
rather than subcells.  Again, assuming a linear
form for $\hat{\rho}_z(x,y)$, we seek to minimize the difference between
$\rho_c$ and $\hat{\rho}_z(x_c,y_c)$, where $\rho_c$ is the neighbor's
value and $\hat{\rho}_z(x_c,y_c)$ is the extrapolation of cell $z$'s
linear function to the position of the neighbor, $(x_c,y_c)$.
More explicitly, we wish to minimize the following equation:
\begin{equation}
\label{eq:least_eq1}
\sum_{c \in \mathcal{N}(z)} \omega^2_{cz} E^2_{zc} \, ,
\end{equation}
where the set of neighbors for cell $z$ is denoted by $c \in \mathcal{N}(z)$,
\begin{equation}
\label{eq:least_eq2}
E^2_{zc} = \left (
-\Delta \rho_{cz} + \rho_{x,z} \Delta x_{cz} + \rho_{y,z} \Delta y_{cz}
\right )^2 \, ,
\end{equation}
$\rho_{x,z}$ and $\rho_{y,z}$ are the $x$ and $y$ components of the
gradient, $\Delta \rho_{cz} = \rho_c - \rho_z$, $\Delta x_{cz} = x_c - x_z$, $\Delta
y_{cz} = y_c - y_z$, and $\omega_{cz}^2 = 1/(\Delta x_{cz}^2 + \Delta y_{cz}^2)$.
Minimizing eq. (\ref{eq:least_eq1}) with respect to the two unknowns, leads to the
following set of linear equations:
\begin{equation}
\begin{array}{rcl}
a \rho_x + b \rho_y & = & d\\
b \rho_x + c \rho_y & = & e \, ,\\
\end{array}
\end{equation}
for each subcell, where
\begin{equation}
\begin{array}{rcl}
a & = & \sum_{c \in \mathcal{N}(z)} \omega^2_{cz} \Delta x^2_{cz}\\
b & = & \sum_{c \in \mathcal{N}(z)} \omega^2_{cz} \Delta x_{cz} \Delta y_{cz}\\
c & = & \sum_{c \in \mathcal{N}(z)} \omega^2_{cz} \Delta y^2_{cz}\\
d & = & \sum_{c \in \mathcal{N}(z)} \omega^2_{cz} \Delta \rho_{cz} \Delta x_{cz}\\
e & = & \sum_{c \in \mathcal{N}(z)} \omega^2_{cz} \Delta \rho_{cz} \Delta
y_{cz} \, .\\
\end{array}
\end{equation}
Solving this linear system with Cramer's rule then gives the least-squares gradients.

After calculating the gradient
we ensure monotonicity using the Barth-Jespersen limiter
\citep{barth97}.  The gradient is limited by a scalar, $\Phi_z$, that
has a range between 0 and 1:
\begin{equation}
\hat{\rho}_z(x,y) = \rho_z + \Phi_z (\vec{\nabla}\rho)_z \cdot (\vec{x} - \vec{x}_z) \, .
\end{equation}
First, we determine the minimum and maximum value of $\rho$ among $\rho_z$ and
its neighbors $\rho_c$:
\begin{equation}
\rho^{\rm min}_z = \min (\rho_z,\rho_c)
\end{equation}
\begin{equation}
\rho^{\rm max}_z = \max (\rho_z,\rho_c) \, .
\end{equation}
We satisfy the requirement that
\begin{equation}
\rho^{\rm min}_z \leq \hat{\rho}(x,y) \leq \rho^{\rm max}_z \, .
\end{equation}
This is accomplished in the following way.
\begin{equation}
\Phi_z = \min ( \Phi^n_z) \, ,
\end{equation}
where $\Phi^n_z$ is a limiter associated with each node of the cell
and is given by
\begin{equation}
\Phi^n_z =
\left \{
\begin{array}{lcr}
\min \left ( 1,\frac{\rho^{\rm max}_z - \rho_z}{ \hat{\rho}(x_n,y_n) - \rho_z} \right ) & \mbox{for} &
\hat{\rho}(x_n,y_n) - \rho_z > 0 \\
\min \left ( 1,\frac{\rho^{\rm min}_z - \rho_z}{ \hat{\rho}(x_n,y_n) - \rho_z} \right ) & \mbox{for} &
\hat{\rho}(x_n,y_n) - \rho_z < 0 \\
1 & \mbox{for} & \hat{\rho}(x_n,y_n) - \rho_z = 0 \, , \\
\end{array}
\right .
\end{equation}
where $\hat{\rho}(x_n,y_n)$ is the unlimited linear function evaluated at the nodes.

With the linear function $\hat{\rho}$ defined for each subcell, we
return to the task of determining the swept mass, $\delta m_e$.  The specific linear function $\hat{\rho}$ for density
used in eq. (\ref{eq:swept_eq2}) depends upon which subcell the edge encroaches upon.  In fact,
it is quite possible that the volume swept by the edge intersects more
than one immediate neighbor of the subcell in question.  Of course, an
accurate swept-edge integration scheme would take into account all
spatial functions in the relevant subcells.  \citet{kucharik03}, and
references therein, have noted that such an accurate scheme may be
cumbersome and computational expensive.  Instead, they suggest an
approximate swept region integration in which only the subcells to the
left and right of the oriented edge need matter in determining the
spatial function of density used for integration.  To this
end, an oriented volume integral is defined:
\begin{equation}
\delta V_e = \oint xdy \, .
\end{equation}
Whether this volume integral is negative or positive, the density
function is taken from either the left or right subcell:
\begin{equation}
\hat{\rho} = \left \{
\begin{array}{lr}
\hat{\rho}_r, & \delta V_e \geq 0 \, ,\\
\hat{\rho}_l, & \delta V_e < 0
\end{array}
\right .
\end{equation}

\subsection{Repair}

While the above scheme is second-order accurate and mass conserving, the
approximations made in the swept-edge algorithm can violate local
bounds.  In practice, the values of the remapped densities should be
bounded by the neighborhood values of the original grid.
Computationally expensive schemes that
find the overlap regions among the old and new grid can be made to preserve the
bounds.  However,
the local bounds can be violated in swept-edge remapping.  Using the
linearity-and-bound-preserving method of \citet{kucharik03}, we conservatively repair quantities by redistributing mass,
energy, momenta, and number to neighbors satisfying the local bounds
of the quantities on the previous Lagrangian grid.  However, the order
in which one processes the cells and subcells influences the specific
values that emerge from the repair process.  Therefore, we employ the order
independent scheme of \citet{loubere06}.  These repair
schemes are designed to repair subcell densities.  For velocities and
composition we wish to preserve the bounds of $\vec{v}$ and $X_i$.
Therefore, we have appropriately adjusted the scheme to conservatively
repair momenta and particle number by preserving the local
bounds of $\vec{v}$ and $X_i$, respectively.

\subsection{Scattering}
\label{section:scatter}
The final remap step is to recover the primary variables for the
next hydrodynamic step.
The new cell-centered and node-centered masses are
\begin{equation}
\tilde{m}_z = \sum_{p \in S(z)} \tilde{m}^z_p
\end{equation}
and
\begin{equation}
\tilde{m}_p = \sum_{z \in S(p)} \tilde{m}^p_z \, ,
\end{equation}
respectively. Consequently, the subcell and cell-averaged densities are
\begin{equation}
\tilde{\rho}^z_p = \frac{\tilde{m}^z_p}{\tilde{V}^z_p}
\end{equation}
and
\begin{equation}
\tilde{\rho}_z = \frac{\tilde{m}_z}{\tilde{V}_z} \, ,
\end{equation}
respectively.

Recovering the velocities at the nodes is slightly more involved.
First, the new subcell-averaged velocity is defined using the remapped
momenta and masses:
\begin{equation}
\tilde{v}^z_p = \frac{\tilde{\mu}^z_p}{\tilde{m}^z_p}
\end{equation}
Then, for each cell we invert the matrix equation that transforms node
velocities into subcell velocities:
\begin{equation}
\label{eq:vmatrix}
\tilde{\bf U}_z = \tilde{\bf M}^{-1}_z \tilde{\bf U}^{SC}_z \, .
\end{equation}
In this scheme, each cell provides its own velocity for a node, which may
be different from another cell's value for the same node.  Therefore, for
each node velocity, we average node velocities resulting from the
inversion of eq. (\ref{eq:vmatrix}):
\begin{equation}
\tilde{v}_p = \frac{1}{\tilde{m}_p} \sum_{z \in S(p)} \tilde{m}^z_p 
\tilde{v}_p(z) \, .
\end{equation}

Finally, we recover the specific internal energy.  Comparing the sum of the old internal energy and
kinetic energy with the sum of remapped internal energy and a kinetic
energy defined by the remapped node velocities will not necessarily
ensure conservation of total energy.  Therefore, the old
kinetic energy is remapped along with the node velocities.  The
discrepancy between these two kinetic energy representations is then
added to the internal energy, ensuring conservation of total energy
during the remap stage:
\begin{equation}
\label{eq:remap_conserve_ene}
\tilde{\mathcal{E}}_z = \sum_{p \in S(z)} \tilde{\mathcal{E}}^z_p
+ \left [
\left ( \sum_{p \in S(z)} \tilde{K}^z_p \right )
- \left ( \sum_{p \in S(z)} \tilde{m}^z_p \frac{|\tilde{v}_p|^2}{2} \right )
\right ] \, .
\end{equation}
The new cell-centered specific internal energy is then obtained using
the modified internal energy, $\tilde{\mathcal{E}}_z$, and the
remapped mass, $\tilde{m}_z$: 
\begin{equation}
\tilde{\varepsilon}_z = \frac{\tilde{\mathcal{E}}_z}{\tilde{m}_z} \, .
\end{equation}

This ensures conservation of total energy at the expense of consistent
remapping of internal energy.  Whichever is desirable depends upon the
problem.  For example, in flows with large kinetic energies and small
internal energies, the discrepancy in kinetic energy remapping could
substantially alter or even dominate the internal energy.
Consequently, we include a flag in BETHE-hydro that determines whether
the remapped kinetic energy differences are added to the internal energy.

\subsection{Remapping Tests}
\label{section:remap_test}

\begin{figure}
\epsscale{.90}
\plotone{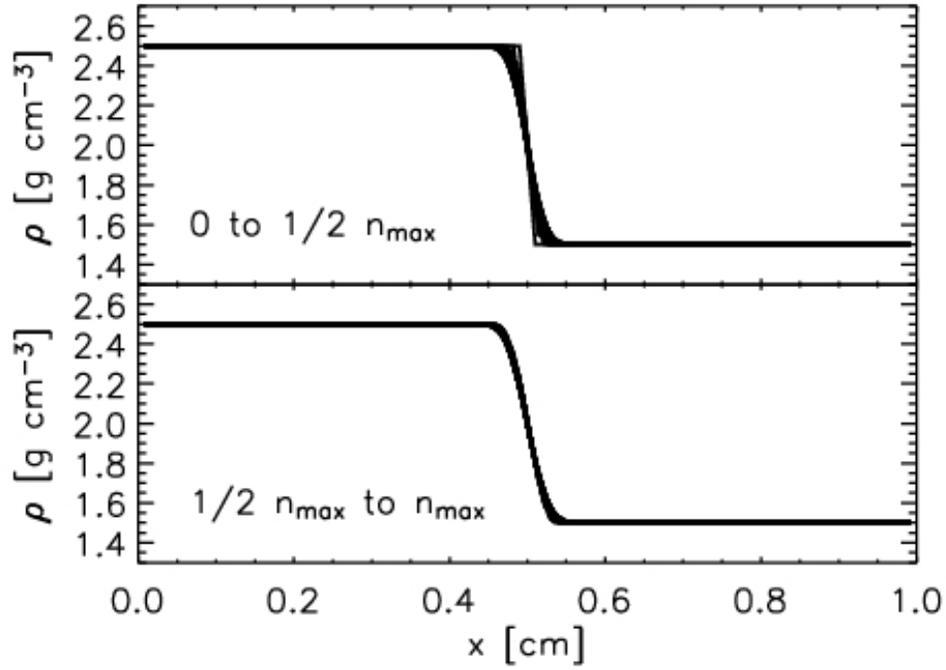}
\caption{Remapping a 1D step function. A step function in density is
  remapped many times with a grid that oscillates for two full cycles.
  There are 51 nodes (50 cells) and $n_{\rm max} = 800$ remapping
  steps.  The top panel displays the density profile for remapping
  steps 0 to $1/2 n_{rm max}$, and the bottom panel shows the profile for
  steps $1/2 n_{\rm max}$ to $n_{\rm max}$.  Initially, the
  discontinuity spreads over a small number of zones ($\sim$4), and in
  later steps (bottom panel) the discontinuity spreads very slowly.  \label{rhoremap_step}}
\end{figure}

Figure \ref{rhoremap_step} demonstrates the basic effectiveness of the remapping
algorithm.  We remap a 1D step function in density many times.  From
$x=0$ to $x=1/2$ cm, the density is $\rho = 2.5$ g cm$^{-3}$, and from $x =
1/2$ to $x = 1$ cm, the density is $\rho = 1.5$ g cm$^{-3}$.  This test has 50 cells
and $N_{nodes} = 51$ nodes.  Indexing each node by $i$, the positions of
the nodes are
\begin{equation}
x_i = (1 - \alpha)  \left ( \frac{i-1}{N_{nodes}-1} \right ) + \alpha \left ( \frac{i-1}{N_{nodes}-1} \right )^2 \, ,
\end{equation}
where
\begin{equation}
\alpha = \frac{1}{4}\sin \left (4 \pi \frac{n}{n_{max}} \right ) 
\end{equation}
and the remapping
step is $n$.  As a result, the grid completes two full cycles in this
remapping test.  Figure \ref{rhoremap_step} shows the result of this
remapping test where $n_{max} = 800$.  The top panel displays the density profile for remapping
steps 0 to $n_{max}/2$, while the bottom panel shows the profile for
steps $n_{max}/2$ to $n_{max}$.  Comparison of the first half (top
panel) with the
second half (bottom panel) indicates that the remapping process diffuses the
discontinuity over a small number ($\sim$4) of zones initially, and
``diffusion'' slows substantially after the initial phase, maintaining
a somewhat consistent width in the discontinuity.

\subsection{Composition Remap}
\label{section:composition_remap}
Barring any nuclear or chemical transformations, the composition
equation, eq. (\ref{eq:advec}), states that $X_i$ is conserved.
The Lagrangian hydro portion of our
algorithm will not change a cell's composition.  All alterations in
composition are, therefore, a result of advection, and in an ALE code
advection is handled by the remapping algorithm.

Equation (\ref{eq:advec}) may be written in Eulerian form:
\begin{equation}
\label{eq:advec_eulerian}
\frac{\partial (\rho X_i)}{\partial t} + \vec{\nabla} \cdot (\rho X_i
\vec{v}) = 0 \, ,
\end{equation}
which implies a close dependence of the
advection of $X_i$ on $\rho$ and $\vec{v}$.  In practice, if the
composition is not advected in a manner entirely consistent with the
advection of mass, then the advection of composition will develop
 peculiarities.  Since the remapping of density is truly a
remapping of the subcell masses, we have designed a scheme for the
remapping of composition which operates on an equivalent subcell quantity.

Specifically, we define new Lagrangian quantities for
composition.   They are the number of species $i$ in the cell,
$N_{i,z}$, and the number of species $i$ in the subcell, $N^z_{i,p}$.
Analogous to the mass density is the number density, $n_{i,z}$ and
$n^z_{i,p}$.  These new quantities are related to the previously defined
quantities by: $n_i = \rho X_i$ and $N_i = n_i V$ or $N_i = m X_i$.

The remap (or advection) of composition follows the scheme
outlined for the advection of mass on the subcell level, with $n_i$
replacing $\rho$ and $N_i$ replacing mass.  The new compositions are
then determined by $\tilde{X}_{i,z} = \tilde{N}_{i,z}/\tilde{m_z}$.
The final minor, but crucial, difference between mass and composition
remapping occurs during the repair process.  While the repair process
maintains the bounds on the number density, we also enforce bound
preservation of the compositions.

While we do not address specific rate equations in this work,
transport and nuclear processes will require consideration of nonzero
terms on the RHS of eqs. (\ref{eq:advec}) and
(\ref{eq:advec_eulerian}).  Since our division of composition into
subcell compositions has consequences for discrete implementations with
rates, we include here some discussion.

Normally, the discrete form of eq. (\ref{eq:advec}) with a nonzero
RHS would be designed to operate at the cell
level.  In other words, it would involve $N_{z,i}$ and $n_{z,i}$.
However, the fundamental Lagrangian subunit is the subcell.
Therefore, we have developed an algorithm to handle the composition
changes, $\Delta X_i$, at the subcell level due to changes at the cell level.

Irrespective of the scheme employed, a condition which must be
satisfied is that
\begin{equation}
\sum_z \Delta N_{z,i} = \sum_z \sum_{p \in S(z)} \Delta N^z_{p,i} \, .
\end{equation}
We'd like to convert the statement of number conservation into the more useful relationship:
\begin{equation}
\label{eq:deltancell}
\Delta N_{z,i} = \sum_{p \in S(z)} \Delta N^z_{p,i} \, .
\end{equation}
When the change in composition due to the rates
is applied in operator-split form to the discrete hydro
equations, we have the following relations:
\begin{equation}
\label{eq:deltaxcriterion1}
\Delta N_{i,z} = \Delta X_{i,z} m_z \, ,
\end{equation}
and
\begin{equation}
\Delta N^z_{i,p} = \Delta X^z_{i,p} m^z_p \, .
\end{equation}
Our dilemma is that the rate equations will determine $\Delta X_{z,i}$,
but they put no constraint on $\Delta X^z_{p,i}$.  A fairly natural and
simple choice is $\Delta X^z_{p,i} = \Delta X_{z,i}$, giving 
\begin{equation}
\label{eq:deltansubcell}
\Delta N^z_{p,i} = \Delta X_{z,i} m^z_p \, .
\end{equation}
Substituting eq. (\ref{eq:deltansubcell}) into
eq. (\ref{eq:deltancell}) we have
\begin{equation}
\Delta N_{z,i} = \sum_{p \in S(z)} \Delta X_{z,i} m^z_p = \Delta
X_{z,i} \sum_{p \in S(z)}
m^z_p = \Delta X_{z,i} m_z \, .
\end{equation}
Hence, by simply stating that $\Delta X^z_{p,i} = \Delta X_{z,i}$, we
have developed discrete rate equations that satisfy conservation of
species number and operates at the subcell level.

\subsection{Angular Velocity Remap}
\label{section:ang_vel_remap}

The remap of the angular momentum deserves special mention.  As in mass remapping, we try
to find 
new subcell
angular momenta, $\tilde{J^z_p}$, based upon an approximate swept-region
approach.  In other words,
\begin{equation}
\tilde{J^z_p} = J^z_p + \sum_{e \in S(z)} \delta J_e \, ,
\end{equation}
which parallels eq. (\ref{eq:swept_eq1}) in form.  A notable
difference, however, is the expression for the swept angular momentum.
It no longer involves a simple volume integral, but an integral
weighted by $r^2$ (see appendix
\ref{section:integrals} for calculating the discrete analog of this integral):
\begin{equation}
\delta J_e = \omega \rho \int r^2 dV \, .
\end{equation}
After finding the new subcell angular
momentum we determine the new cell angular momentum,
\begin{equation}
\tilde{J_z} = \sum_{p \in S(z)} \tilde{J^p_z} \, ,
\end{equation}
and node angular momentum,
\begin{equation}
\tilde{J_p} = \sum_{z \in S(p)} \tilde{J^p_z} \, ,
\end{equation}
and in turn determine the new angular velocity;
\begin{equation}
\tilde{\omega}_p = \frac{\tilde{J_p}}{\tilde{\mathcal{I}_p}} \, .
\end{equation}

\section{Code Tests}
\label{section:tests}

In this section, we characterize BETHE-hydro's 
performance using several test problems.  First, we demonstrate that
this code produces 2$^{nd}$-order accurate solutions for
self-gravitating flows. To assess the accuracy of high Mach-number simulations, we use the Sod shock tube problem, Sedov blast
wave, and Noh implosion problem, which all have analytic solutions.
Furthermore, we simulate the Saltzman and
Dukowicz piston problems to test the code's ability to simulate piston-driven shocks using oblique meshes.  In simulating two important
hydrodynamic instabilities, the Rayleigh-Taylor and Kelvin-Helmholtz
instabilities, we further demonstrate this
code's strengths and limitations.  Demonstrating BETHE-hydro's ability to simulate
astrophysical phenomena, we conclude with a core-collapse supernova simulation.

\subsection{2$^{\rm nd}$-Order Accuracy}
\label{section:2ndorder}

\begin{figure}
\epsscale{0.7}
\plotone{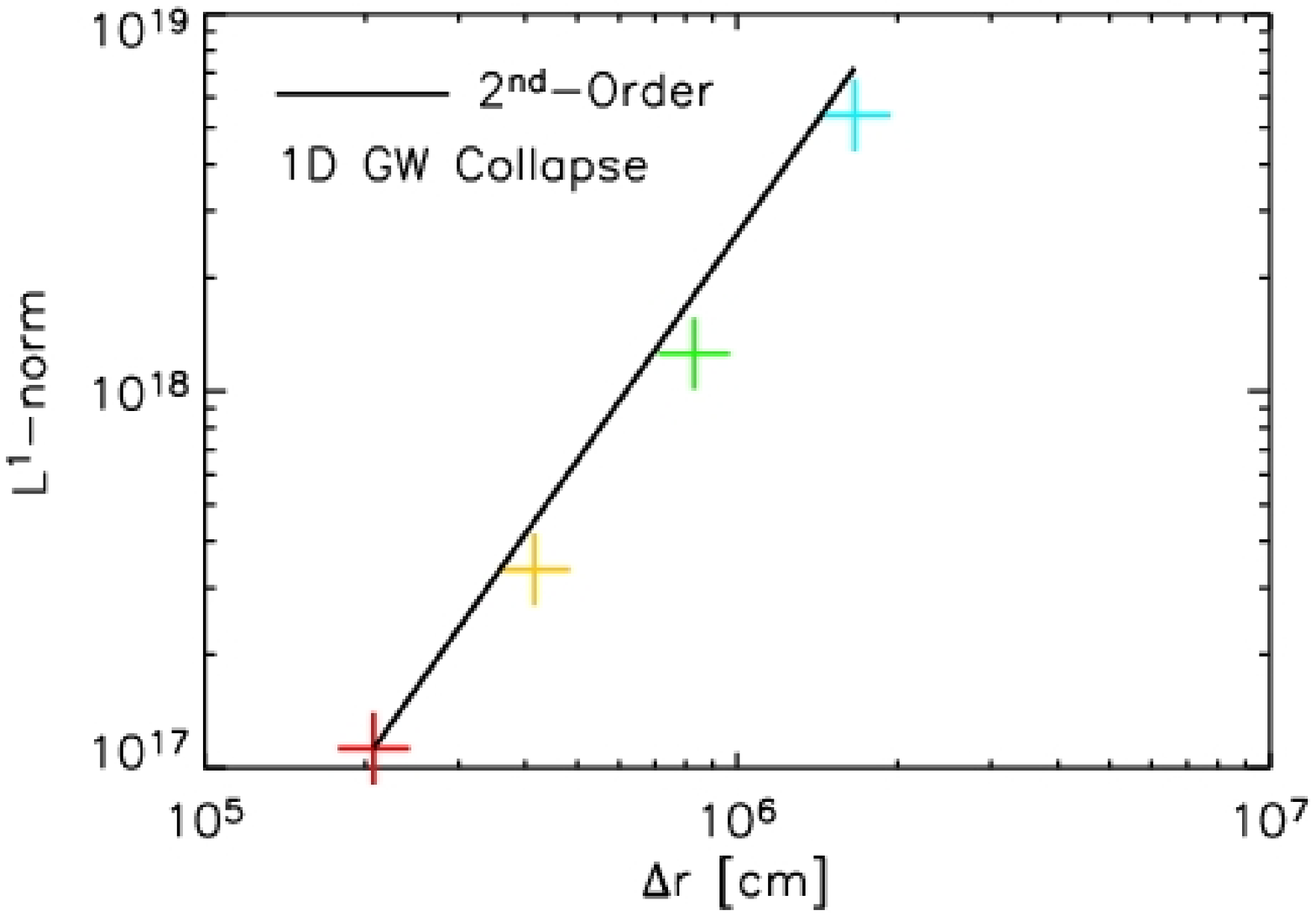}\\
\plotone{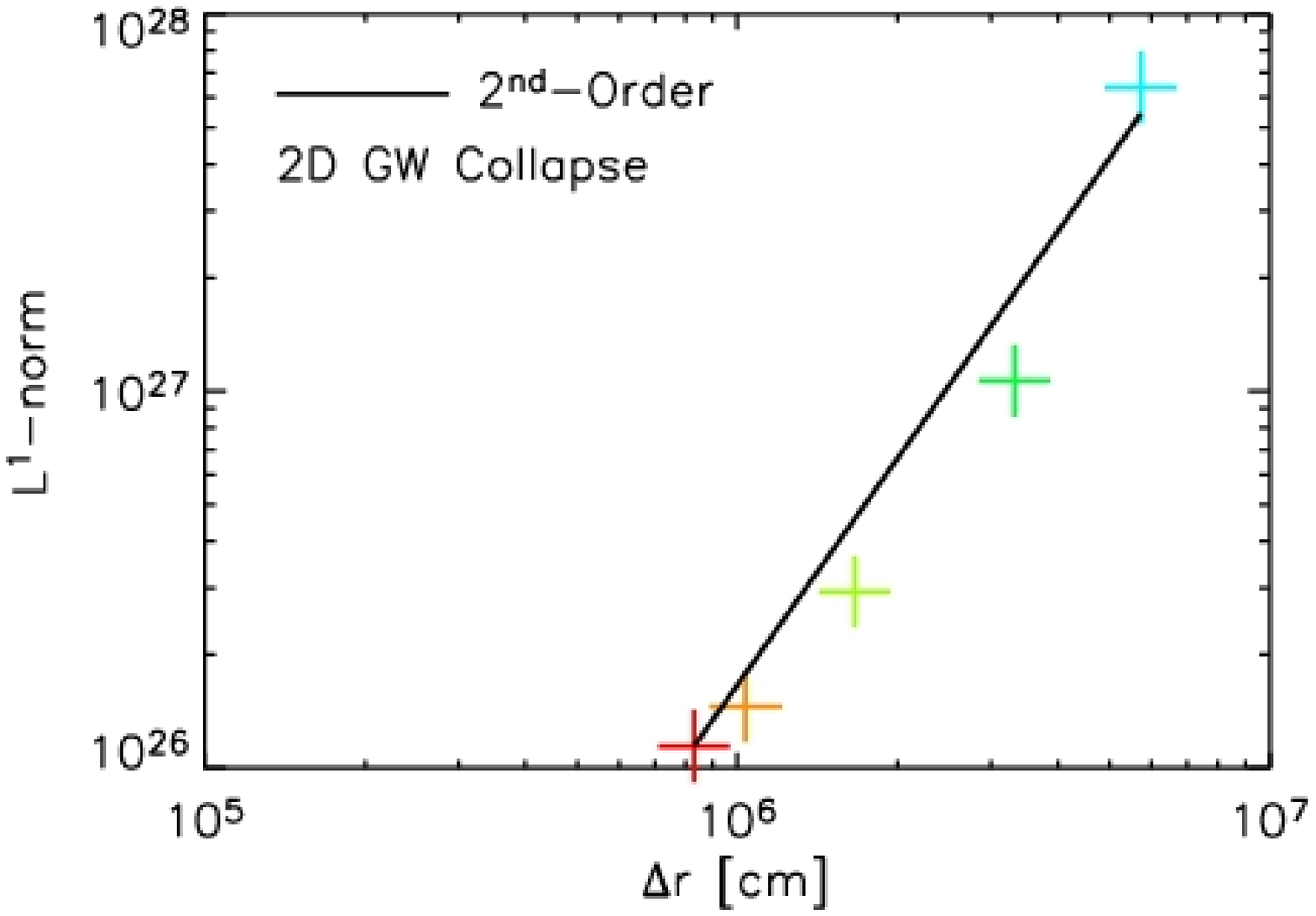}
\caption{
The $L^1$-norm as a function of
$\Delta r$ (crosses), the initial zone size, is plotted for Goldreich-Weber simulations in 1D
(top panel) and 2D (bottom panel).  The $L^1$-norm is calculated at t
= 130 ms for both simulations.  Both 1D and 2D simulations (crosses)
converge with roughly 2$^{\rm nd}$-order accuracy (solid line).  See
\S \ref{section:2ndorder}.
\label{gw_resstudy}}
\end{figure}

To verify the 2$^{\rm nd}$-order character of BETHE-hydro, a smooth
hydrodynamics flow is required.  The
Goldreich-Weber self-similar collapse
(see \S \ref{section:gravity_tests} and Fig. \ref{gwplot}) satisfies this requirement and is
ideal to check convergence of the hydrodynamic and gravity solvers and
their coupling.  We use $L^1$-norms of the error:
\begin{equation}
\label{eq:l1norm}
L^1 = \sum |e_z \Delta r^d| \, ,
\end{equation}
where the error is $e_z = \rho_{\rm ana}(\vec{x_z}) -\rho_z$,
$\rho_{\rm ana}$ is the analytic density at $t=130$ ms, $\Delta
r$ is the zone size, and $d$ is 1 for 1D and 2 for 2D.  Strictly
speaking, the simulations are Lagrangian and have time-varying zone sizes.
Fortunately, this collapse problem is self-similar, implying a direct
correlation between the starting resolution and the resolution at a
later time.  Therefore, we use the initial zone size in eq. (\ref{eq:l1norm})
and in Fig. \ref{gw_resstudy}.  The $L^1$-norm as a function of
$\Delta r$ (crosses) is plotted in Fig. \ref{gw_resstudy}.  As the
figure demonstrates, both 1D simulations (top panel) and 2D simulations (bottom panel)
converge with roughly 2$^{\rm nd}$-order accuracy.

\subsection{Sod Shock Tube Problem}
\label{section:sod}

The Sod shock tube problem is a standard analytic test that assesses the code's
ability to simulate three characteristic waves:  a shock, a
rarefaction wave, and a contact discontinuity.
A simple gamma-law EOS is employed with $\gamma =
1.4$.  Initially, the computational domain is divided into left and
right static
regions with different densities and pressures.
Specifically, the left and right regions have densities of 1.0 g
cm$^{-3}$ and $0.125$ g cm$^{-3}$ and pressures of 1.0 erg cm$^{-3}$ and
0.1 erg cm$^{-3}$, respectively.  This gives rise to a self-similar
solution involving a shock propagating to the right, a rarefaction
wave propagating to the left, and a contact discontinuity in between.

\begin{figure}
\epsscale{.80}
\plotone{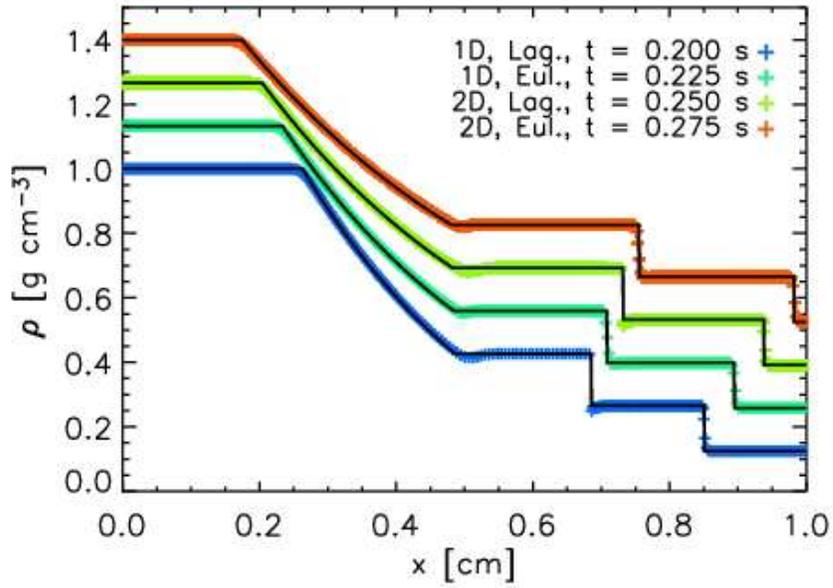}
\caption{Sod shock tube test.  For a gamma-law EOS and $\gamma = 1.4$,
  we compare the results of the 1D Lagrangian (bottom), 1D Eulerian
  (2nd from the bottom), 2D Lagrangian (3rd), and 2D Eulerian (top) Sod
  shock tube tests with the analytic result (solid lines).  Other than
  the 1D Lagrangian results, the density profiles have
  been shifted vertically to distinguish features.  The profiles
  are further separated by displaying them at different times: 1D
  Lagrangian ($t = 0.2$ s), 1D Eulerian
  ($t = 0.225$ s), 2D Lagrangian ($t = 0.25$ s), and 2D Eulerian ($t =
  0.275$ s).  The 1D calculations are
  resolved with 400 zones, and the 2D tests are resolved with 400x10
  zones.  See \S \ref{section:sod} for a discussion.\label{sodplot}}
\end{figure}

In Fig. \ref{sodplot}, we display the Sod shock tube test results for 1D Lagrangian, 1D
Eulerian, 2D Lagrangian, and 2D Eulerian configurations.
Cell-centered densities and locations are marked with plus signs, while
the analytic results are denoted by solid dark lines.  The 1D Lagrangian
result shows the appropriate density profile, while the other profiles
have been shifted vertically so that distinguishing features are more
easily compared.  In addition, the profiles are
further distinguished by displaying them at different times: 1D Lagrangian
($t = 0.2$ s), 1D Eulerian ($t = 0.225$ s), 2D Lagrangian ($t = 0.25$
s), and 2D Eulerian ($t = 0.275$ s).  The 1D calculations are resolved
with 400 zones. Similarly, the 2D tests are resolved along the
direction of shock propagation with 400 zones, and they are resolved
in the perpendicular direction by 10 zones.

The overall features of the shock, post-shock material, contact
discontinuity, and rarefaction wave are reproduced. For all scenarios,
the shock is resolved within a few zones.  For the 1D Eulerian and Lagrangian
calculations the post-shock density is good to $\sim 0.05$\% and $\sim
0.01$\%, respectively.  Upon viewing the
profiles in greater detail, there is a
noticeable departure from the analytic solution at the tail of the
rarefaction wave where the density dips below the expected value.
This error in density at the tail of the rarefaction wave is $\sim 2.5$\% for the
Lagrangian simulation and $\sim 1$\% for the Eulerian simulation.  Since the
contact discontinuity moves with the flow speed, both Lagrangian
calculations resolve the contact discontinuity exactly from one zone to the next and
maintain this resolution throughout the simulation.  The Eulerian
calculations, not surprisingly, distribute the contact discontinuity
over several ($\sim$4) zones.

\subsection{Sedov Blast Wave}
\label{section:sedov}

A classic test, the Sedov blast wave problem provides a quantitative
measure of a code's ability to simulate spherical explosions.
Initial conditions are 
set so that the total energy of the Sedov blast is 0.244816 ergs,
$\rho_0 = 1.0$ g cm$^{-3}$, $\varepsilon_0 = 1 \times 10^{-20}$ ergs
g$^{-1}$, and the gamma-law EOS has $\gamma = 5/3$.  
We compare the simulations with the analytic result (solid lines) for the following
scenarios: 1D Lagrangian, 1D Eulerian, 2D Lagrangian using the butterfly
mesh, and 2D Lagrangian using the spiderweb mesh.

\begin{figure}
\epsscale{.60}
\plotone{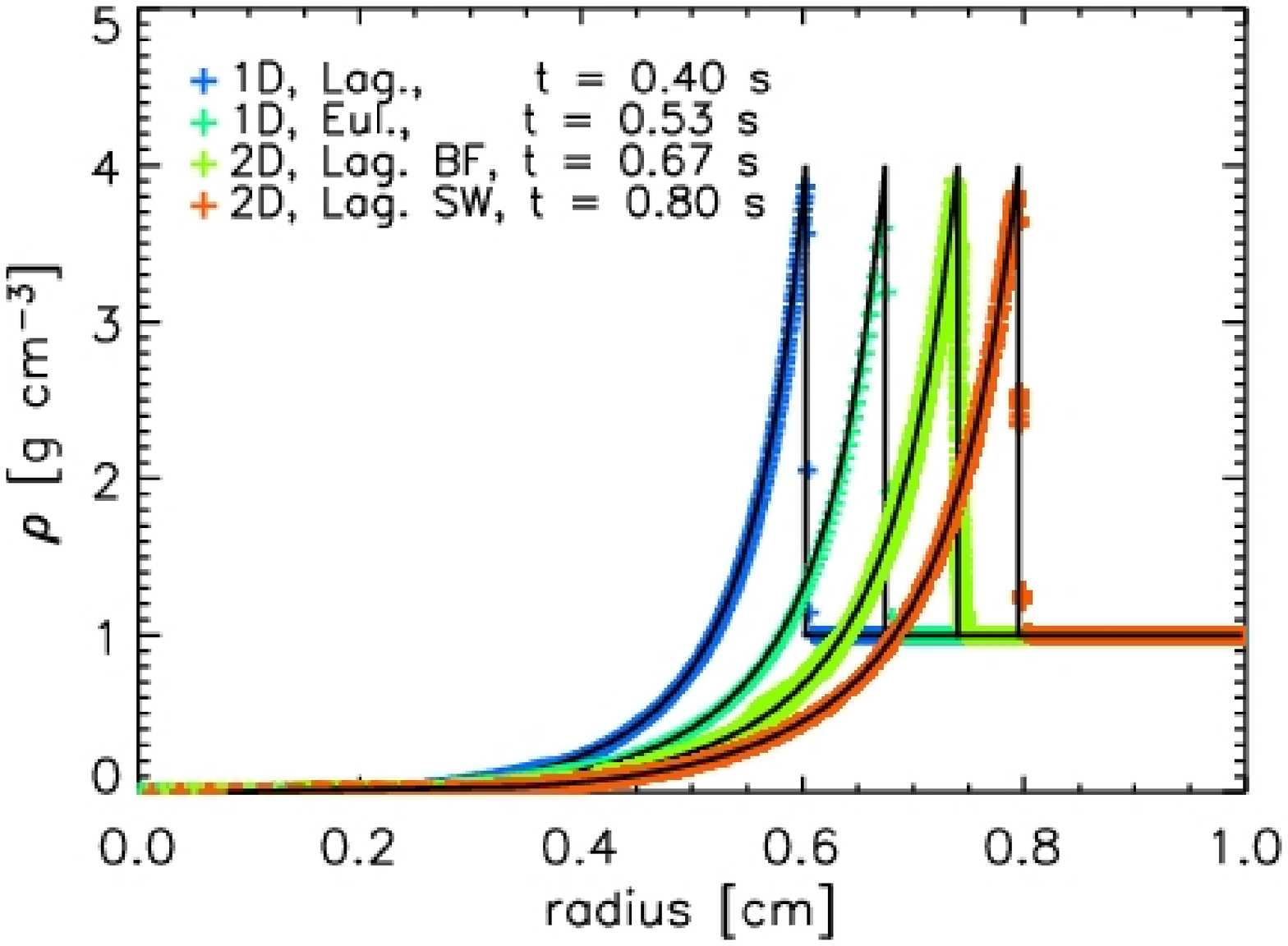}\\
\plotone{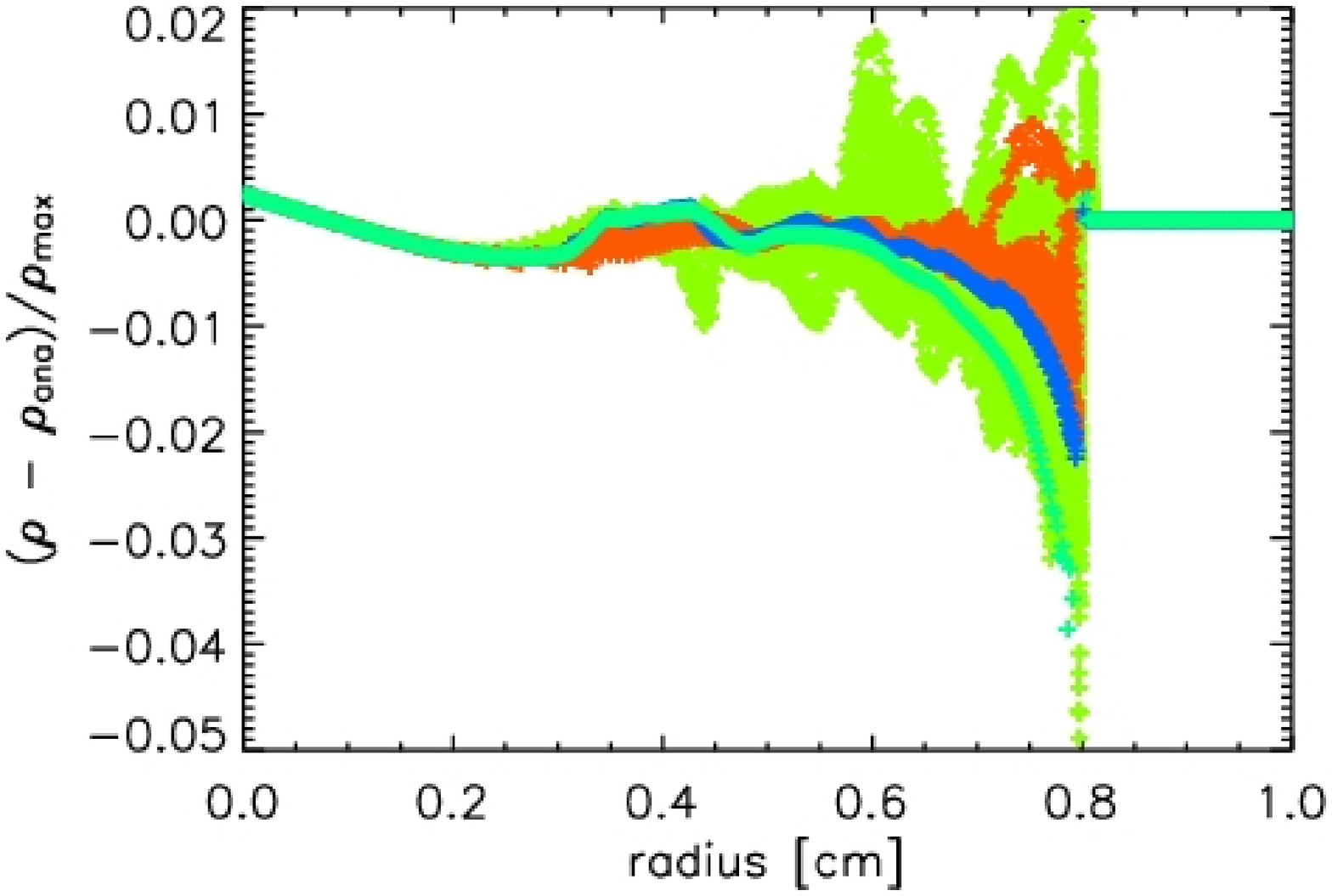}
\caption{
Density profiles of the Sedov blast wave test.  The Sedov blast energy
is 0.244816 ergs.  Initial conditions are 
$\rho_0 = 1.0$ g cm$^{-3}$, $\varepsilon_0 = 1 \times 10^{-20}$ ergs
g$^{-1}$, and the gamma-law EOS has $\gamma = 5/3$.  
Top panel compares the simulations (crosses) with the analytic result
(solid lines) for 1D Lagrangian ($t = 0.4$ s), 1D Eulerian ($t = 0.53$ s), 2D Lagrangian using the butterfly
mesh ($t = 0.66$ s), and 2D Lagrangian using the spiderweb mesh ($t =
0.80$ s).
1D calculations are resolved with 400
zones. The spiderweb test has a total of 12,381 zones with 200 radial
zones and a maximum of 64 angular zones.  The butterfly mesh has a
total of 35,000 zones with effectively 200 radial
and 200 angular zones.
Plotted in the bottom panel are the relative errors of the density,
$(\rho - \rho_{\rm ana})/\rho_{\rm max}$, vs. radius, where $\rho$
is the simulated density profile, $\rho_{\rm ana}$ is the analytic
profile, and $\rho_{\rm max}$ is the maximum density of the analytic profile.
\label{sedovplot}}
\end{figure}

Results of the Sedov explosion are plotted in Fig. \ref{sedovplot}.
The top panel compares the analytic and numerical density profiles, while
the bottom panel shows the relative differences between the numerical
solutions and the analytic profile.
To easily
distinguish different runs and details, we plot the 1D Lagrangian calculation
at $t = 0.4$ s, the 1D Eulerian at $t = 0.53$ s, the 2D Lagrangian using the
butterfly mesh at $t = 0.66$ s, and the 2D Lagrangian using the spiderweb
mesh at $t = 0.80$ s.  The 1D calculations are resolved with 400
zones. The spiderweb test has a total of 12,381 zones with 200 radial
zones and a maximum of 64 angular zones, while the butterfly mesh has a
total of 35,000 zones with effectively 200 radial
and 200 angular zones.
Qualitatively, all simulations reproduce the overall structure and
position of the shock and post-shock flows.

There are some practical issues about formulating the Sedov runs that are worth
mentioning.  For the 1D simulations, we found that simply placing the
initial explosion energy in a small number of inner zones was
adequate.  On the other hand, the 2D simulations using
non-spherical grids required more care.  Simply depositing all the
energy within a small number of zones near the center led to severe
grid tangling and distortions.  We
remedied this by initiating all profiles with the analytic solution at
$t = 0.001$ s.

The bottom panel of Fig. \ref{sedovplot} emphasizes the quantitative accuracy of the
post-shock solutions.  Plotted are the relative errors of the density,
$(\rho - \rho_{\rm ana})/\rho_{\rm max}$, versus radius, where $\rho$
is the simulated density profile, $\rho_{\rm ana}$ is the analytic
profile, and $\rho_{\rm max}$ is the maximum density of the analytic profile.  The 1D Lagrangian simulation gives the best
results with a maximum relative error near the shock of $\sim 2$\%.  In
comparison, the 1D Eulerian test gives $\sim 4$\% deviation near the
shock.  The 2D Lagrangian simulation using the spiderweb mesh has a
maximum deviation similar to the 1D Lagrangian simulation, even though
the 1D simulation has radial zones with half the zone size.  The simulation using
the spiderweb mesh shows some slight departure from spherical
symmetry.  Specifically, for
most of the post-shock region, the peak-to-peak variation of density is less than
$\sim 1$\%, and near the shock the peak-to-peak variation of density reaches
$\sim 3$\%.  The 2D Lagrangian simulation using the butterfly mesh has
a maximum density deviation of
$\sim 4$\% near the shock and a deviation from
spherical symmetry of similar magnitude.

\subsection{Noh Implosion Problem}

Initial conditions for the Noh Problem are uniform density $\rho_0 =
1$ g cm$^{-3}$, zero (or
very small) internal energy $\varepsilon_0 = 0$ ergs, a convergent velocity
field with magnitude $v_0 = 1$ cm s$^{-1}$, and a gamma-law EOS with
$\gamma = 5/3$.  Subsequent evolution produces a symmetric
self-similar flow including supersonic accretion, an accretion shock, and
stationary post-shock matter in which kinetic energy has been
converted into internal energy.  The analytic solution for this problem
is
\begin{equation}
\{ \rho, \varepsilon, v\} =
\left \{
\begin{array}{ll}
\left \{ \rho_0 \left ( \frac{\gamma + 1}{\gamma - 1} \right )^d
,\frac{1}{2}v^2_0,0  \right \} &
{\rm if} \, r < r_s \\
\left \{ \rho_0 \left ( 1 - \frac{v_0 t}{r} \right )^{d-1}
,0,v_0  \right \} &
{\rm if} \, r > r_s \\
\end{array}
\right .
\end{equation}
where $v$ is the velocity magnitude, the shock position is $r_s = u_s
t$, the shock velocity is $u_s =
\frac{1}{2}(\gamma-1) v_0$, and $d$ is 2 for 2D Cartesian, 3 for 2D
simulations using
cylindrical coordinates, and 3 for 1D spherically symmetric simulations.

\begin{figure}
\epsscale{.80}
\plotone{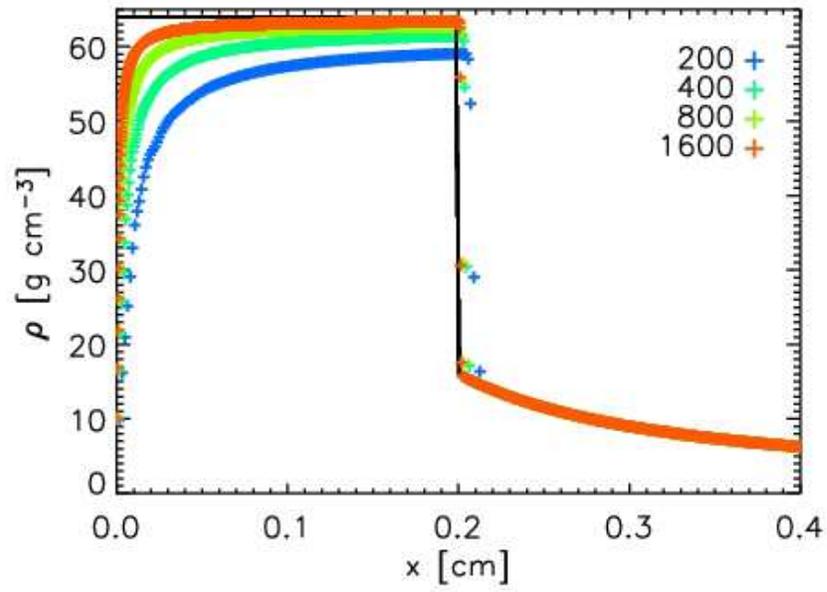}
\caption{Density profile for the Noh Problem at $t = 0.2$ s.  The solid line is the analytic solution for $\gamma
  = 5/3$, and the crosses show the numerical results for resolutions of
  200, 400, 800, and 1600 zones.  The downturn of the density profile near
  the center is ``wall heating,'' a common problem for Lagrangian
  schemes \citep{rider00}.  For the post shock material, the higher
  resolution runs capture the analytic solution.  The upstream
  flow matches the analytic solution for all resolutions.
\label{noh1dres}}
\end{figure}

In Fig. \ref{noh1dres}, we present the density profiles at $t = 0.2$ s
for 1D, spherically
symmetric simulations of the Noh problem using four
resolutions.  The initial grid spans the range $x \in \{ 0: 1\}$ cm
and is evenly divided into 200, 400, 800, and 1600 zones.
The solid line is the analytic solution for $\gamma
  = 5/3$.  Higher resolution simulations more accurately capture the shock
  position and the post-shock density profile.  However, all resolutions depart significantly from the analytic
  solution near the center.  This is ``wall heating'' and is a common problem for Lagrangian
  schemes \citep{rider00}.

\begin{figure}
\epsscale{1.1}
\plottwo{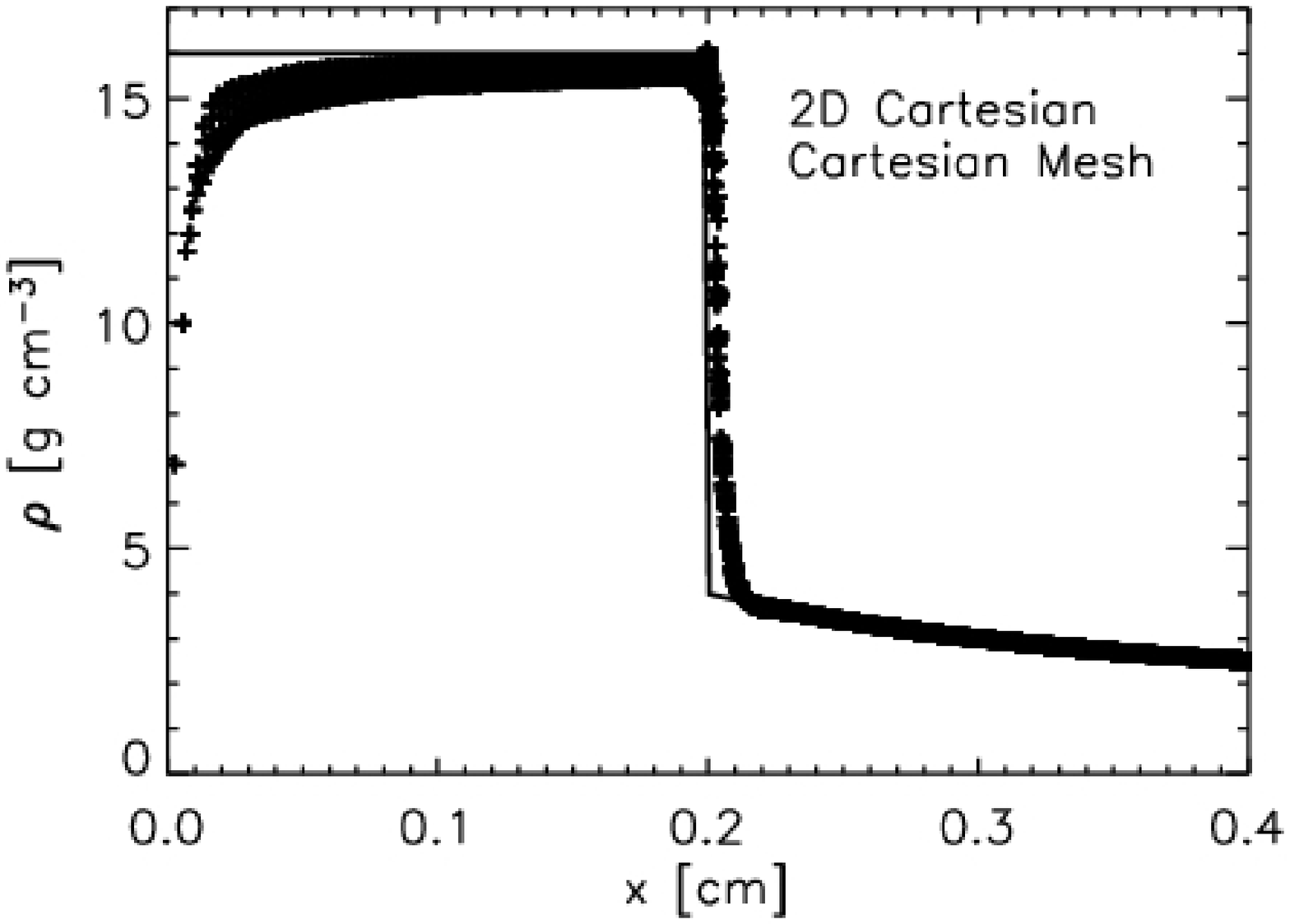}{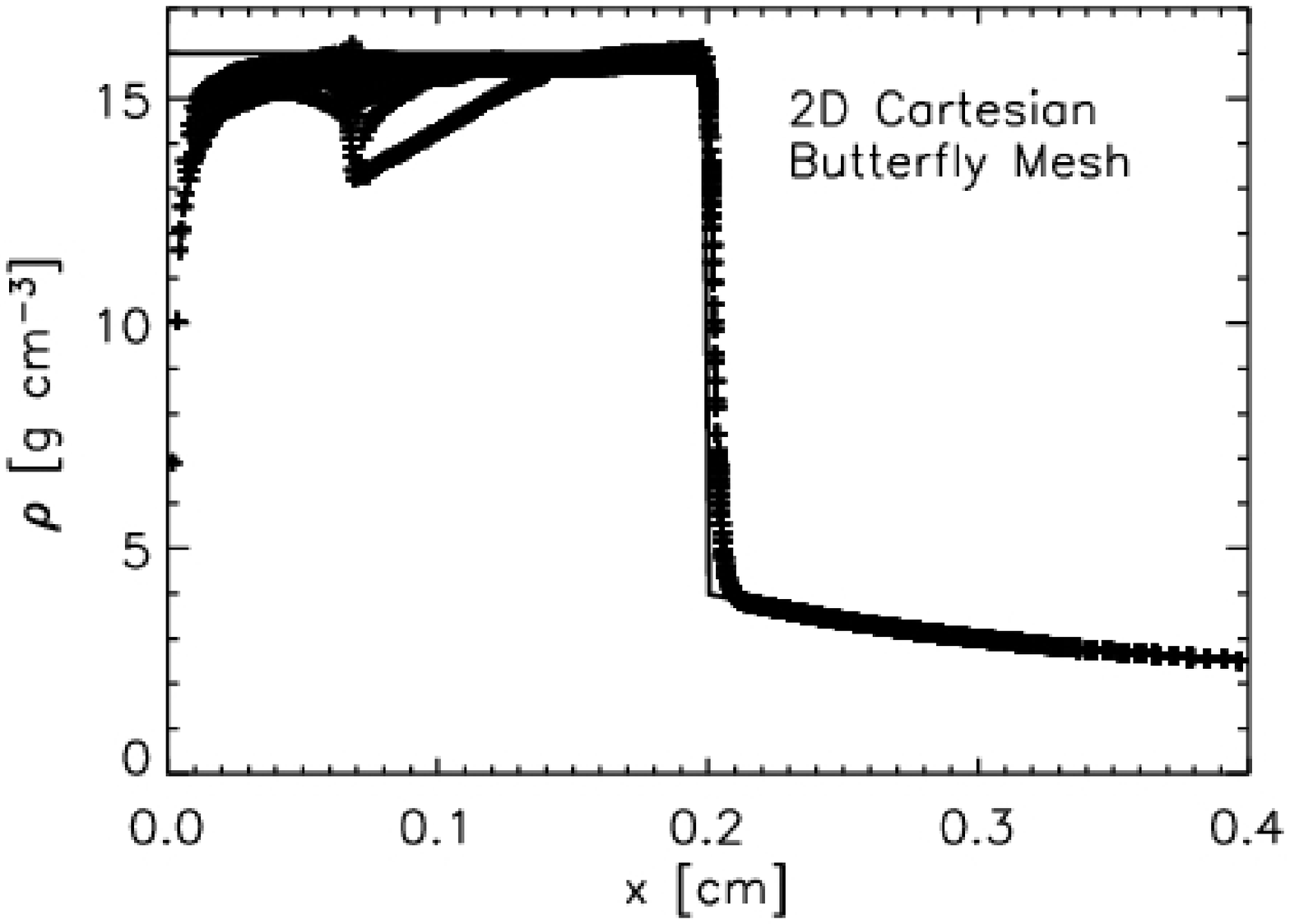}\\
\plottwo{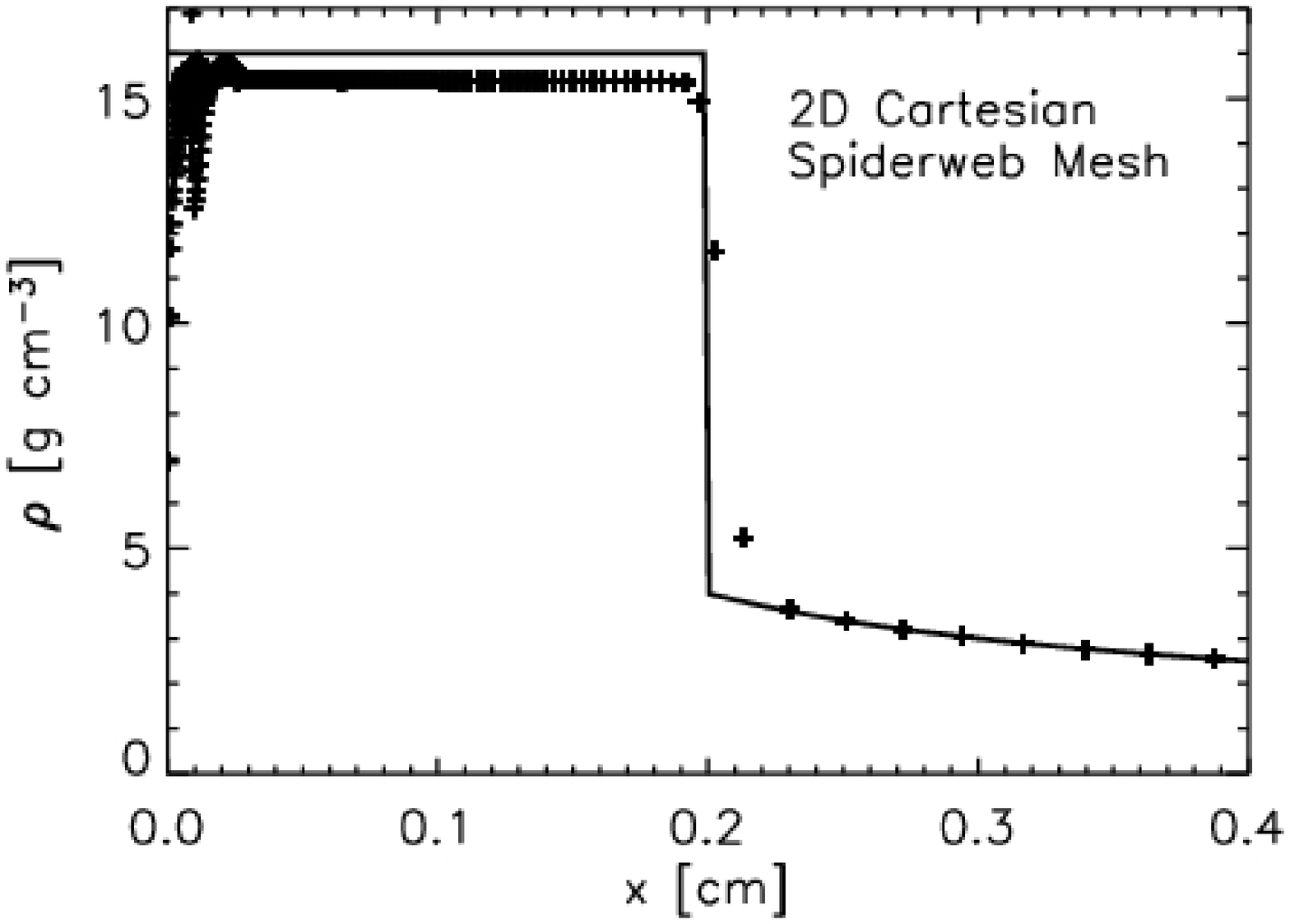}{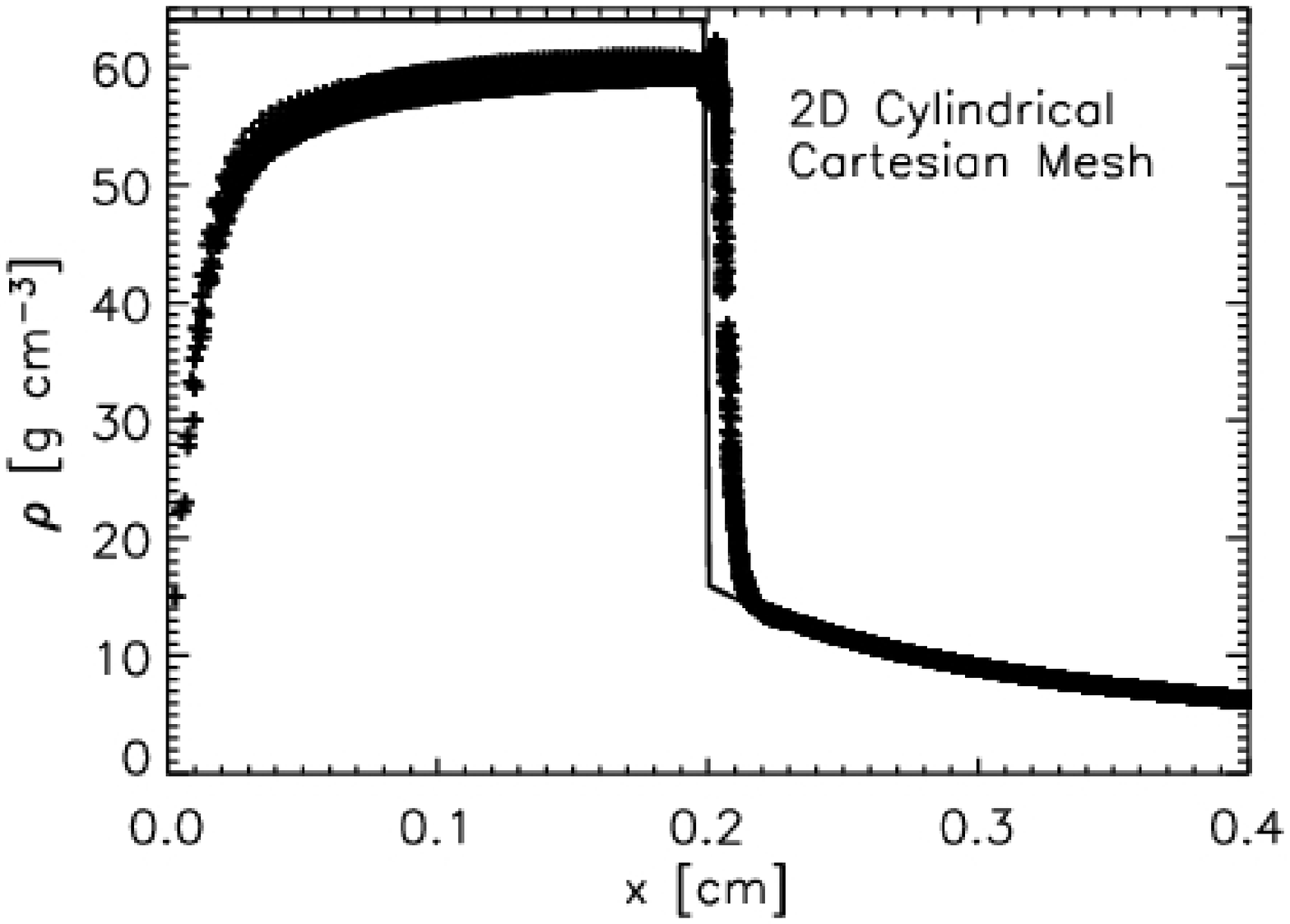}
\caption{Similar to Fig. \ref{noh1dres}, except here we present the
  results of 2D tests.  All but the lower-right show results using 2D
  Cartesian coordinates.  The lower-right shows results using 2D cylindrical
  coordinates.  The grids used are: a Cartesian
grid with 100$\times$200 zones (top-left and lower-right), a butterfly
mesh with 22,400 zones (top-right), and a spiderweb mesh with 8550
zones (lower-left).  Both the top-left and
bottom-right panels indicate that using the Cartesian mesh produces fairly smooth results, with some ($\sim$7\%)
asymmetry in the post-shock region.  Using the spiderweb mesh produces
perfectly symmetric solutions except near the center where the
deviation from symmetry is as large as $\sim$25\%.  Using the butterfly
mesh produces similar mixed accuracy in symmetry.
\label{nohplot_2d}}
\end{figure}

The four panels of Fig. \ref{nohplot_2d} present the density profiles
for 2D simulations.  Three of the four panels show the results using Cartesian
coordinates.  The top-left panel shows the results using a Cartesian
grid with 100$\times$200 zones, the top-right panel shows the results
using the butterfly mesh with 22,400 zones, and the lower-left panel shows the results
using the spiderweb mesh.  The fourth panel, lower-right, shows the
results using 2D cylindrical coordinates and a Cartesian mesh with
100$\times$200 zones.  Obvious is the fact that the mesh
employed has consequences for the solution.  Both the top-left and
bottom-right panels indicate that using the Cartesian mesh for this
problem produces fairly smooth results, with some
asymmetry ($\sim$7\%) in the post-shock region.  The lower-left panel
shows that using the spiderweb mesh produces perfectly symmetric
solutions except near the center where the deviation from symmetry is
as large as $\sim$25\%.  Finally, simulations using the butterfly
mesh, top-right panel, shows a similarly mixed capacity to preserve
symmetry.  In this light, it is important to choose a grid that
minimizes numerical artifacts for the problem at hand.  A task easily
accomplished with the use of ALE methods.

\subsection{Saltzman Piston Problem}
\label{section:saltzman}

\begin{figure}
\epsscale{0.8}
\plotone{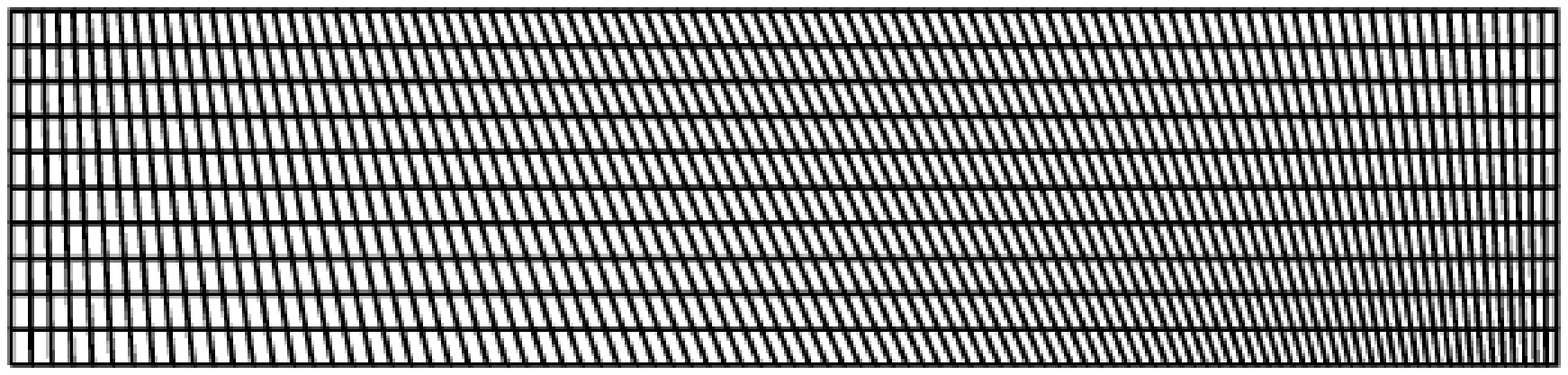} \\
\plotone{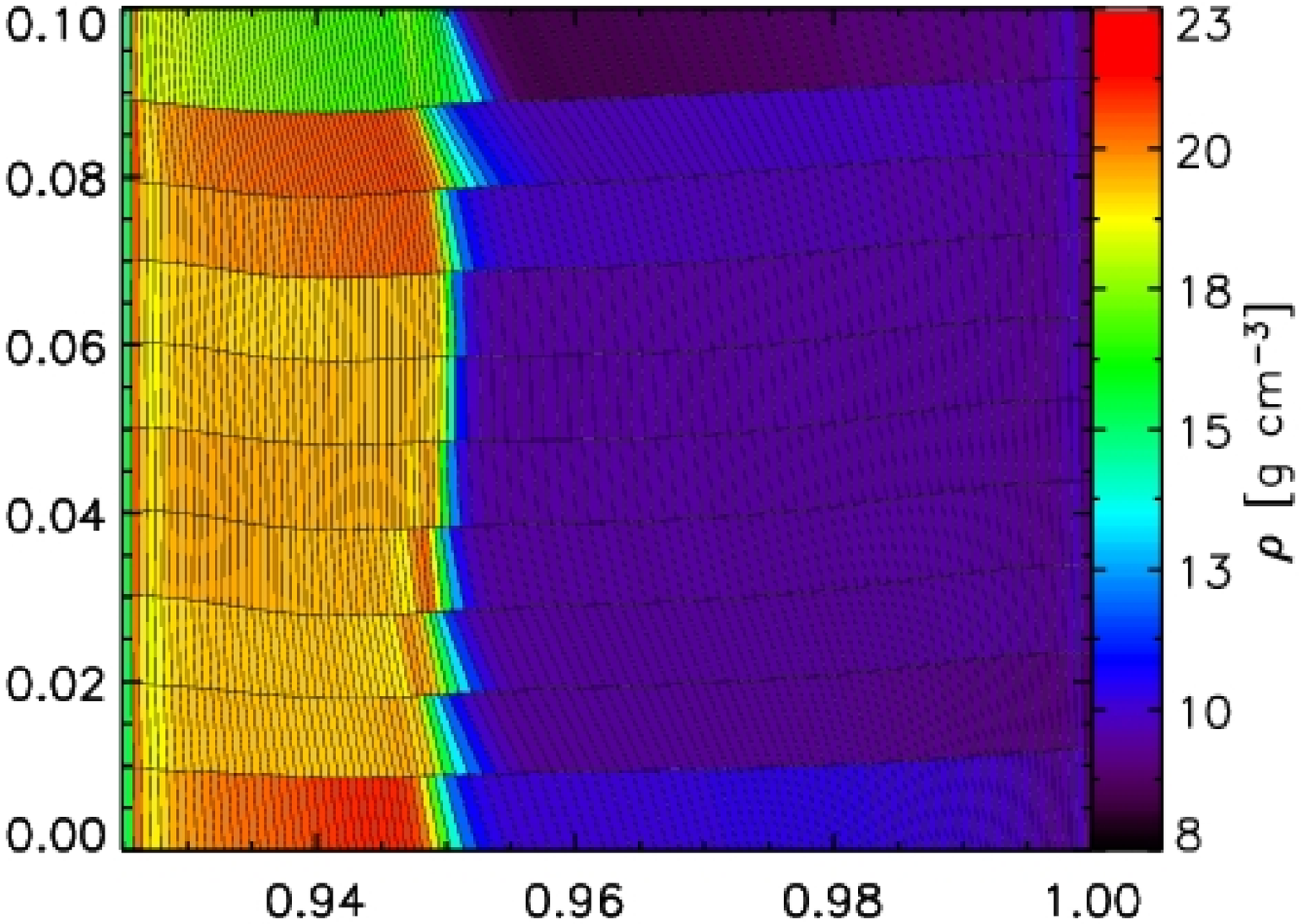}
\caption{
Saltzman piston problem. This problem tests the code's ability to
resolve shocks that are oblique to the orientation of the grid. The
top panel shows the initial  grid with $100 \times 10$ zones. The left
wall is a piston moving at a constant velocity, 1.0 cm s$^{-1}$, to
the right.  Initially, the density and internal energy are set equal
to 1.0 g cm$^{-3}$ and 0.0 ergs, respectively, and we use a gamma-law
EOS with $\gamma = 5/3$.  The grid and
the density colormap are shown in the bottom panel at $t = 0.925$
s. See \S \ref{section:saltzman} for a discussion of the results.
\label{saltzman}}
\end{figure}

A standard test for arbitrary grid codes, the Saltzman piston problem
\citep{margolin88} addresses the ability of a code to simulate a simple
piston-driven shock using a grid with mesh lines oblique to the shock normal.  The top panel of Fig. \ref{saltzman} shows the
grid with $100 \times 10$ zones.  For a grid with $N_x \times N_y$ nodes within a domain defined
by $x \in \{ 0 : x_{\rm max} \}$ and $y \in \{ 0 : y_{\rm max} \}$,
where $x_{\rm max} = 1.0$ and $x_{\rm max} = 0.1$, the $x$ positions of the
nodes are given by
\begin{equation}
  x = (i-1) \frac{x_{\rm max}}{N_x - 1}
      + (N_y-j) \sin \left ( \pi \frac{(i-1)}{(N_x-1)} \right)
     \frac{y_{\rm max}}{N_y - 1} \, ,
\end{equation}
where $i \in \{ 1 : N_x \}$ and $j \in \{ 1 : N_y \}$.
At the piston, the left wall is
  moving at a constant velocity, 1.0 cm s$^{-1}$, to the right.
  Initially, the density and internal energy are set equal to 1.0 g
  cm$^{-3}$ and 0.0 ergs, respectively, and we use a gamma-law EOS with $\gamma = 5/3$.

In the bottom panel of Fig. \ref{saltzman}, we show the grid and the
density colormap at $t = 0.925$ s after the shock has traversed
the domain twice, reflecting off the right and left walls once.
In this snapshot, the shock is traveling to the right.  Our results are
to be compared with Figs. 15 and 16 of \citet{campbell01}.  Figure 15 of
\citet{campbell01} depicts a simulation with severe grid buckling, which we
do not observe in our simulations.  Instead, our
Fig. \ref{saltzman} shows reduced grid buckling and appropriate
densities in accordance with the results of Fig. 16 of \citet{campbell01}.

\subsection{Dukowicz Piston Problem}
\label{section:dukowicz}

\begin{figure}
\epsscale{0.8}
\plotone{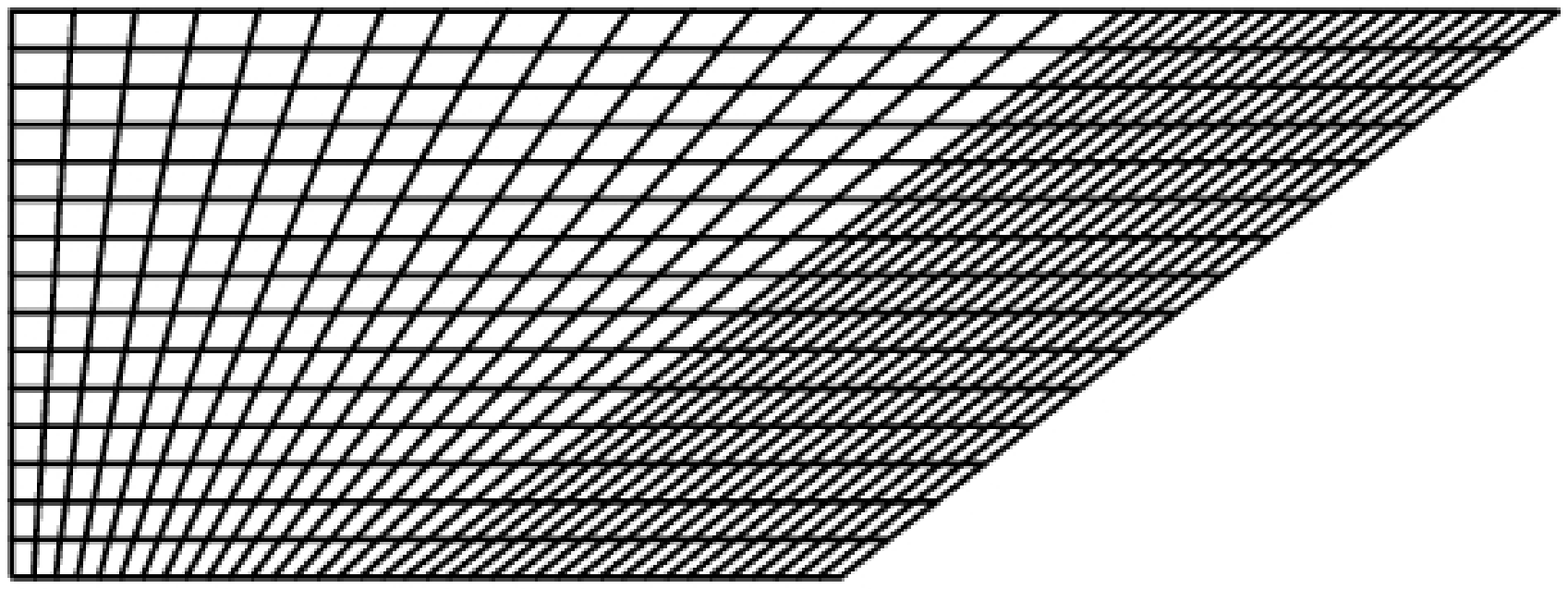}\\
\plotone{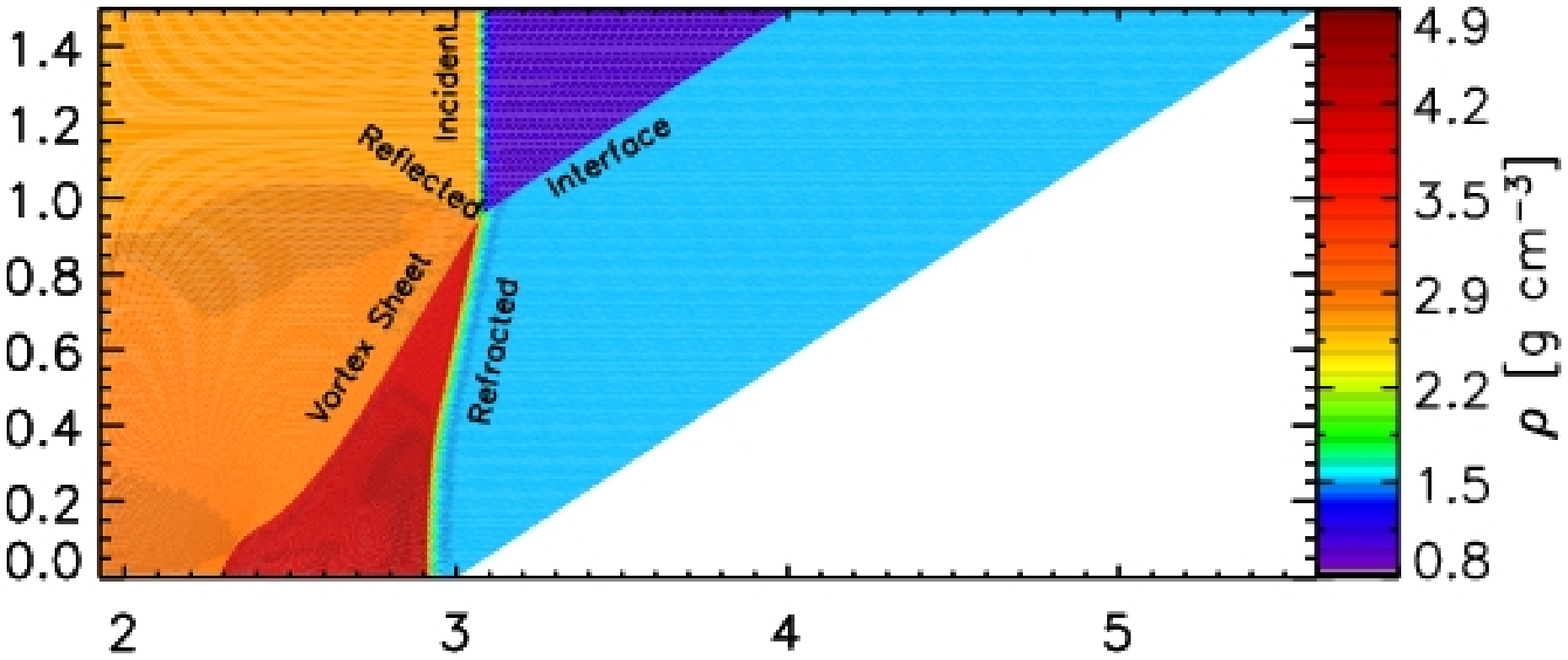}
\caption{
The Dukowicz piston problem, another test
using an oblique mesh.  The initial setup involves two
regions having an interface with a 60$^{\circ}$ orientation with the
vertical.  The top panel shows a low-resolution example of the grid used.  Region 1 has a density of 1 g cm$^{-3}$ and is resolved with
$144 \times 120$ zones, and region 2 has a
density of 1.5 g cm$^{-3}$ and is gridded with a $160 \times 120$ mesh,
with the vertical mesh lines uniformly slanted at 60$^{\circ}$.
Initially both regions are in equilibrium with $P = 1.0$ erg
cm$^{-3}$.  The left boundary is a piston with a velocity in the positive $x$
direction and a magnitude of 1.48 cm s$^{-1}$.
The piston-driven shock
  travels from left to right, and encounters the interface.
  The incident shock continues to the lower density region, a
  transmitted/refracted shock propagates into the higher density
  region, a vortex sheet develops behind the transmitted shock, and a
  reflected shock propagates into the incident shock's post-shock
  flow.   The orientations of the flow are reproduced accurately (see
  \S \ref{section:dukowicz} and Table \ref{table:dukowicz}).
\label{dukowicz}}
\end{figure}

The Dukowicz piston problem \citep{dukowicz92} is another test using an oblique mesh.  The initial setup involves two
regions.  Region 1 has a density of 1 g cm$^{-3}$ and is resolved with
$144 \times 120$ zones.  In the vertical domain, $y \in \{ 0 : 1.5 \}$ cm,
and the mesh lines evenly partition the space into 120 sections.  Dividing
region 1 horizontally involves mesh lines with changing orientation.
The leftmost mesh line is parallel to the vertical, while the rightmost mesh line is
oriented 60$^{\circ}$ relative to the vertical.  The mesh lines in
between smoothly transition from 0$^{\circ}$ to 60$^{\circ}$.  Region 2 has a
density of 1.5 g cm$^{-3}$ and is gridded with a $160 \times 120$ mesh,
with the vertical mesh lines uniformly slanted at 60$^{\circ}$.
Including region 1 and region 2, the bottom boundary spans the range
$x \in \{ 0: 3 \}$ cm and is evenly divided into 304 segments.
Initially, both regions are in equilibrium with $P = 1.0$ erg
cm$^{-3}$.  The top, bottom, and right boundaries are reflecting, and
the left boundary is a piston with a velocity in the positive $x$
direction and a magnitude of 1.48 cm s$^{-1}$ (see the top panel of Fig. \ref{dukowicz} for a low
resolution example of this grid).

A piston-driven shock
  travels from left to right, and encounters the density jump at an
  angle of 60$^{\circ}$, producing a rich set of phenomena.
  The incident shock continues into the lower density region, a
  transmitted/refracted shock propagates into the higher density
  region, a vortex sheet develops behind the transmitted shock, and a
  reflected shock propagates into the incident shock's post-shock flow
  (see the labels in the bottom plot of Fig. \ref{dukowicz} for visual
  reference).  The results are to be compared with the semi-analytic
  shock-polar analysis presented by \citet{dukowicz92} (see Figure 13
  and Table I of \citet{dukowicz92}).  In their paper, angles subtended by the five regions
  are presented.  In Table \ref{table:dukowicz}, we recast this information as
  the angles that the transmitted shock, vortex sheet, reflected
  shock, and incident shock have with $x$-axis.  The first row gives
  the analytic values, which should be compared with the simulated orientations in
  the second row.  Strikingly, despite the fact that the incident
  shock has traversed a grid with an oblique mesh, the simulated and
  analytic orientations differ very little.  The simulated and
  analytic reflected-shock orientations agree to within $\sim 0.5$\%,
  and the orientations of the transmitted shock and vortex sheet agree
  to within 2\%.

\subsection{Rayleigh-Taylor Instability}
\label{section:rt}

Here, we simulate the growth of a single mode of the Rayleigh-Taylor Instability.
A heavy fluid is placed on top
of a lighter fluid in the presence of a constant gravitational
acceleration.  The top and bottom densities are
$\rho_1 = 2.0$ g cm$^{-3}$ and $\rho_2 = 1.0$ g cm$^{-3}$, respectively, and the gravitational
acceleration points downward with magnitude $g = 0.1$ cm s$^{-2}$.
The domain is a rectangle with $x \in \{ -0.25 : 0.24 \}$ cm and $y \in \{ -0.75
: 0.75 \}$ cm and $100 \times 300$ grid zones.
The top and bottom boundaries are reflecting while the left and right
boundaries are periodic.  For such configurations, small perturbations
of the interface between the heavy and light fluids are unstable to exponential
growth.  Assuming that the boundaries are far from the
interface, the exponential growth rate of a perturbation with
wavenumber $k$ is
\begin{equation}
\omega = \sqrt{\frac{k g (\rho_1 - \rho_2)}{\rho_1 + \rho_2}} \, .
\end{equation}
For all single-mode Rayleigh-Taylor simulations we initiate the
perturbation by setting the vertical component of the velocity equal to $v_y
= 2.5 \times 10^{-3} (1 + \cos(2 \pi x / \lambda)) (1 + \cos(3 \pi
y))$, where the wavelength, $\lambda$, is $0.5$ cm.
Therefore, we simulate one wavelength of the mode, and the exponential
growth rate ($\omega$) should be  0.65 s$^{-1}$. 

\begin{figure}
\epsscale{1.0}
\plotone{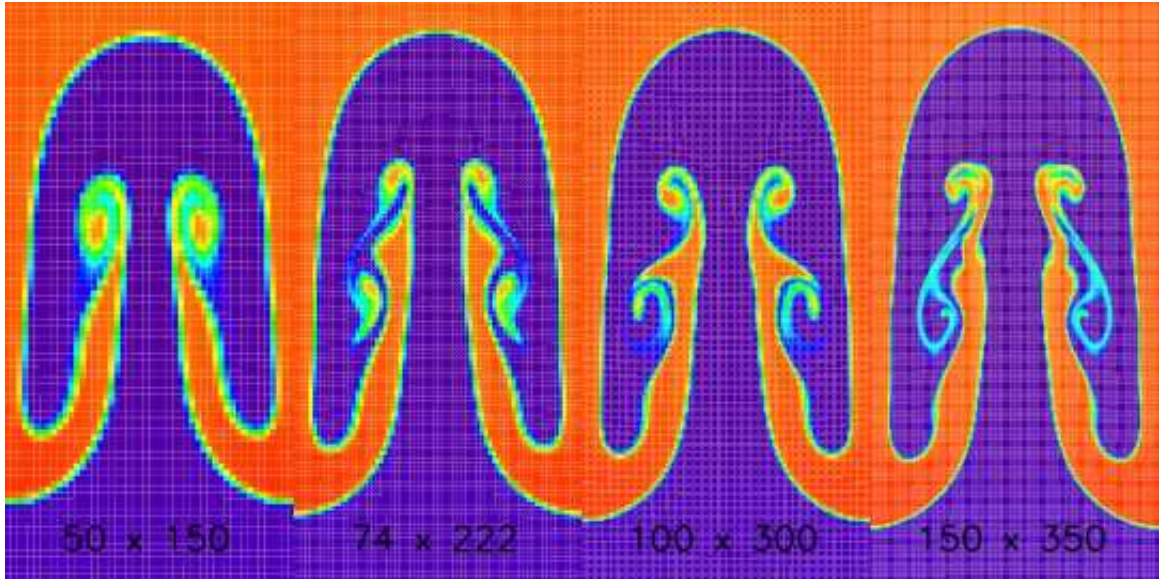}
\caption{
Nonlinear phase of a single mode of the Rayleigh-Taylor instability at
$t = 12.75$ s.
The top and bottom densities are
$\rho_1 = 2.0$ g cm$^{-3}$ and $\rho_2 = 1.0$ g cm$^{-3}$, respectively, and the gravitational
acceleration points downward with magnitude $g = 0.1$ cm s$^{-2}$.
The top and bottom boundaries are reflecting while the left and right
boundaries are periodic.  From left to right, grid sizes are $50 \times 150$, $74 \times
222$, $100 \times 300$ (this is the resolution used in Fig. \ref{rt_stills_fmerit}), and $150 \times 450$.  See \S \ref{section:rt} for discussions
comparing these results with those in other works.
\label{rt_stills_res}}
\end{figure}

Figure \ref{rt_stills_res} shows the evolution of a single-mode
Rayleigh-Taylor instability  at $t = 12.75$ s for four resolutions.
From left to right, the grid sizes are $50 \times 150$, $74 \times
222$, $100 \times 300$, and $150 \times 450$.  Gross features
compare well, with all the resolutions differing by only a few
percent in the maximum and minimum position of the interface.
However, higher resolution simulations manifest greater complexity for the Kelvin-Helmholtz rolls.
One can compare the third panel to the results of Fig. 4.5 of
  \citet{liska03}.  In a general sense, these authors conclude that
  fewer features imply more dissipation.  However, it could be that
  some of the Kelvin-Helmholtz rolls are seeded by grid noise in some schemes\footnote{See results from Jim Stone's Athena for a more
  favorable comparison and a discussion of the grid noise issue
  (http://www.astro.princeton.edu/~jstone/tests/rt/rt.html).}.

\begin{figure}
\epsscale{.80}
\plotone{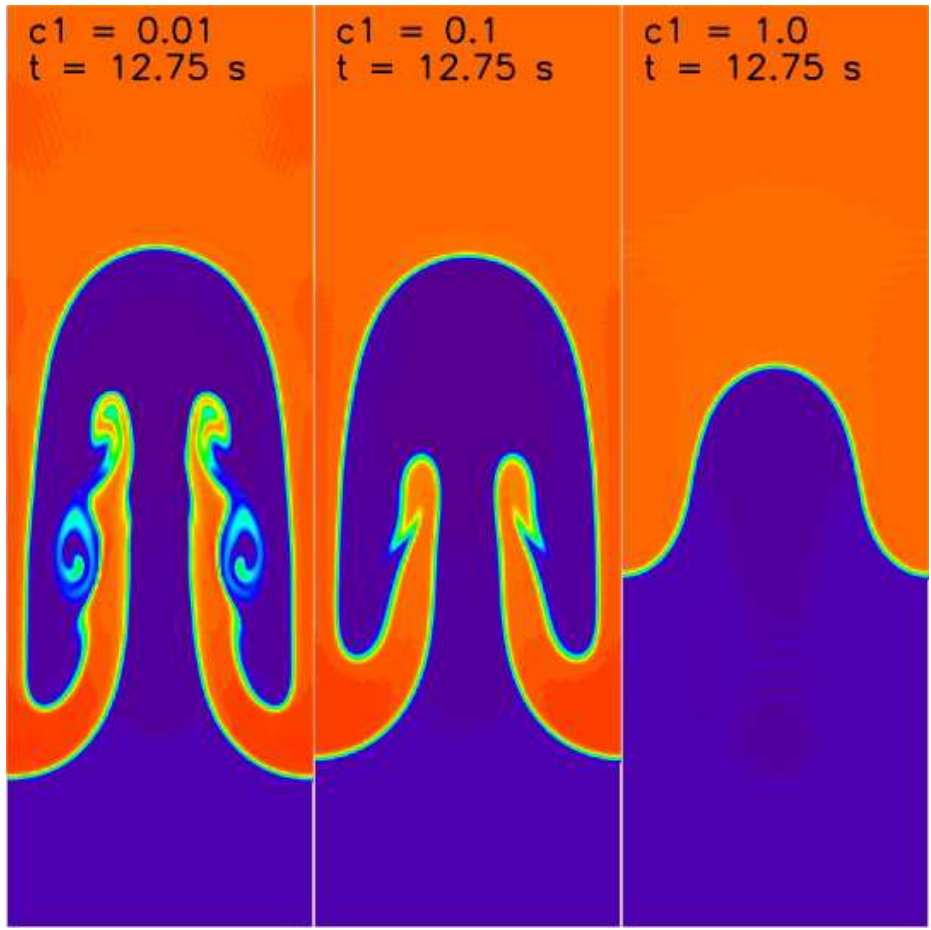}
\caption{Single-mode Rayleigh-Taylor instability: investigating the
  artificial viscosity parameter.  The same resolution, setup, and
  time are shown as presented in Fig. \ref{rt_stills_fmerit} and the third panel
  of Fig. \ref{rt_stills_res}.  Each panel shows results for
  different values of the
  artificial viscosity parameter, $c_1$, that is the coefficient of $c_s (\vec{\nabla}
  \cdot \vec{v})$ (see \S \ref{section:artificial_viscosity}).
  The artificial viscosity parameters are $c_1 = 0.01$ (left panel), $c_1 =
  0.1$ (center panel) and, $c_1 = 1.0$ (right panel).  From this analysis, it would seem
  that one should choose the lower values of $c_1$ to accurately
  represent low Mach number Rayleigh-Taylor flows.  However,
  higher values help to eliminate unwanted ringing in post
  shock flows.
\label{rt_stills_visc}}
\end{figure}

Next, we discuss the effects of artificial viscosity on the
single-mode Rayleigh-Taylor flow.  As stated in \S
\ref{section:artificial_viscosity}, artificial viscosity is a
requirement for ALE algorithms in order to simulate shocks.  In
the current formulation, there are two parameters of the artificial
viscosity scheme.  One parameter, $c2$, is the coefficient for
$(\vec{\nabla} \cdot \vec{v})^2$, which is largest in
shocks.  Effectively, this term provides shock resolution.  The second
parameter, $c_1$, is the coefficient of the term that
is proportional to $c_s (\vec{\nabla} \cdot \vec{v})$ and is designed
to reduce the amount of post-shock ringing in the solution.  Typical
values suggested for both are 1.0 \citep{campbell01}.  For the
single-mode Rayleigh-Taylor test, the
flows are subsonic, so the term multiplied by $c_1$ has the
greatest impact on the magnitude of the artificial viscosity forces.
Each panel of Fig. \ref{rt_stills_visc} presents the nonlinear
Rayleigh-Taylor flow after 12.75 s for different values of the
artificial viscosity parameter, $c_1$.  Clearly, $c_1 = 0.01$ (left
panel) reproduces the expected results for this low-Mach-number flow.  On the
other hand, the $c_1 = 0.1$ run (center panel) displays significant departures for
the Kelvin-Helmholtz rolls, while the overall progression of the
plumes remains similar.  Unfortunately, the model with $c_1 = 1.0$ (right panel)
completely suppresses the Kelvin-Helmholtz rolls on these scales and
plume progression is severely retarded.  

\begin{figure}
\epsscale{.80}
\plotone{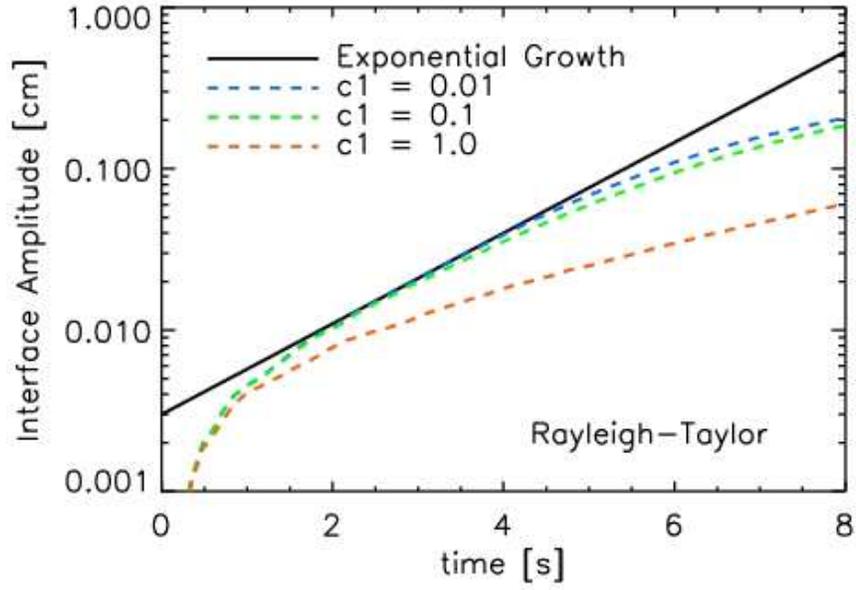}
\caption{
The interface
amplitude vs. time for the single-mode Rayleigh-Taylor instability
test.  Resolutions and setup are similar to those presented in Fig. \ref{rt_stills_visc}.
We compare the analytic exponential growth rate (solid line) with
simulation results (dashed lines) for viscosity
  parameters, $c_1$, of 0.01, 0.1, and 1.0.  Simulations with $c_1 = 0.01$ and 0.1 manifest exponential growth for several
  e-folding times, while the run with $c_1 = 1.0$ seems
  to follow the linear phase for  only 1 s ($\sim$ 1/2 e-folding).
\label{rt_linear}}
\end{figure}

In Fig. \ref{rt_linear}, the interface perturbation
amplitude is plotted as a function of time (dashed lines).
The
  solid line is the analytic exponential growth rate scaled to the simulation
  results.  The dashed lines show the simulation results for viscosity
  parameters, $c_1$, of 0.01, 0.1, and 1.0.  Three distinct phases are apparent: an early transient phase, a phase in which
the slope most closely matches the exponential growth rate, and the
subsequent nonlinear phase.  The
  simulations with $c_1 = 0.01$ and 0.1 manifest exponential growth for several
  e-folding times.  The run with $c_1 = 1.0$, on the other hand, seems
  to follow the linear phase for  only 1 s ($\sim$ 1/2 e-folding), if at
  all.  Around 5 s, the evolution of the interface amplitude enters the
  nonlinear phase.

For the artificial viscosity scheme that we employ, the Rayleigh-Taylor
instability test indicates that lower values of $c_1$ are preferred.
However, tests of the Sod shock-tube problem including a parameter
study of $c_1$ indicate that unwanted post-shock ringing is diminished only for
values above $\sim0.5$.  This represents the primary weakness of
BETHE-hydro to simulate flows with both shocks and Rayleigh-Taylor
instabilities.  To be clear, this does not represent a weakness of ALE
methods in general, but of the tensor artificial viscosity
algorithm designed to mitigate the hourglass instability that we
employ.  Other artificial viscosity schemes such as edge viscosity
(see \citet{campbell01} for references) allow for proper development
of hydrodynamic instabilities, but do little to suppress the hourglass
instability (Milan Kucharik, private communication).
Of great interest to users of ALE is a methodology that suppresses the
hourglass instability and post-shock ringing while enabling proper
evolution of hydrodynamic instabilities.

\subsection{Kelvin-Helmholtz Instability}

Another important phenomenon we explore is the Kelvin-Helmholtz shear
instability.  \citet{agertz07} have shown that SPH has trouble
simulating the Kelvin-Helmholtz instability when extreme density
contrasts are involved.  We find that this instability is reasonably
well handled in BETHE-hydro, and that the evolution during the small amplitude regime
is accurately characterized by analytic linear analysis.  The
calculational domain covers the square region bounded by
$x \in \{ 0 : 1 \}$ cm and $y \in \{ 0 : 1 \}$ cm and has $256 \times 256$ zones.
The top and bottom boundaries are reflecting, while the left and right
boundaries are periodic.
For $y<0.5$ cm, $\rho_{\rm b} = 1.0$ g cm$^{-3}$, and for $y>0.5$ cm, $\rho_{\rm t}
= \rho_{\rm b}/\chi$, where 
$\chi = 8$.  These regions are in pressure equilibrium with $P =
1.0$ erg cm$^{-3}$.  The shearing velocity, $v_{\rm shear}$ is scaled with respect
to the slowest sound speed (sound speed in the bottom region, $c_b$).
This relative shearing velocity is split between the top region,
which flows to the left with speed $\frac{1}{2} v_{\rm shear}$, and the
bottom region, which flows to the right with the same speed.  For this
test, the linear coefficient in the viscosity, $c_1$, is set to 0.01.  A gamma-law
EOS is used with $\gamma = 5/3$.
The initial
perturbation (see Agertz et al. 2007 \nocite{agertz07}) is placed in a small band
centered on the interface and is a perturbation in velocity given by 
\begin{equation}
\label{eq:kh_pert}
v_y = \delta v_y v_{\rm shear}  \sin \left ( \frac{2 \pi x}{\lambda} \right ) \, {\rm for}
\, \left | y - 0.5 \right | \le 0.025 \, ,
\end{equation}
where $\lambda = 1/3$ cm.  It should be noted that this perturbation
is not an eigenmode of the instability.  Therefore, simulations have a transient phase at the
beginning in which this perturbation settles into one or more modes of the
instability.  We have found that using amplitudes suggested by
\citet{agertz07}, $\delta v_y = 1/80$ and 1/40, produces strong transients that complicate the
interpretation of the linear regime.  To avoid strong long-lasting
transients, we set $\delta v_y = 1/160$.
These initial conditions should produce a perturbation of the interface that
grows exponentially in magnitude with an e-folding time given by
\begin{equation}
\tau_{\rm KH} = \frac{\lambda (\rho_{\rm top} + \rho_{\rm bot})}
{2 \pi v_{\rm shear} \sqrt{\rho_{\rm top} \rho_{\rm bot}}  } \, .
\end{equation}
For the simulations presented here, we consider two shearing
velocities: $v_{\rm shear} = \frac{1}{2} c_b = 0.6455$ cm s$^{-1}$ and $v_{\rm shear} =
\frac{1}{4} c_b = 0.32275$ cm s$^{-1}$.  Hence, $\tau_{\rm KH}$ is
0.262 s and 0.523 s, respectively.

\begin{figure}
\epsscale{.70}
\plotone{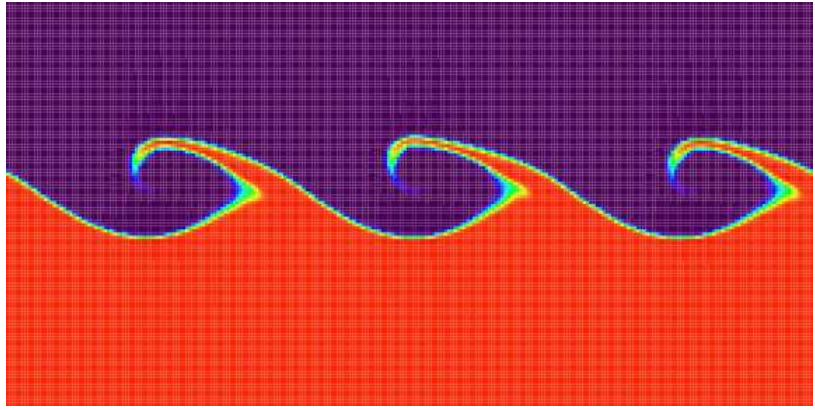}\\
\plotone{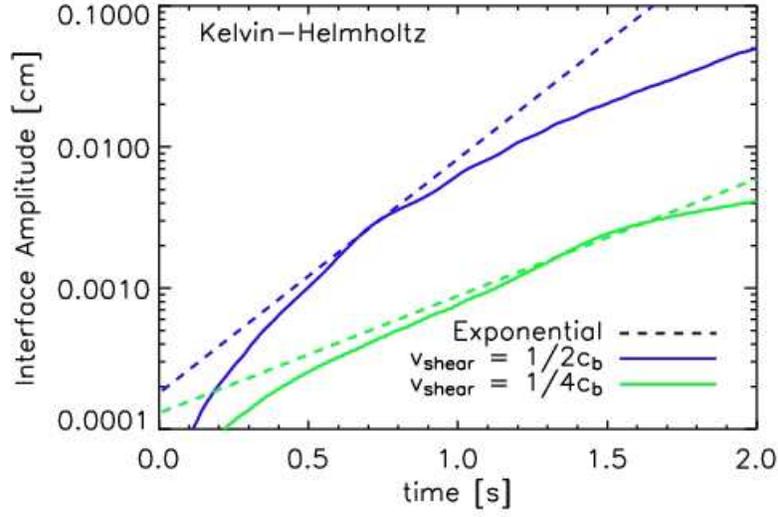}
\caption{
Kelvin-Helmholtz shear instability.  The domain is resolved with $256
\times 256$ zones.  The bottom region has $\rho_{\rm b} = 1.0$, and
the top region has $\rho_{\rm t} = \rho_{\rm b}/\chi$, where $\chi =
8$, and both regions have $P = 1.0$ erg cm$^{-3}$.  A gamma-law EOS is
used with $\gamma = 5/3$.  The top panel shows the development of Kelvin-Helmholtz rolls at
$t= 5.5$ s for $v_{\rm shear} = \frac{1}{4} c_b$ and $\tau_{\rm KH} $
0.523 s.
The bottom plot of Fig. \ref{kh_still} shows the interface amplitude
(solid line) vs. time and compares to the expected exponential growth
(dashed line) for a simulation with $v_{\rm shear} = \frac{1}{4}c_b$
(green) and $v_{\rm shear} = \frac{1}{2}c_b$ (blue).  There are three
distinct phases in the log-linear plot: an early transient phase, a
phase in which the slope most closely matches the exponential growth
rate, and the subsequent nonlinear phase.
\label{kh_still}}
\end{figure}

The top panel of Fig. \ref{kh_still} shows the evolution at
$t= 5.5$ s for $v_{\rm shear} = \frac{1}{4} c_b$.  The first set of
nonlinear Kelvin-Helmholtz rolls have appeared.  Qualitatively, the
morphology of the rolls is
similar to the results presented in Figure 13. of
\citet{agertz07} labeled by ``grid 1'' at $t = 2 \pi \tau_{\rm KH}$ (there is a factor of $2 \pi$ difference in the definition of
$\tau_{\rm KH}$ between our work and theirs).  There are two main
differences.  For one, we simulate with a factor two larger wavelength
to adequately resolve
the wavelength for linear analysis.  Secondly, the time at which we
present the results in the top panel of Fig. \ref{kh_still}
corresponds to  $t = 1.67 \times (2 \pi \tau_{\rm KH})$, not $t = 2
\pi \tau_{\rm KH}$.  We have noticed similar discrepancies in the time
required to achieve similar evolution using $\lambda = 1/6$ and $v_{\rm shear} = \frac{1}{2}c_b$.  Additionally, we analyze the
growth rate of the interface during the linear regime.  We determine the
interface amplitude by generating a
contour for $\rho = 0.5 (\rho_{\rm t} + \rho_{\rm b})$, measuring the
peak to peak amplitude, and dividing by two.  The bottom
plot of Fig.
\ref{kh_still} shows this amplitude (solid line) versus time and
compares to the expected exponential growth (dashed line) for
a simulation with $v_{\rm shear} = \frac{1}{4}c_b$ (green) and $v_{\rm
shear} = \frac{1}{2}c_b$ (blue).  There are three distinct phases
in the log-linear plot: an early transient phase, a phase in which
the slope most closely matches the exponential growth rate, and the
subsequent nonlinear phase.  While the growth rate during the linear
regime differs slightly ($\sim$10\%) from theory, the bottom panel of
Fig. \ref{kh_still} demonstrates that the correct
dependence of the growth rate on $v_{\rm shear}$ is obtained.

\subsection{Core-Collapse Test}
\label{section:corecollapse}
\begin{figure}
\plotone{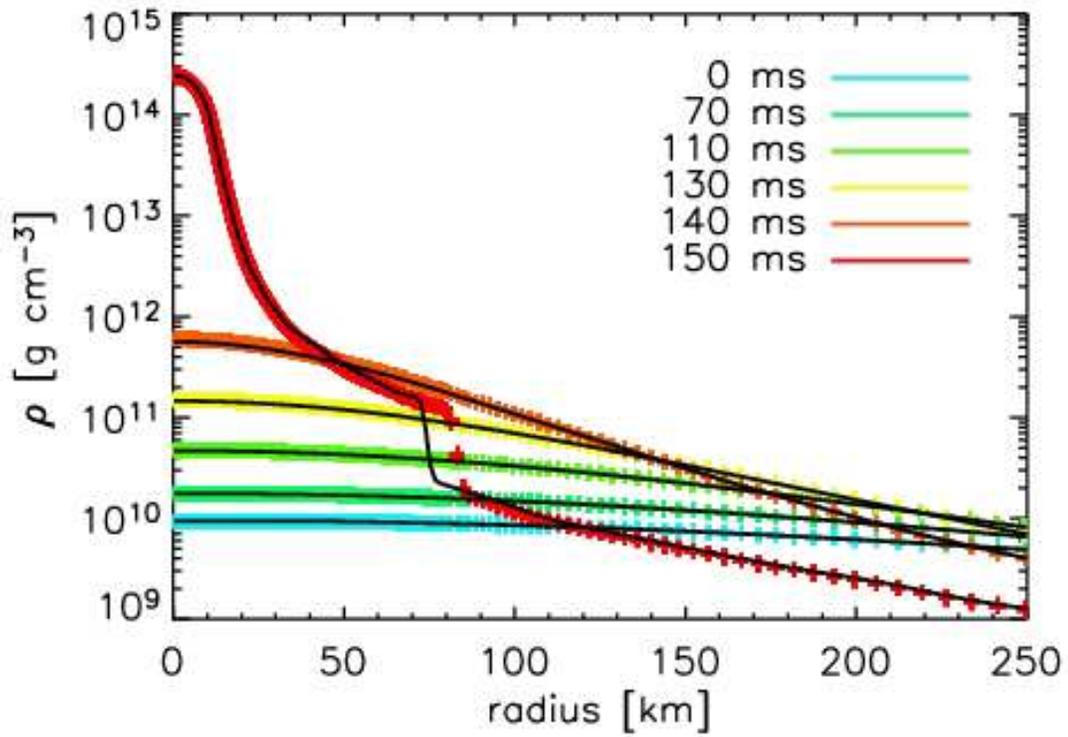}
\caption{Density vs. radius for the core collapse of a 15-M$_{\sun}$ star.  The model is the s15
  model of Weaver \& Woosley.  Density vs. radius for 1D (lines) and
  2D (crosses) at times 0, 70, 110, 130, 140, and 150 ms after the
  start of the calculation.  Core bounce occurs at 148 ms.  The 2D
  grid is composed of a butterfly mesh in the interior with a minimum
  cell size of $\sim$0.5 km and extends to 50 km.  A spherical grid
  extends the domain out to 4000 km.  There are 23,750 cells in total,
  with an effective resolution of $\sim$250 radial and $\sim$100 angular
  zones.  The 1D grid has
  250 zones, with similar resolution to the 2D run at all radii.  For the
  most part, the 1D and 2D calculations track one another quite well.
  There is roughly a $\sim$10\% difference in the shock radii at 150 ms.
\label{cc_rhoplot}}
\end{figure}

\begin{figure}
\plotone{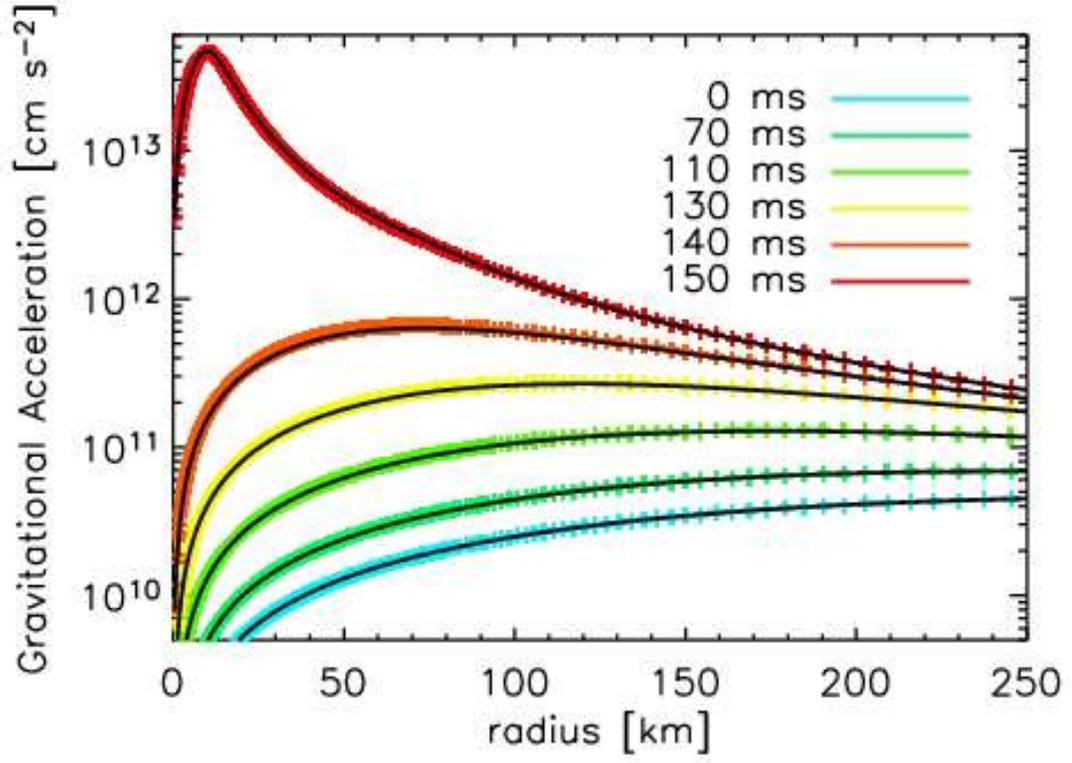}
\caption{Radial component of the gravitational acceleration during the
  core collapse of a 15-M$_{\sun}$ star.  Similar to
  Fig. \ref{cc_rhoplot}.  Note the good match between the
  1D and 2D results.
\label{cc_gaccplot}}
\end{figure}

Incorporating all aspects of BETHE-hydro, we simulate 1D
and 2D hydrodynamic core-collapse of a 15
M$_{\sun}$ star.  We use the s15s7b2 model (S15) of \citet{woosley95} and the Shen EOS \citep{shen98} and initiate
collapse using a $Y_e$-$\rho$ parameterization \citep{liebendorfer06}.  While there are no analytic solutions for such a
test, we test to see whether the density profiles, timescales, and shock
radii, etc. all match experience with other codes and results
published \citep{liebendorfer01a,liebendorfer01b,rampp02,buras03,thompson03,liebendorfer05}.  For 2D simulations,
  the grid is composed of a butterfly mesh in the interior with a minimum
  cell size of $\sim$0.5 km and extends to 50 km where a spherical grid
  carries the domain out to 4000 km.  In total, there are 23,750 cells,
  with an effective resolution of $\sim$250 radial and $\sim$100 angular
  zones.  For the best comparison, the 1D grid has
  250 zones, mimicking the effective radial resolution of the 2D simulation.  Figure \ref{cc_rhoplot} depicts the density
vs. radius for 1D (lines) and
  2D (crosses) at times 0, 70, 110, 130, 140, and 150 ms after the
  start of the calculation.  Core bounce occurs at 148 ms.    Other
  than a $\sim$10\% difference in the shock radii at 150 ms, 1D and 2D
  calculations track one another quite well.  Furthermore,
  Fig. \ref{cc_gaccplot} shows an exceptional correspondence between the
  1D and 2D gravitational accelerations.

Using this core-collapse test to best represent the conditions during
astrophysical
simulations, we describe some timing results of this problem.  A standard measure is the average CPU time spent per cell per
cycle.  Using one core of a Dual-Core AMD Opteron\texttrademark 2.8 GHz
processor for this test, the average time spent is $8.125 \times
10^{-5}$ CPU seconds/cycle/cell.  80\% of which is spent in AMG1R6,
the multigrid solver.  For this problem, the iterative multigrid
solver usually takes $\sim$5 to $\sim$10 cycles to achieve a
fractional residual of $\sim$$10^{-7}$ to $\sim$$10^{-9}$, where the
fractional residual of the linear system $A\vec{x} = \vec{b}$ is $(A\vec{x}-\vec{b})/\vec{b}$.

\begin{figure}
\plotone{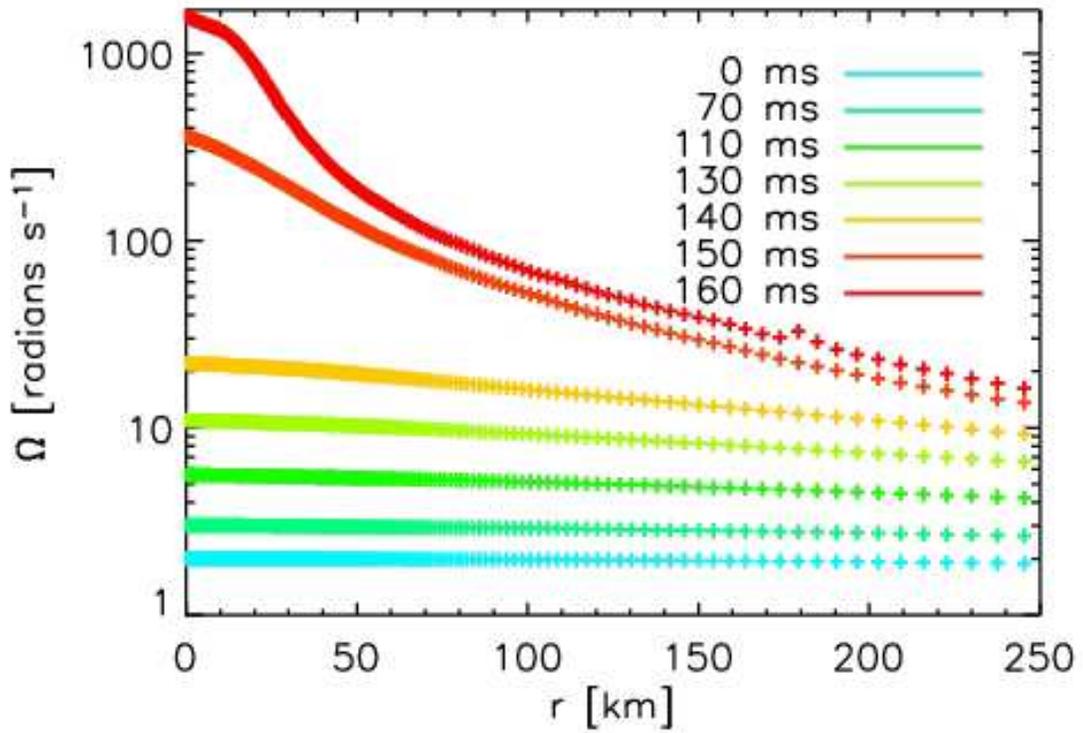}
\caption{
Angular velocity evolution during the core collapse of a rotating 15-M$_{\sun}$ star.  This test problem was calculated on the same grid
  as in Fig. \ref{cc_rhoplot}. The initial angular velocity profile is
  constant on cylinders and is given by
  $\Omega(r) = 1/(1 + (r/A)^2)$, where $A = 1000$ km and $\Omega_0 =
  2$ radians s$^{-1}$.  As expected, the central
  angular velocity, $\Omega_c$ is proportional to $\rho_c^{2/3}$, where $\rho_c$ is
  the central density.  The central density compresses from
  $\sim$10$^{10}$ g cm$^{-3}$ at $t = 0$ ms to $\sim$2.2 $\times 10^{14}$ g
  cm$^{-3}$ at $t = 160$ ms.  Therefore, at $t = 160$ ms, $\Omega_c$ should be
  $\sim$1600 radians s$^{-1}$, consistent with results shown
  in this figure.  Other than a slight,
  but noticeable, glitch at the location of the shock ($\sim$180 km), the angular velocity
  evolves smoothly with no evidence of axis effects.
\label{cc_angvelplot}}
\end{figure}

We re-simulate the
collapse of the S15 model, but with an angular velocity profile of the form
  $\Omega(r) = 1/(1 + (r/A)^2)$, where $A = 1000$ km and $\Omega_0 =
  2$ radians s$^{-1}$.  The $\Omega$ vs. $r$ plot is shown in Fig. \ref{cc_angvelplot}.  
Core
  bounce occurs at 151 ms.  First of all, total
  angular momentum is conserved to machine accuracy.  Since there are no analytic descriptions for
  the evolution of angular velocity in core collapse scenarios, we
  cannot validate the numerical results via analytic theory.  However, the central
  angular velocity, $\Omega_c$ is proportional to $\rho_c^{2/3}$, where $\rho_c$ is
  the central density.  Since the central density compresses from
  $\sim$10$^{10}$ g cm$^{-3}$ at $t = 0$ ms to $\sim$2.2 $\times 10^{14}$ g
  cm$^{-3}$ at $t = 160$ ms, $\Omega_c$ should be
  $\sim$1600 radians s$^{-1}$ at $t = 160$ ms.  This is consistent with results shown
  in Fig. \ref{cc_angvelplot}.  Furthermore, the angular velocity
  evolves smoothly with no evidence of axis effects.  On the other hand, there is a slight,
  but noticeable, glitch at the location of the shock.

\section{Discussion and Conclusions}
\label{section:discussion}

In this paper, we have presented the algorithms employed in BETHE-hydro, a new code
for 1D and 2D astrophysical hydrodynamic simulations.  The
hydrodynamic core is an ALE algorithm, and its most
striking feature is the ability to use
arbitrary, unstructured grids.
With finite-differencing based upon the support-operator method
\citep{shashkov95} of \citet{caramana98c} and \citet{caramana98a},
energy is conserved to roundoff error in the absence of rotation and
gravity, and momentum is strictly conserved using Cartesian
coordinates.  For all other circumstances, energy and momentum are
conserved accurately, if not precisely.  We use a subcell remapping scheme that
conservatively remaps mass, momentum, energy, and number density.  For
2D calculations using cylindrical
coordinates, we include rotational terms in the Lagrangian solver, and
develop a remapping algorithm that conservatively remaps angular
momentum while minimizing unwanted features in the angular velocity near
the axis.  To provide shock resolution and grid stability, we use the
tensor artificial viscosity of \citet{campbell01}, and to minimize
hourglass instabilities, we have developed a subcell pressure method
that is a close derivative of the scheme developed by
\citet{caramana98a}, but avoids pathological problems when used
in conjunction with the subcell remapping algorithm.  Finally, we
have developed a gravity solver for arbitrary grids that uses a
support-operator technique for elliptic equations \citep{morel98} and
an iterative multigrid-preconditioned
conjugate-gradient method \citep{ruge87} to solve the system of linear
equations.

Overall, BETHE-hydro offers many unique and useful features for
astrophysical simulations.  For one, by using ALE techniques, the
structure of BETHE-hydro is straightforward, enabling
simple inclusion of a variety of additional physics packages.
Examples, which will be discussed in future papers, are nuclear
networks and time-dependent radiation transport.  In contrast with
other techniques, such as higher-order Godunov
methods, no assumptions are made in ALE techniques about characteristic
waves nor the relationships among important thermodynamic variables.  Hence, one of its useful features is an ability to use a general EOS.

Most important among BETHE-hydro's strengths is the ability to
solve self-gravitating hydrodynamic flows on arbitrary
grids.  This is achieved primarily because the foundation of the Lagrangian
hydrodynamics solver is an arbitrary, unstructured polygonal
grid. Furthermore, grids may be time-dependent since the hydrodynamic
flow from one timestep is remapped to another arbitrary grid for subsequent
evolution.  Consequently, simulations may be executed using a purely
Lagrangian, purely Eulerian, or an arbitrarily defined time-dependent
grid.  With BETHE-hydro, simulations have great flexibility, tailoring
the grid to minimize numerical error and suiting the grid to the
computational challenge.

Ironically, this flexibility leads to BETHE-hydro's most prominent
weakness, which is shared among all ALE codes, the hourglass
instability and grid buckling.  We use subcell pressure and tensor artificial viscosity algorithms to mitigate these
numerical artifacts, but the resolution is imperfect.  In \S
\ref{section:hourglass}, we present a subcell pressure algorithm that
is compatible with the subcell remapping algorithm, but the efficacy
of hourglass elimination is slightly compromised.  Although the tensor
artificial viscosity does well to mitigate grid buckling, it can be
more resistive to the proper development of hydrodynamic instabilities
compared to other artificial viscosity schemes used in
ALE (Milan Kucharik, private communication).  Hence, we
diminish the effects of the hourglass instability, but with some unwanted
side effects.  While mitigation in 2D is tractable, the
instability in 3D has many more modes and is not as easily eliminated
(Guglielmo Scovazzi, private communication ),
making long 3D ALE simulations a challenge.

Throughout this paper, we have not only demonstrated BETHE-hydro's
flexibility, but have shown that it produces accurate and 2$^{nd}$-order
convergent solutions.  With density distributions having analytic
potentials, we have shown that the 2D gravity solver gives accurate
spherically symmetric potentials when a non-spherical grid is used, it
produces accurate non-spherical potentials, and solutions
converge with 2$^{nd}$-order accuracy.  Further tests demonstrated
accurate solutions for self-gravitating hydrostatic and dynamic
problems.  We have shown that our hourglass elimination algorithm
minimizes the hourglass instability and does not present problems when
used in conjunction with subcell remapping.  In addition, we have
quantified the accuracy of hydrodynamic simulations by simulating
problems with analytic solutions.  Simulating piston-driven shocks
using oblique meshes, we confirmed we can obtain accurate solutions in the context
of arbitrary grids.  Verifying the code's ability to capture basic hydrodynamic
instabilities, we simulate the Rayleigh-Taylor and Kelvin-Helmholtz
instabilities.  Concluding the tests, we simulated a supernova core collapse, which demonstrates the ability to simulate complex
astrophysical phenomena.

Simulating hydrodynamic flow is fundamental to understanding most
astrophysical objects, and despite the long tradition of
hydrodynamic simulations many puzzles remain.  This is due primarily to the need
to address time-dependent gravitational potentials, complicated
equations of state (EOSs), flexible grids, multi-D shock structures,
and chaotic and turbulent flows.  Therefore, with BETHE-hydro, we introduce
a uniquely flexible and functional tool for advancing the theory of complex
astrophysical phenomena.

\acknowledgments
We would like to thank Mikhail Shashkov, Raphael Loub{\`e}re, Milan
Kucharik, and Burton Wendroff for valuable conversations about ALE
techniques.  Discussions concerning the 2D gravity solver with Ivan Hubeny, Jim
Morel, David Moulton, and David Keyes of TOPS were very 
fruitful.
 We acknowledge support for this work                                           
from the Scientific Discovery through Advanced Computing                        
(SciDAC) program of the DOE, under grant numbers DE-FC02-01ER41184              
and DE-FC02-06ER41452, and from the NSF under grant number AST-0504947.         
J.W.M. thanks the Joint Institute for Nuclear Astrophysics (JINA) for           
support under NSF grant PHY0216783.
We thank Jeff Fookson and Neal Lauber of the Steward Computer Support
Group for their invaluable help with the local Beowulf cluster, Grendel.

\appendix

\section{Hydrodynamic equations:  Eulerian form}
\label{section:eulerian_hydro}
In Eulerian form, the equations for conservation of mass, momentum,
and energy are
\begin{equation}
\label{eq:mass_eul}
\frac{\partial \rho}{\partial t} + \vec{\nabla} \cdot ( \rho \vec{v})
= 0 \, ,
\end{equation}
\begin{equation}
\label{eq:mom_eul}
\frac{\partial (\rho \vec{v})}{\partial t} 
+ \vec{\nabla} \cdot ( \rho \vec{v} \vec{v}) + \vec{\nabla} P = -\rho
\vec{\nabla} \Phi \, ,
\end{equation}
and
\begin{equation}
\label{eq:ene_eul}
\frac{\partial (\rho \varepsilon + 1/2 \rho v^2) }{\partial t}
+ \vec{\nabla} \cdot [ (\rho \varepsilon + P + 1/2 \rho v^2) \vec{v}]
= - \rho \vec{\nabla} \Phi \cdot \vec{v} \, .
\end{equation}

\section{Volume and Weighted-Volume Integrals in 2D}
\label{section:integrals}

In this appendix, we give the exact analytic volumes and
weighted-volume integrals in discrete form.  Since we are dealing with
2D simulations all integrals are 2D integrals, and can be further
reduced using Green's formula to 1D boundary integrals.  For example,
the volume of zone $z$ in Cartesian coordinates is
\begin{equation}
\int_{V_z}dV = \oint_{\partial V_z} x dy
= \sum_e \frac{1}{2}(x_1 + x_2)(y_2-y_1)
\end{equation}
or alternatively,
\begin{equation}
\int_{V_z}dV = - \oint_{\partial V_z} y dx
= - \sum_e \frac{1}{2}(y_1 + y_2)(x_2-x_1) \, ,
\end{equation}
where the direction of the boundary integral is clockwise, $e$ indicates an edge of the cell, and subscripts 1 and 2
represent the endpoints of edge $e$.  In cylindrical coordinates, the
volume integral is
\begin{equation}
\int_{V_z}dV = 2\pi \frac{1}{2}\oint_{\partial V_z} r^2 dz
= 2\pi \sum_e \frac{1}{6}(r^2_1 + r_1r_2 + r^2_2)(z_2-z_1) \, .
\end{equation}

Throughout BETHE-hydro, integrals of linear functions produce volume integrals
weighted by $x$ or $y$.  The discrete form of these integrals are
\begin{equation}
\int_{V_z}x dV = \frac{1}{2}\oint_{\partial V_z} x^2 dy
= \sum_e \frac{1}{6}(x^2_1 + x_1x_2 + x^2_2)(y_2-y_1)
\end{equation}
and
\begin{equation}
\int_{V_z}y dV = - \frac{1}{2}\oint_{\partial V_z} y^2 dx
= - \sum_e \frac{1}{6}(y^2_1 + y_1y_2 + y^2_2)(x_2-x_1) \, .
\end{equation}
Using cylindrical coordinates these are
\begin{equation}
\int_{V_z}rdV = 2\pi \frac{1}{3}\oint_{\partial V_z} r^3 dz
= 2\pi \sum_e \frac{1}{12}(r^3_1 + r^2_1r_2 + r_1r^2_2 + r^3_2)(z_2-z_1)
\end{equation}
and
\begin{equation}
\begin{array}{lcl}
\int_{V_z}z dV & = & 2\pi \frac{1}{2}\oint_{\partial V_z} r^2 z dz \\
& = & 2 \pi \sum_e \frac{1}{6}\left [
(\frac{3}{4}r^2_2 + \frac{1}{2}r_2r_1 + \frac{1}{4}r^2_1)(z_2 - z_1)
+
(r^2_2 + r_2r_1 + r^2_1)z_1
\right ] (z_2 - z_1) \,.
\end{array}
\end{equation}

Finally, for angular momentum remapping, we require
integrals in cylindrical coordinates weighted by $r^2$:
\begin{equation}
\int_{V_z}r^2dV = 2\pi \frac{1}{4}\oint_{\partial V_z} r^4 dz
= 2\pi \sum_e \frac{1}{20}(r^4_1 + r^3_1r_2 + r^2_1r^2_2 + r_1r^3_2 + r^4_2)
(z_2-z_1) \, .
\end{equation}


\clearpage


\begin{deluxetable}{ll}
\tabletypesize{\scriptsize}
\tablecaption{The predictor-corrector procedure.  The operations to
  determine predicted and corrected values are listed in the left and
  right columns, respectively.  See \S
  \ref{section:predictor_corrector} for a detailed
  discussion. \label{table:predictor-corrector}}
\tablewidth{0pt}
\tablehead{ \colhead{Predictor} & \colhead{Corrector}}
\startdata
$P^n \, \& \, \vec{S}^n \longrightarrow \vec{f}^{n}$ &
$P^{n+1/2} \, \& \, \vec{S}^{n+1/2} \longrightarrow \vec{f}^{n+1/2}$
\\
$\vec{f}^n, \, \vec{f}^n_{\rm{visc}}, \, \vec{A}^n_p,
  \, \& \, \vec{g}^n \longrightarrow
  \vec{v}^{n+1,pr}$ &
$\vec{f}^{n+1/2}, \, \vec{f}^n_{\rm{visc}}, \, \vec{A}^{n+1/2}_p,
  \, \& \, \vec{g}^{n+1/2} \longrightarrow
  \vec{v}^{n+1}$
\\
$\vec{v}^{n+1/2} = \frac{1}{2}(\vec{v}^{n+1,pr} + \vec{v}^n)$ &
$\vec{v}^{n+1/2} = \frac{1}{2}(\vec{v}^{n+1} + \vec{v}^n)$
\\
$\vec{x}^{n+1,pr} = \vec{x}^n + \Delta t \vec{v}^{n+1/2}$ &
$\vec{x}^{n+1} = \vec{x}^n + \Delta t \vec{v}^{n+1/2}$
\\
$\vec{f}^n \, \& \, \vec{v}^{n+1/2} \longrightarrow
\varepsilon^{n+1,pr}$ &
$\vec{f}^{n+1/2} \, \& \, \vec{v}^{n+1/2} \longrightarrow
\varepsilon^{n+1}$
\\
$\vec{x}^{n+1,pr} \longrightarrow V^{n+1,pr} \longrightarrow
  \rho^{n+1,pr}$ &
$x^{n+1} \longrightarrow V^{n+1} \longrightarrow
  \rho^{n+1}$
\\
calculate $\vec{g}^{n+1,pr} \, \& \, \vec{A}^{n+1,pr}_p$ &
\\
$e^{n+1,pr} \& \rho^{n+1,pr} \longrightarrow P^{n+1,pr}$ &
\\
$\vec{x}^{n+1/2} = \frac{1}{2}(\vec{x}^{n+1,pr} + \vec{x}^n)$ &

\\
$P^{n+1/2} = \frac{1}{2}(P^{n+1,pr} + P^n)$ &

\\
$\vec{g}^{n+1/2} = \frac{1}{2}(\vec{g}^{n+1,pr} + \vec{g}^n)$ &

\\
$\vec{A}^{n+1/2}_p = \frac{1}{2}(\vec{A}^{n+1,pr}_p
+ \vec{A}^n_p)$ &

\\
\enddata
\end{deluxetable}

\begin{deluxetable}{lll}
\tabletypesize{\scriptsize}
\tablecaption{
Relative error in density, $(\rho_z - \rho_{\rm ana})/(\max(\rho_{\rm ana} ) )$,
at the center for the Goldreich-Weber self-similar collapse test.  The
first column is the time in milliseconds, and the second and third
columns are the error between the simulation, $\rho_z$, and the analytic solution,
$\rho_{\rm ana}$, scaled by the maximum density of the analytic
solution $\max(\rho_{\rm ana})$ for 1D simulation and 2D simulations,
respectively.  See \S \ref{section:gravity_tests} for a discussion.
\label{table:gw}
}
\tablewidth{0pt}
\tablehead{ \colhead{Time (ms)} & \colhead{1D} & \colhead{2D}}
\startdata
0   & $-9.2 \times 10^{-8}$ & $-1.2 \times 10^{-7}$ \\
20  & $-7.2 \times 10^{-4}$ & $-1.4 \times 10^{-3}$ \\
40  & $-1.2 \times 10^{-3}$ & $-2.2 \times 10^{-3}$ \\
60  & $-1.9 \times 10^{-3}$ & $-3.6 \times 10^{-3}$ \\
80  & $-3.2 \times 10^{-3}$ & $-6.0 \times 10^{-3}$ \\
100 & $-6.2 \times 10^{-3}$ & $-1.1 \times 10^{-2}$ \\
120 & $-1.6 \times 10^{-2}$ & $-2.8 \times 10^{-2}$ \\
130 & $-3.6 \times 10^{-2}$ & $-6.3 \times 10^{-2}$ \\
\enddata
\end{deluxetable}

\begin{deluxetable}{lllll}
\tabletypesize{\scriptsize}
\tablecaption{The analytic and simulated orientations of features for the Dukowicz
  problem.  While \citet{dukowicz92} published the angles subtended by
  various regions, we list the angles of each feature
  with respect to the $x$-axis.\label{table:dukowicz}}
\tablewidth{0pt}
\tablehead{\colhead{} & \colhead{Transmitted Shock} & \colhead{Vortex Sheet} & \colhead{Reflected Shock} & \colhead{Incident shock}}
\startdata
Analytic (deg.) &-101.02 & -122.92 & 146.12 & 90 \\
Simulation & -103.21 & -124.42 & 146.92 & 90.32 \\
\enddata
\end{deluxetable}


\clearpage

\clearpage

\clearpage

\clearpage

\clearpage

\clearpage

\clearpage

\clearpage

\clearpage

\clearpage

\clearpage

\clearpage

\clearpage

\clearpage

\clearpage

\clearpage

\clearpage

\clearpage

\clearpage

\clearpage

\clearpage

\clearpage

\clearpage

\clearpage

\clearpage

\end{document}